\documentclass[preprint2]{aastex62}

\usepackage{graphicx}
\usepackage{enumitem}
\UseRawInputEncoding
\newcommand{\Hii}{H \textsc{ii}}

\newcommand{\nthp}{$\rm N_{2}H^+$}
\newcommand{\cth}{$\rm C_{2}H$}
\newcommand{\htcop}{$\rm H^{13}CO^+$}
\newcommand{\hcop}{$\rm HCO^+$}
\newcommand{\hcn}{$\rm HCN$}
\newcommand{\htco}{$\rm H_{2}CO$}
\newcommand{\httco}{$\rm H_{2}^{13}CO$}

\newcommand{\hto}{$\rm H_{2}O$}
\newcommand{\chtoh}{$\rm CH_{3}OH$}
\newcommand{\lam}{$\lambda$ Orionis}

\shorttitle{A chemical probe of stellar feedback on cores in the \lam\ cloud}
\shortauthors{Yi et al.}

\begin{document}

\title{Planck Cold Clumps in the $\lambda$ Orionis Complex. III. \\ A chemical probe of stellar feedback on cores in the \lam\ cloud}

\author[0000-0002-0786-7307]{Hee-Weon Yi}
\author{Jeong-Eun Lee}
\affiliation{Kyung Hee University, School of Space Research, Seocheon-Dong, Giheung-Gu, Yongin-Si, Gyeonggi-Do, 446-701, KOREA}

\author{Kee-Tae Kim}
\affiliation{Korea Astronomy and Space Science Institute 776, Daedeokdae-ro, Yuseong-gu, Daejeon, Republic of Korea}
\affiliation{University of Science and Technology, Korea (UST), 217 Gajeong-ro, Yuseong-gu, Daejeon 34113, Republic of Korea}

\author{Tie Liu}
\affiliation{Shanghai Astronomical Observatory, Chinese Academy of Sciences, 80 Nandan Road, Shanghai 200030, P. R. China}

\author{Beomdu Lim}
\affiliation{Kyung Hee University, School of Space Research, Seocheon-Dong, Giheung-Gu, Yongin-Si, Gyeonggi-Do, 446-701, KOREA}

\author{Ken'ichi Tatematsu}
\affiliation{Nobeyama Radio Observatory, National Astronomical Observatory of Japan, National Institutes of Natural Sciences, 462-2 Nobeyama,
Minamimaki, Minamisaku, Nagano 384-1305, Japan}
\affiliation{Department of Astronomical Science, SOKENDAI (The Graduate University for Advanced Studies), 2-21-1 Osawa, Mitaka, Tokyo
181-8588, Japan}

\author{JCMT Large Program ``SCOPE'' collaboration}

\begin{abstract}
Massive stars have a strong impact on their local environments.
However, how stellar feedback regulates star formation is still under debate.
In this context, we studied the chemical properties of 80 dense cores in the Orion molecular cloud complex
composed of the Orion A (39 cores), B (26 cores), and \lam\ (15 cores) clouds
using multiple molecular line data taken with
the Korean Very Long Baseline Interferometry Network (KVN) 21-m telescopes. 
The \lam\ cloud has an H \textsc{ii} bubble surrounding the O-type star $\lambda$ Ori, 
and hence it is exposed to the ultraviolet (UV) radiation field of the massive star.
The abundances of \cth\ and HCN, which are sensitive to UV radiation, appear to be higher in the cores in the \lam\ cloud than those in the Orion A and B clouds,
while the HDCO to \htco\ abundance ratios show an opposite trend, indicating
a warmer condition in the \lam\ cloud.
The detection rates of dense gas tracers such as the \nthp,\ \hcop,\ and \htcop\ lines are also lower in the \lam\ cloud. 
These chemical properties imply that the cores in the $\lambda$ Orionis cloud are 
heated by UV photons from $\lambda$ Ori.
Furthermore, the cores in the \lam\ cloud do not show 
any statistically significant excess in the infall signature of \hcop\ (1 -- 0), unlike the Orion A and B clouds.
Our results support the idea that feedback from massive stars impacts star formation in a negative way
by heating and evaporating dense materials, as in the \lam\ cloud.

\end{abstract}


\section{Introduction} \label{sec:intro}
Stars are formed by the gravitational collapse of dense cores (n $\geq 10^{5}\, \rm cm^{-3}$) \citep{BT07,MB83,Shu87}.
As a result, the physical and chemical properties of dense cores can provide clues 
to the initial conditions of star formation.
Collapsing cores hosting central protostars are called ``protostellar cores", while cores without any sign of star formation, such as embedded infrared sources or outflows/jets, are known as ``starless cores".

Many recent observational studies have discovered a large number of dense cores
in different star-forming environments \citep{Eden19, Liu18, Tang18, Tate17, Yi18, Yuan16}.
Some of these cores are exposed to the strong far-ultraviolet (FUV) and extreme-ultraviolet (EUV) radiation fields of massive stars.
Under such harsh environments the internal structure and chemical abundances of their natal clouds are modified.
FUV radiation (6 eV $\leq$ $h\nu$ $\leq$ 13.6 eV) dissociates 
molecules in the surroundings and then produces different chemical species.
This radiative feedback can reduce the amount of neutral molecular gas
through ionization, heating, and photo-evaporation,  
which can reduce the star formation rate (SFR) or 
the star formation efficiency (SFE) \citep{Dale12, Dale13, Murray11, Solomon79}.
The negative impact on star formation can eventually terminate star formation activity altogether \citep{Mathieu08}.
\citet{Kruijssen19} recently reported that stellar feedback can shorten 
the cloud lifetime and lower star formation efficiencies to about 2 to 3 \%. 
Furthermore, the radiative and mechanical feedback from the O-type massive stars can increase pressure
(e.g., pressure by radiation, ionized gas, and stellar wind), and it can impact star formation in a negative way by decreasing SFR per unit area \citep{McLeod19}.
A number of theoretical studies have also predicted negative consequences from stellar feedback. 
Simulations by \citet{Dale12,Dale13} found that the overall effect of ionization feedback
at molecular clouds scales was most likely negative, by lowering the SFE over a given time-scale, resulting in fewer stars being born.
In an idealized cloud, photoionization seems to increase the SFE at early times, however, at later times, 
the intense radiation depresses total star formation efficiency via the photoevaporation of dense gas \citep{Shima17}.

In contrast, massive stars can trigger
the formation of new generations of stars in surrounding material, when it is compressed 
by the expansion of an H \textsc{ii} region \citep{EL1977},  
or when pre-existing dense clumps in their strong FUV radiation field become gravitationally unstable \citep{Sandford82,LL94}.
A large number of observational studies have reported the signposts of 
triggered star formation, e.g., the presence of young stellar objects (YSOs)
at the border of an H \textsc{ii} bubble and their age gradient 
\citep{Deharveng03,Deharveng05,Koenig08,Lim14,Liu12a,Sicilia04,Xu17,Zavagno06}.
Recently, a subgroup of YSOs receding away from ionizing sources was found
in the young open cluster NGC 1893 \citep{Lim18}.
However, these signposts are still insufficient to fully explain the physical causality 
between ionizing sources and new generations of stars.

There are several challenges to validating the effects of stellar feedback on star formation.
The objects formed by the feedback are spatially mixed with spontaneously formed objects, and thus it is difficult to distinguish them by observation \citep{Dale15}. 
\citet{Dale15} also demonstrated that variation in SFE may not be a proxy for triggered star formation because the measured gas mass is only a fraction of the cloud mass.
High values of apparent SFE may occur in two different ways, either by efficient star formation or by the efficient removal of molecular gas by photoevaporation and stellar winds \citep{Getman12}.
For these reasons, the impact of stellar feedback on star formation is still under debate.

In this context, 
the Orion molecular cloud complex (OMC) is an ideal testbed to examine 
the effects of stellar feedback on surroundings and subsequent star formation. 
This star-forming complex is composed of three giant molecular clouds (diameter $>$ 30 pc),
the Orion A, B, and $\lambda$ Orionis clouds, located
at similar distances of 350 -- 450 pc \citep{Kounkel17,Getman19,Zucker19}. 
The Orion A and B clouds are well-known active star-forming regions on a scale of 15 pc and contain a large number of low- and high-mass stars \citep{Megeath12} 
with a total mass greater than 2 $\times$ 10$^5$ $M_{\sun}$ \citep{Wilson05}. 
The $\lambda$ Orionis cloud hosts one of the most prominent OB associations 
in the OMC and its total mass is about 1.4 $\times$ 10$^4$ $M_{\sun}$ \citep{Lang2000}. 
The main ionizing source is the O8 III star $\lambda$ Ori.
The high UV energy from $\lambda$ Ori created a giant H \textsc{ii} region Sh 2-264 with a radius of 34.5 pc \citep{BN05,MM87}.
A ring structure surrounding the \Hii\ region was shaped by a supernova explosion, 
which is considered to be caused by a binary companion of $\lambda$ Ori about several million years ago \citep{CS96, DM02,Hernandez10,Kounkel18}.
\citet{Lang2000} found a low-level of star-forming activity along the inner ring of Sh 2-264 and 
concluded that the ring may not be a site of triggered star formation
because only a small number of YSOs are associated with this ring.
\citet{Mathieu08} argued that the supernova explosion terminated star formation in the \lam\ cloud, 
and that the intense radiation from remaining massive stars destroys accretion disks around YSOs.
Thus, the $\lambda$ Orionis cloud appears to be a star-forming region dominated by negative stellar-feedback, 
while the Orion A and B clouds are free from the negative effects of feedback, since they have different star formation histories.

In our previous paper  \citep[hereafter Paper I]{Yi18}, the physical properties (size, column density, mass, and number density)
 of 119 cores in the OMC were studied using the 850 \micron \ dust continuum image.
We found evidence that core formation was suppressed in the $\lambda$ Orionis cloud; 
specifically, the core formation efficiency (CFE) in the $\lambda$ Orionis cloud (0.10) is lower  
than those in the Orion A (0.12), and B (0.20) clouds.
Recently, the Nobeyama single-pointing survey observed 207 dense cores in five different environments 
($\lambda$ Orionis, Orion A, Orion B, the Galactic plane, and high latitudes) to study the initial conditions of star formation \citep{Kim2020}.
The deuterium fraction of starless cores in the \lam, Orion A and B clouds was not significantly different, suggesting there is no systematic difference in the chemical properties among the three clouds \citep{Kim2020}.
In this study, we aim to further investigate the chemical differences in the context of the negative effects of stellar feedback 
on 80 cores, using multiple molecular line data 
obtained with the Korean Very Long Baseline Interferometry (VLBI) Network (KVN) 21­-m telescopes.

Spectral lines detected in cores are useful tools for probing different star-forming conditions.
Deuterium fractions (D/H ratios) and the abundances of various molecules in the three clouds 
were compared to study the environmental effects on dense cores.
In addition, we probed the signatures of ongoing star formation (jets driven by protostars and infall motions in cores).
Sample selection and observations are described in Section \ref{sec:Observation}. 
In Section \ref{sec:Results}, we present the detection rates, line widths, and abundance ratios of some molecules.
In Section \ref{sec:infall}, we investigate the infall signature, which is another tracer of star formation activities, in each cloud.
In Section \ref{sec:Discussion} we discuss the results of this work and our previous study in terms of the effects of negative stellar feedback,  
and summarize our main results in Section \ref{sec:Summary}.

\section{Data} \label{sec:Observation}
\subsection{Sample selection} \label{subsec:sample}
In Paper I, we identified a total of 119 cores using 850 \micron\ dust continuum data.
The list of the cores was used to select targets of observation. 
The Orion A and B clouds are active star-forming regions and several cores are clustered within the filamentary clumps \citep{Yi18}. 
We thus selected cores with separation larger than 32$\arcsec$ from the others considering the beam sizes of
21-m telescopes at about 90 GHz (Table \ref{tab:1}).
A total of 39, 26, and 15 cores are selected in the Orion A, B, and \lam\ clouds, respectively.
Our sample contains a total of 80 cores in the three clouds. 

We divided these cores into two groups, protostellar and starless cores. 
The former has infrared sources that were classified using 
their spectral index ($\alpha$) 
and bolometric temperature ($T_{\rm bol}$) (see section 3.3.3 of Paper I).
The latter does not show any observational sign of star formation.
 The number of protostellar and starless cores are 5 and 10 in the \lam\ cloud, respectively. 
The sample for the Orion A cloud contains 13 protostellar cores and 26 starless cores.
In the Orion B cloud, 7 protostellar cores and 19 starless cores were selected.
The number ratios of protostellar to starless cores are roughly 0.5 in the  \lam\ and Orion A clouds 
and 0.4 in the Orion B cloud.

\subsection{Observation and Data reduction of KVN}
We carried out single-dish observations with the KVN 21-m telescopes 
at the Yonsei, Ulsan, and Tamna stations \citep{Kim2011, Lee2011}.
A multi-frequency receiving system is attached to each telescope with the 22, 44, 86, and 129 GHz bands  \citep{Han08}.
The observations were made from 2016 August to 2017 March using several receivers simultaneously. 
All the 80 cores were observed in single-pointing mode with 
the rms level of $T_{A}^{*}$ lower than 0.1 K at a velocity of $\sim$0.1 km/s (channel width is 0.05 km/s).
The $J=1-0$ transitions of  five molecules (\nthp, HCO$^+$, H$^{13}$CO$^+$, C$_2$H, and HCN), including SiO thermal line ($v=0, \, J= 1-0$), 
two masers (CH$_3$OH  7$_{0} \rightarrow 6_{1} \, A^+$ and  
H$_2$O  6$_{16} \rightarrow 5_{23}$), and  H$_2$CO (2$_{1,2} \rightarrow 1_{1,1}$), 
H$_2$$^{13}$CO (2$_{1,2} \rightarrow 1_{1,1}$),
HDCO  (2$_{0,2} \rightarrow 1_{0,1}$) lines were observed in the position switching mode. 
We conducted multi-frequency observations in four receivers with 64 MHz bandwidth and 4096 channels,
which give a spectral channel width of 15.625 kHz.
Table \ref{tab:1} lists the observed molecular lines, beam sizes, main-beam efficiencies ($\eta_{\rm mb}$), and system temperatures. 
The observed antenna temperature, $T_{A}^*$, 
was converted to main-beam brightness temperature as $T_{\rm mb} = T_{A}^ *$$ / \eta_{\rm mb}$. 
The focus and pointing were checked every two to three hours 
by observing the strong SiO maser source Orion KL.  
All the spectra were reduced with the GILDAS/CLASS package. 
A first-order polynomial baseline subtraction was performed for the majority of the observed sources, except for some sources, which need a higher order of polynomial baseline fitting.
For these sources, we carried out a second-order polynomial baseline subtraction. 
A Hanning smoothing technique was then applied 
to the spectra to achieve a velocity resolution of 0.1 km s$^{-1}$.
To characterize the spectral features, 
we fit a Gaussian profile to each observed spectrum.
The peak temperature at the $T_{A}^*$ scale, systemic velocity, and full width at half maximum (FWHM) inferred by Gaussian fitting to each spectrum are listed in Table \ref{tab:2}-\ref{tab:3}.

\subsection{James Clerk Maxwell Telescope (JCMT/SCUBA-2)}
As part of the legacy survey, Submillimetre Common-User Bolometer Array 2 (SCUBA-2) 
Continuum Observations of Pre-protostellar Evolution (SCOPE) project observed
dust continuum emission of about 1000 of dense clumps (N(H$_{2}$) $>$ 5$\times$10$^{21}$ cm $^{-2}$) 
at 850 $\micron$ \citep{Liu18,Eden19}. 
The list of the clumps in this survey was taken from the catalog of the Planck Galactic Cold Clumps (PGCCs),
which contains all the cold dust clumps detected by the $\it Planck$ mission, 
in combination with the $\it IRAS$ 100 $\micron$ data \citep{Planck16}.
From the SCOPE project, we obtained the 850 $\micron$ dust continuum emission maps
to estimate column densities of H$_{2}$ molecules toward the 80 cores in our sample.
The main-beam size of SCUBA-2 at 850 $ \micron$ is about 14$\arcsec$. 
To derive abundances of given molecules relative to molecular hydrogen 
(X(molecule)=[N(molecule)/N(H$_{2}$)]),
the 850 $ \micron$ continuum maps were convolved to the same main-beam size of the KVN telescopes $\sim 30\arcsec$ for 
\nthp, \hcop, \htcop, HCN, and \cth, and $\sim$ 20$\arcsec$ for \htco\ and HDCO line data
using the starlink kappa package.
The detailed explanations for observations and data reduction of the original SCUBA-2 data can be found in Paper I.

\subsection{Source classification} \label{subsec:classify}
In Paper I, we divided cores into starless and protostellar cores using 
the Wide-field Infrared Survey Explorer $(\it WISE)$ photometric data provided by the AllWISE data release \citep{Cutri14}, 
and then classified the evolutionary stages of the individual protostellar cores. 
Since the sensitivity and spatial resolution of the $\it Herschel$ data are better than those of the $\it WISE$,
we used the $\it Herschel$ Orion Protostar Survey 
(HOPS) catalog of cores in the Orion A and B clouds to confirm 
the previously classified protostellar and starless cores.

In the case of the \lam\ cloud, the census of YSOs is somewhat incomplete due to the absence of 
high-sensitivity and high spatial resolution mid-infrared data.
A SiO thermal emission line can be an alternative way to classify protostellar cores
in the $\lambda$ Orionis cloud  
because this emission line traces protostellar jets \citep{Lee2017}.
The SiO emission line was detected only in two cores classified as starless in Paper I: G191.90-11.21N and G192.32-11.88N.
These cores are located in the quiescent and isolated region of the cloud,  
and therefore there is no contamination source that can drive SiO emission.
Thus, G191.90-11.21N and G192.32-11.88N were re-classified as 
protostellar cores containing Class 0 YSOs (appendix \ref{subsec:SiO}).

In the Orion A and B clouds, 
there is no starless core that has been re-classified by the molecular line data
because none of them showed SiO thermal or maser emission.
Using the HOPS catalog, we re-classified four starless cores 
(G208.68-19.20N1, G208.68-19.20S, G209.55-19.68N1, and G210.97-19.33S2)
as protostellar cores in the Orion A cloud because these cores host 
either Class 0 or Class I objects.
In the case of the Orion B cloud, four starless cores 
(G205.46-14.56M1, G206.12-15.76, G206.93-16.61W3, and G206.93-16.61W6) 
were re-classified as protostellar cores with Class 0 YSOs. 

In summary, a total of ten starless cores were re-classified as protostellar cores
in the \lam\ (2), Orion A (4), and B (4) clouds. 
The ratio of protostellar cores to starless cores is 0.9 in the $\lambda$ Orionis, 0.8 in the Orion A, and 0.7 in the Orion B clouds. 
These comparable core ratios allow us to perform an unbiased study of the chemical properties of cores.
Table \ref{tab:4} summarizes the numbers of protostellar, starless, and total cores in the three clouds.

\section{Results} \label{sec:Results}
We investigated various molecular tracers and abundance variations in cores within the three clouds
to understand the chemical properties of the cores under different levels of stellar feedback.
In this section, the detection rates, line widths, and abundance ratios of the observed molecules are presented.

\subsection{Detection rates}   \label{subsec:detection}
\begin{figure}[ht!]
\plotone{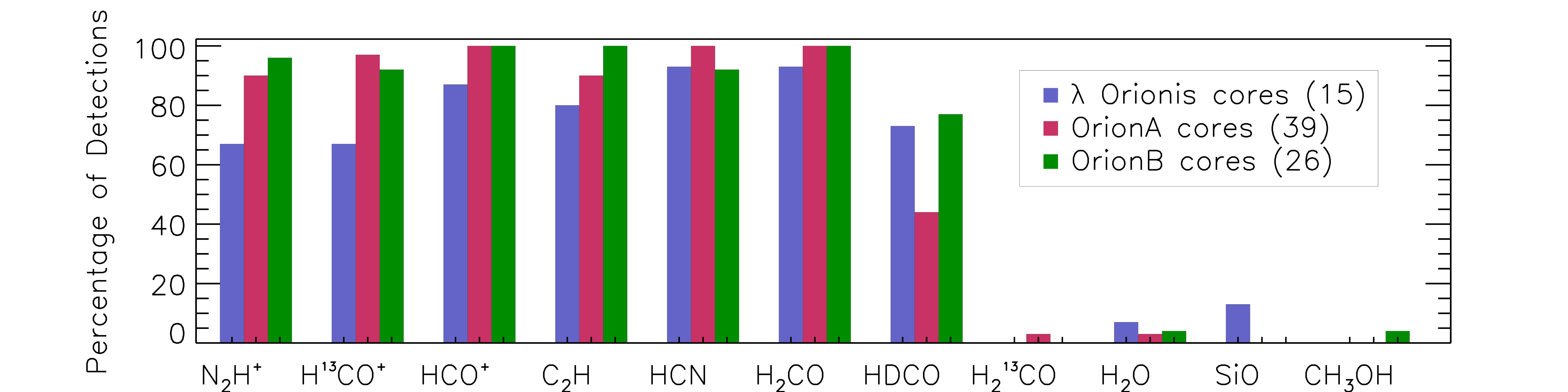}
\caption{Detection rates of the observed molecular lines toward 15 cores in the $\lambda$ Orionis cloud, 
39 cores in the Orion A, and 26 cores in the Orion B clouds.
Different colors denote different regions (blue: $\lambda$ Orionis cloud, magenta: Orion A cloud, green: Orion B cloud).
\label{fig:1}}
\end{figure}

A total of 11 molecular spectral lines were observed for 80 dense cores in the OMC.
We considered the line detected only when the signal-to-noise ratio (SNR) of a spectrum was greater than 3.
The detection rate of a molecular line is defined by the number of cores with the line detection relative to the total number of cores (N$_{\rm detection}$/N$_{\rm total}$).
Figure \ref{fig:1} displays the detection rates of the individual lines in the three clouds (see also Table \ref{tab:5}).
Cores in the \lam\ cloud tend to have lower detection rates than those in the other two clouds.

\hcop, \htcop, and \nthp\ are well-known tracers of dense gas \citep{Ikeda07,Kauffmann17,Lee04,Pety17}. 
\nthp\ in particular is not easily depleted at low temperature and high density.
The mean H$_{2}$ column density of cores in the \lam\ cloud is lower  than those of
the Orion A and B clouds (Paper I), and thus the detection rates of these molecular lines are also expected to be lower in the \lam\ cloud.
Their detection rates are indeed the lowest in the \lam\ cloud (87, 67, and 67 \% for the \hcop, \htcop, and \nthp\ lines, respectively).

The two \cth\ and HCN lines have been used to trace photodissociation regions (PDRs) \citep{Cuadrado15,Fuente93,Watanabe12}, 
where FUV photons govern the structure, dynamics, chemistry and thermal balance of the cloud as well as the star formation \citep{TH85, Hollenbach91,HT99}.
Massive stars in the \lam\ cloud emit high energy UV photons, which can easily photoionize or photodissociate molecules.
The detection rates of \cth\ and HCN lines are thus expected to be the highest in the \lam\ cloud.
However, the detection rate of the \cth\ emission line was found to be higher in the Orion A (90 \%) and B (100 \%) clouds 
than in the \lam\ cloud (80 \%), while the detection rate for HCN 
was very high and comparable in all three clouds (\lam: 93 \%, Orion A: 100 \%, Orion B: 92 \%).
Recent studies have suggested that \cth\ and HCN are prevalent molecules  in molecular clouds and dense interstellar medium \citep{LL01,Loison14,Liszt18}.
The ubiquity of the two molecules probably makes it difficult to find their origin of emission.
The abundances of \cth\ and HCN and their origins are discussed in Section \ref{subsubsec:PDR1}.

 H$_{2}$CO is abundant in the general interstellar medium.
 The detection rates of this molecular line are about 93\% in the $\lambda$ Orionis and 100 \% in the Orion A and B clouds.
The optically thin isotopologue line, \httco\ 2$_{1,2} - 1_{1,1}$ has been detected toward only one core.
The derived \httco\ column density of the core is 7.56 $\times10^{11}$ cm$^{-2}$.
Given the detection limit of spectral lines (SNR $>$ 3), all the non-detected cores may have \httco\ column densities smaller than $\sim$ 5$\times$10$^{11}$ cm$^{-2}$.
The HDCO line, a deuterated species of H$_{2}$CO, shows variations in detection rates among the three clouds. 
Its detection rates are 44\% and 77\% in the Orion A and B clouds, individually, while 73\% in the \lam\ cloud.

We detected an \hto\ maser emission toward three cores (G196.92-10.37, G210.49-19.79W, and G205.46-14.56S) and a \chtoh\ maser emission in one core (G205.46-14.56S).
The 22 GHz \hto\ and 44 GHz class I \chtoh\ masers are usually associated with inner jets and outflow lobes, respectively \citep[e.g.,][]{Kurtz04,Kim18}. 
The three \hto\ maser-emitting cores are located in the three clouds, respectively, and the sole \chtoh\ maser-detected core, 
which also emits \hto\ maser emission, is located in the Orion B cloud.
The SiO thermal line ($v=0, \, J= 1-0$) was detected in two cores 
(G191.90-11.21N and G192.32-11.88N) 
within the $\lambda$ Orionis cloud, while
the line was not detected in either of the Orion A and B clouds.
We discuss the detailed properties of these five cores in Appendix \ref{sec:maser}.

\subsection{Line widths}    \label{subsec:widths}
Since the internal motion of gas broadens spectral lines, 
line width provides information about velocity dispersion 
caused by both thermal and non-thermal motions of gas in the cloud. 
Most molecular clouds have velocity dispersions greater than 
the predicted thermal velocity dispersion ($\sigma_{\rm T}$) \citep{WW11}.
This fact implies the presence of non-thermal motions related to 
turbulence and bulk motions within the cloud.
In cores with specific radii (0.01 $< R_{\rm core} < $ 0.1 pc), 
non-thermal velocity dispersion ($\sigma_{\rm NT}$) is greater than $\sigma_{\rm T}$ \citep{WW11}.
Most of the cores in our sample were smaller than 0.1 pc (Paper I), and the line widths may be governed by non-thermal motions. 

The non-thermal velocity dispersion is calculated using the equation below.
\begin{equation}
\sigma_{\rm NT} = (\sigma_{\rm obs}^{2} - \sigma_{\rm T}^2)^{1/2} = \left[\frac{\Delta V^{2}}{8\ln(2)} - \frac{kT_{\rm ex}}{m_{\rm obs}}\right]^{1/2}, 
\end{equation}
where $\Delta V$, $m_{\rm obs}$, and $T_{\rm ex}$ represent the FWHM of the molecular line profile, the molecular mass, and the excitation temperature of the gas, respectively. 
The line widths (FWHM) are obtained from the best-fit Gaussian profile and $T_{\rm ex}$ is assumed to be equal to the dust temperature $T_{\rm d}$ (see Appendix \ref{sec:CD}).   
All cores have much larger $\sigma_{\rm obs}$ than $\sigma_{\rm T}$, indicating that they have greater non-thermal motions than thermal motions, as shown by previous works \citep{Liu12b, Servajean19, TH20}.
The median values of  $\sigma_{\rm NT}$ and $\sigma_{\rm T}$ are summarized in Table \ref{tab:6}.

Figure \ref{fig:fig2} displays the line widths (${\Delta V}$) of six frequently detected molecular lines 
(\hcop, \htcop, \cth, HCN, \htco, and HDCO) with respect to that of N$_2$H$^+$. 
The ${\Delta V}$ of the $\rm HCO^+$, HCN, \cth, and H$_{2}$CO lines show large spreads,
while those of \htcop\ and HDCO exhibit relatively small dispersions (Figure \ref{fig:fig2}).
We also present the median value of $\Delta V$ ($\Delta V_{\rm mol, med}$) of the seven molecular lines 
for the three clouds in Table \ref{tab:7}.
The Orion A and B clouds show a larger $\Delta V$ in four molecular lines (\hcop, \cth, HCN, and \htco)
than in the \lam\ cloud.

The $\Delta V_{\rm mol, med}$ of HDCO ($\Delta V_{\rm HDCO, med}$) and \nthp\ ($\Delta V_{\rm N_{2}H^{+}, med}$) 
are the same (0.5 km s$^{-1}$) in the \lam\ cloud (Table \ref{tab:7}).
The two values are indistinguishable in the Orion A and B clouds, as well, considering the velocity resolution of 0.1 km s$^{-1}$.
Figure \ref{fig:fig2} shows a tight correlation between the HDCO and N$_2$H$^+$ line widths in all three clouds.
Since the \nthp\ line traces cold and dense gas, the HDCO line also seems to trace the cold and dense core centers.
Although $\Delta V_{\rm H^{13}CO^+, med}$ is slightly larger than those of \nthp\ and HDCO in each cloud,
the $\Delta V_{\rm H^{13}CO^+, med}$ among the three clouds did not show a significant difference.
Thus, the \htcop\ line emission seems to be a good tracer for the dense part of cores
\citep{Ikeda07, Onishi02, Sanhueza12, Yoo18}.
In summary, \nthp, HDCO, and \htcop\ had similar line widths in the three clouds (Table \ref{tab:7}), suggesting that  
these emission lines are less affected by turbulence and trace the densest part of cores.

The $\Delta V_{\rm mol, med}$ of four other molecules (\hcop, HCN, \cth, and H$_{2}$CO) were
larger than that of \nthp\ in all three clouds.
This is probably because the four molecular lines are optically thick, and they trace different regions from \nthp.
In particular, \cth\ molecules are abundant in PDRs because
this molecule can form through the photodissociation of acetylene
(C$_{2}$H$_{2}$): C$_{2}$H$_{2}$ + $h\nu$ $\rightarrow$  C$_{2}$H + H  \citep{Fuente93,Cuadrado15}. 
These observational results suggest that
the \cth\ molecule traces a PDR or the outer regions of cores rather than the dense gas of core centers.
The \hcop\ and HCN emission lines are good tracers of star-forming activities (e.g., outflow or jet) 
\citep{Michiyama18,Rawlings04, Riffel10,WS14}. 
Star-forming activities can generate turbulence \citep{Plunkett15}, and this can then increase the $\Delta V$ of the two molecular lines,
particularly in the Orion A and B clouds.
The HCN line shows the largest difference in $\Delta V_{\rm med}$ among the three clouds (Table \ref{tab:7}).
In addition, the \htco\ emission line is considered to be a tracer of outflow or jet \citep{Bachiller97}, and the $\Delta V_{\rm med}$  of \htco\ is also larger in the Orion A and B clouds than those in the \lam\ cloud. 
These facts further support that the star-forming activities contribute to the line broadening of \hcop\ and HCN molecules.
We will discuss more detailed properties of \hcop, H$_{2}$CO, \cth, and HCN molecules in Section \ref{sec:Discussion}.

\subsection{Chemical abundances} \label{subsec:abundance}
The chemical abundances of cores and their evolution from starless to protostellar stages are greatly affected by the intense radiation field produced by massive stars.
The FUV radiation forms PDRs in surrounding molecular clouds, where the gas is heated up via the photoelectric effect \citep{HT99}.
In order to study how UV radiation changes the chemical abundances of cores, we calculate the molecular abundances and abundance ratios of selected species.

\subsubsection{Abundances of PDR tracers} \label{subsubsec:PDR1}
We focus on the effects of stellar FUV radiation on cores in each cloud in this section.
In PDRs, the abundance of some specific molecules, such as \cth\ and HCN, are enhanced \citep{Hasegawa01,Sternberg95}
by photodissociation or dissociative recombination. 
To test whether the UV radiation induces abundance enhancements of those molecules, 
we compared the abundances of \cth\ (X(C$_{2}$H)), HCN (X(HCN)), and \hcop\ (X(\hcop)) in the \lam\ cloud with those in the other two clouds.
For the abundance calculations, 
the H$_{2}$ column density was derived from the 850 $\micron$ dust continuum emission obtained from the JCMT/SCUBA-2 images (see Paper I for detail). 
The median and mean abundance of the observed molecules are summarized in Table \ref{tab:8}.

Several FUV-involved pathways are available for \cth\ formation.
The final step in the formation of \cth\ can be either the photodissociation of $\rm C_{2}H_{2}$ or 
dissociative recombination of $\rm C_{2}H_{3}^{+}$ ($\rm C_{2}H_{3}^{+}$ + $e^{-}$ $\rightarrow$ \cth\ + H$_{2}$ ) and 
$\rm C_{2}H_{2}^{+}$ ($\rm C_{2}H_{2}^{+}$ + $e^{-}$ $\rightarrow$ \cth\ + H) \citep{Bergin16,Nagy15,Fuente93}. 
In any case, the FUV radiation plays a key role in providing photons and generating electrons.
The median X(\cth) in the $\lambda$ Orionis cloud is three times higher (4.5 $\times$ 10$^{-8}$) than those in the Orion A (1.6 $\times$ 10$^{-8}$) and B (1.5 $\times$ 10$^{-8}$) clouds.
Since the extinction of the UV radiation by H$_{2}$ is expected to be less effective in the \lam\ cloud (because of the lower H$_{2}$ column density, Paper I),
the high X(\cth) in the $\lambda$ Orionis cloud is probably caused by the active photolysis of $\rm C_{2}H_{2}$.

The HCN molecule can form in three ways
through dissociative recombination 
(HCNH$^+$ + $e^-$ $\rightarrow$ HCN + H), 
charge transfer (HCN$^+$ + H $\rightarrow$ HCN + H$^+$), and 
neutral-neutral reaction (CH$_{2}$ + N $\rightarrow$ HCN + H)
 \citep{Sternberg95,Hirota98,Hasegawa00,Boger05}.
In the dissociative recombination, once HCNH$^+$ is recombined with free electrons, then HCN abundance rapidly increases.
The sequence of HCN formation also depends on the abundance of ``free carbon'', either as C$^+$ atomic ions or C atoms.
The incident FUV field maintains an extended cloud layer where most of the carbon-bearing molecules are singly ionized.
This layer extends deeper into the cloud than does the H \textsc{i} zone because of the effective self-shielding of hydrogen molecules.
Thus, C$^+$ atomic ions can exist in condition where the hydrogen is fully molecular at cloud depths of $A_{V}$ = 0.7 to 1.7 mag \citep{Sternberg95}.
The $A_{V}$ through the \lam\ cloud is about 1.25 mag \citep{Goldsmith16}, which means that the C$^+$ atomic ion can exist all over the cloud.
The C$^+$ ions also play an important role in the formation of HCN$^+$ and CH$_{2}$ \citep[][see Section 3.1 and 3.2]{Boger05}, which are directly involved in the formation of  HCN.
The median values of X(HCN) are 7.6 $\times$ 10$^{-10}$, 3.3 $\times$ 10$^{-10}$, and 4.0 $\times$ 10$^{-10}$ in the \lam, Orion A, and B clouds, respectively.
All these chemical reactions are possible explanations for the enhanced HCN abundance in the \lam\ cloud.
Another exothermic neutral-neutral reaction with an activation barrier of about 760 K accounts for 50 to 70\% of the total HCN production in PDRs: CN + H$_{2}$  $\rightarrow$ HCN + H ($E_{a} =$ 760 K) \citep{Hasegawa00}.
This process can drive a high abundance of HCN in the \Hii\ region but is not valid in cores due to their low temperatures.

The cumulative distributions of X(C$_{2}$H) and X(HCN) are plotted in Figure \ref{fig:3}.
The high abundances of \cth\ and HCN in the \lam\ cloud are consistent with those
anticipated from the active photodissociation and dissociative recombination processes.
The correlation between the integrated intensity of the \cth\ and HCN lines can tell whether they have formed through a similar process.
The HCN integrated intensity is obtained using the weakest $F$ = 0 $\rightarrow$ 1 hyperfine component to avoid a self-absorption dip at the systemic velocity.
Figure \ref{fig:4} shows a clear correlation between the two integrated intensities of the \cth\ and HCN lines, 
suggesting UV radiation is involved in the formation of the two molecules.

Another representative species of PDRs is  \hcop\ \citep{Hasegawa00,Hasegawa01}.
However, the \hcop\ abundance in the three clouds appears to be similar, unlike the \cth\ and HCN abundances (see Table \ref{tab:8}).
The \hcop\ formation is initiated by singly ionized carbon and OH \citep{Hasegawa00}.
The main channels of formation and destruction of \hcop\ are predicted to be
\[
\rm CO^+ + H_2 \rightarrow \rm HCO^+ + H   \quad (formation) \]
\[
\rm HCO^+ + e^- \rightarrow \rm CO + H  \quad  (destruction)\]
\[
\rm HCO^+ + h\nu \rightarrow \rm CO^+ + H  \quad (destruction). \]

In the \lam\ cloud, the destruction process of \hcop\ could be dominated by photodissociation.
Star-forming activities such as the jets/outflows of adjacent protostars can also increase \hcop\ abundance, as was suggested in previous studies \citep{Michiyama18,Rawlings04, Riffel10,WS14}.
In the Orion A and B clouds, the abundance of \hcop\ is enhanced by the star-forming activities in the protostellar cores.
The similarity of \hcop\ abundance among three clouds implies that
the \hcop\ line is not a reliable tracer of PDRs, unlike the \cth\ and HCN molecular lines.

\subsubsection{D/H ratios of protostellar and starless cores} \label{subsubsec:D/H_1}
In this section, we examine the deuterium fractions of \htco\ in the three clouds.
The \htco\ molecule is already formed during the pre-collapse phase (n $< 5\times10^{4}$ cm$^{-3}$) \citep{Aikawa12}.
Deuterated \htco\ (HDCO) molecules form in the ice-mantles of dust grains via the hydrogenation of CO ice \citep{Fontani15, Walsh15}, 
and the deuterium fractions in the cores are mostly settled before their collapse begins \citep{Aikawa12,Roberts02, Turner1990}.
Thus, the [HDCO]/[\htco] ratio is mostly
maintained during  the early evolutionary phase of cores, 
where the dust temperature is lower than the sublimation temperatures of the molecules \citep{Codella12}.
However, the UV photons from massive (proto)stars can also desorb molecules from dust grains via photodesorption \citep{TH82, Hasegawa92, Walsh10}.

Other molecules, such as \nthp\ and HNC, 
and their deuterated isotopologues ($\rm N_{2}D^{+}$ and DNC) have also been used to measure the D/H ratio in massive star-forming regions \citep{Fontani15} and the Orion region \citep{Kim2020}.
However, \citet{Kim2020} did not find significant differences in the D/H ratios of starless cores in the Orion region (the \lam, Orion A, and B clouds).
The $\rm N_{2}D^{+}$ and DNC molecules can form in the gas phase, mainly from a route involving the $\rm H_{2}D^{+}$ ion \citep{Gerlich02,Parise09}.
However, the $\rm H_{2}D^{+}$ ion  can be destroyed by electrons and then dissociate into the H$_{2}$ molecule and D atom \citep{Datz95,Lee03}.
The deuterium atom (produced by the dissociative recombination of deuterated ions) then becomes frozen onto grain surfaces to form molecular ices, resulting in a high abundance of HDCO via grain-surface hydrogenation \citep{Tielens1983, Aikawa12, LB15}.
Thus, the different formation processes of deuterated molecules can lead to relevant differences in their D/H ratios \citep{Aikawa12}; 
the freezing of deuterium onto grain surfaces could reduce the efficiency of deuteration in a molecule in the gas phase \citep{LB15}.

We present the mean D/H ratio of \htco\ in protostellar and starless cores in Table \ref{tab:9}.
The mean values in protostellar cores are similar in the three clouds.
In starless cores, the mean D/H ratio is higher by a factor of 1.4 in the Orion A (0.079 $\pm$ 0.006) and B (0.081$\pm$ 0.006) clouds compared to the \lam\ cloud (0.056 $\pm$ 0.005).
The low D/H ratio of starless cores in the \lam\ cloud indicates that they are probably warmer
than the starless cores in the other two clouds.

Figure \ref{fig:5} shows the distributions of the D/H ratios of \htco\ in each cloud.
The ratios range from 0.01 to 0.26, which are consistent with values derived from 
the low-mass Class 0 protostars from 0.03 to 0.31 in the Orion A and B clouds \citep{Kang15}.
The red dotted lines represent the D/H ratio of 0.10; four starless cores exhibit the D/H ratios $>$ 0.1 in each
the Orion A (G207.36-19.82N3, G209.55-19.68S2, G209.94-19.52S1, and G211.16-19.33N3) and 
B (G203.21-11.20E1, G203.21-11.20W1, G204.4-11.3A2W, and G206.69-16.60N) clouds, 
while none of starless cores show a high ratio in the \lam\ cloud. 
The fractions of cold starless cores with a D/H ratio greater than 0.1 are 0 (0/6), 0.36 (4/11), and 0.27 (4/15) in the \lam, Orion A, and B clouds, respectively.

The low D/H ratio of \htco\ derived from the starless cores in the \lam\ cloud is a consistent result with its higher temperature (16 K) compared to
the temperatures (13-14 K) in the Orion A and B clouds (Paper I). 
This low D/H ratio, resulting from the warm condition prior to the collapse phase, has been hinted in hot molecular cores, 
where the molecules are sublimated thermally from the grain surfaces around massive young stars \citep{Loinard02}

\subsubsection{Relation of D/H ratios and $L_{\rm bol}$, $T_{\rm bol}$, and M$_{\rm core}$/$L_{\rm bol}$} \label{subsubsec:D/H_2}
To examine the chemical evolutionary stages of protostellar cores,
we investigated variations in D/H ratios with bolometric luminosity ($L_{\rm bol}$), temperature ($T_{\rm bol}$), and 
mass-to-luminosity ratio ($M_{\rm core}/L_{\rm bol}$).
CO molecules are highly depleted in the early evolutionary stage under the low temperature.
In the more evolved stage of protostellar cores, CO and \htco\ are released
from the dust grains by shocks and sublimation at $T \gtrsim$ 17 K \citep{MD14}.
However, deuterium fractionation is not affected by shocks, mantle evaporation, or accretion after the Class 0 stage \citep{Lee04,Roberts02}.

The $L_{\rm bol}$ is directly related to the accretion rate of a central protostar in the early embedded evolutionary stage.
Since the accretion shocks heat the envelope, an anti-correlation is expected between the D/H ratios and $L_{\rm bol}$.
In contrast, the core mass decreases with time,
so that the mass-to-luminosity ratio would decline as protostars evolve. 
$T_{\rm bol}$ is also a useful evolutionary indicator of young stellar objects.
We have calculated the Pearson correlation coefficient to investigate the relation between the D/H ratio of \htco\ and those physical parameters ($L_{\rm bol}$, $T_{\rm bol}$, and $M_{\rm core}/L_{\rm bol}$).

Figure \ref{fig:6} shows the D/H ratios of \htco\ for 15 protostellar cores in the three clouds with respect to $L_{\rm bol}$, $T_{\rm bol}$, and $M_{\rm core}/L_{\rm bol}$.
There is no clear correlation between the ratio and $T_{\rm bol}$ as suggested by \citet{Kang15} and \citet{Roberts02}. 
The D/H ratio shows an anti-correlation with the $L_{\rm bol}$, while
it reveals a linear correlation with the $M_{\rm core}/L_{\rm bol}$. 
Most of the detected cores follow the trends within a standard deviation of 0.01. 
If the mass-to-luminosity ratio is a good tracer of evolutionary stage, the D/H ratio seems to well trace the early stage of star formation.
The protostellar cores  in the \lam \ cloud  have a relatively larger $M_{\rm core}/L_{\rm bol}$
than those in the Orion A and B clouds.
This indicates that the five protostellar cores in the \lam \ cloud are likely in the early stage of star formation.
Since the intense FUV photons heat up the molecular gas and dust particles, 
the transition of cores from prestellar to protostar stages is impeded in the \lam\ cloud.
Therefore, the chemical timescales in the prestellar or protostellar stage in the \lam\ cloud might differ from those in the Orion A and B clouds because of the different dynamical timescale, 
as well as the direct effect of the enhanced FUV radiation environment.
We further explore the chemical timescale of cores in each cloud, depending on the FUV radiation field in the next Section \ref{subsubsec:N/O}.

\subsubsection{\nthp\ to \hcop\ ratios}  \label{subsubsec:N/O}
To estimate the chemical ages of the cores, we probed the \hcop\ and \nthp\ abundances in the cores.
The CO molecules are desorbed from grain mantles under high temperature ( $>$ 20 K) conditions.
The CO desorption causes the significant destruction of $\rm  N_{2}H^+$ molecules, 
leading to the formation of $\rm HCO^+$ \citep{Jorgensen04, Lee04}. 
Several previous studies have  suggested the $ N(\rm N_{2}H^{+})$/$N(\rm HCO^+)$ ratio as a chemical clock because 
the HCO$^+$ abundance increases with respect to $\rm  N_{2}H^{+}$ as a core evolves to a warmer phase \citep{TT1977, Sanhueza12, Ohashi14}.
From this reaction, we expect a higher $ N(\rm N_{2}H^{+})$/$N(\rm HCO^+)$ ratio in the \lam \ cloud because 
the timescale of the prestellar stage in the cloud is considered to be longer than those in the Orion A and B clouds.

Figure \ref{fig:7} shows the relation between the column densities of \nthp\ and \hcop\ in each cloud.
The median values of the $ N(\rm N_{2}H^{+})$/$N(\rm HCO^+)$ ratio in the starless cores are 0.27, 0.17, and 0.13, and those in protostellar cores 
are 0.26, 0.14, and 0.04 in the \lam, Orion A, and B clouds, respectively. 
The ratio in the protostellar cores of the \lam\ cloud is higher than 
those in the Orion A and B clouds by a factor of two and six, respectively. 
In the \lam\ cloud, the difference between the median values (0.01) in the starless and protostellar cores is negligible, 
while the difference is significant (0.09) in the Orion B cloud.
These results imply that the protostellar cores in the \lam\ cloud might be chemically very young, not much different from 
the starless cores.

We also investigated the $ N(\rm N_{2}H^{+})$/$N(\rm HCO^+)$ ratios in protostellar cores harboring 
different evolutionary stages of YSOs. 
The median ratio for Class 0 objects (a total of 13 in the three clouds) is about 0.14, while Class I sources (11) and 
YSOs showing a flat-spectrum (5) have slightly smaller values of 0.09.
The median values suggest that the $N(\rm N_{2}H^{+})$/$N(\rm HCO^+)$ ratios act as a chemical clock, decreasing with the evolution of cores.
The results from this and
other previous studies demonstrate that the $ N(\rm N_{2}H^{+})$/$N(\rm HCO^+)$ ratio is, indeed, 
a good chemical evolutionary tracer, from prestellar to protostellar stages.
This high $ N(\rm N_{2}H^{+})$/$N(\rm HCO^+)$ ratio in the \lam \ cloud suggests
that the lifetime of prestellar cores in the \lam \ cloud is longer than those in the Orion A and B clouds.
This longer prestellar stage timescale can result in enhanced abundances of nitrogen-bearing molecules
via the reaction with N$_{2}$ gas \citep{Lee03}.
The median abundance of \nthp\ is the highest in the \lam\ cloud (9.6 $\times10^{-9}$, see Table \ref{tab:8}), consistent with our scenario.

This scenario might be also supported by the different median line widths, which are affected by turbulence (Section \ref{subsec:widths}).
Turbulence can promote the evolution of cores and shorten the timescale of cores in the prestellar stage,
because it can generate density and velocity fluctuations which develop instabilities and consequently contribute to the formation of filaments \citep{V2019}.
The turbulent motions in the Orion A and B clouds appear to be larger than those measured in the \lam\ cloud
(see Table \ref{tab:7}).
In addition, most clumps and cores in the \lam\ cloud are isolated and compact, 
while the clumps in the Orion A and B clouds are extended along filaments, and the cores are very clustered (Paper I).
The mean column density of cores in the \lam\ cloud is 0.9$\times$10$^{23}$ cm$^{-2}$, which is two to four times 
lower than those in the Orion A (2.3$\times$10$^{23}$ cm$^{-2}$) and B clouds (3.8$\times$10$^{23}$ cm$^{-2}$, Paper I). 
The CFE is much lower in the \lam\ cloud, as discussed further in Section \ref{sec:Discussion}. 
Consequently, prestellar cores in the \lam\ cloud are less affected by turbulence,
and thus have a longer timescale to reach the protostellar stage, resulting in more abundant nitrogen-bearing molecules.

In summary, according to the $ N(\rm N_{2}H^{+})$/$N(\rm HCO^+)$ ratios, 
the protostellar cores in the \lam\ cloud are chemically younger than those in the Orion A and B clouds.
This result is consistent with the fact that
the formation and evolution of cores seems to be retarded by stellar feedback
in the \lam \ cloud, as discussed in Section \ref{subsubsec:D/H_2}.
Thus, the relative abundance of 
$\rm  N_{2}H^{+}$ to $\rm HCO^+$ 
is the result of chemical evolution as well as the different star-forming environments, especially the FUV radiation.

\section{Infall motions in the three clouds} \label{sec:infall}
Gravitational collapse is one of the key signatures of on-going star formation.
According to the ``inside-out'' collapse model and observations of protostellar cores
\citep{Shu77, Myers96, Evans99, Evans03},
an asymmetric double-peaked velocity profile with a higher peak intensity 
in the blueshifted part can be a tool to probe infall motions.
To detect this signature in the three clouds, 
we used both optically thick HCO$^+$ ($J=1-0$) and optically thin $\rm H^{13}CO^+$ ($J=1-0$) lines. 
The HCO$^+$ molecular line has been widely used to investigate infall motions \citep[e.g.,][]{Fuller05, Wu07,Rygl13,Yoo18}.
A combination of HCO$^+$ and $\rm H^{13}CO^+$ molecular lines is, 
in general, employed to disentangle an asymmetric ``blue profile''. 
The HCN line is also a good tracer of gravitational inflow \citep{Wu03,Jin16}, 
but this line was not used in this study because its optically thin isotope line (H$^{13}$CN) was not observed in our KVN observations.
The HCN line has hyperfine components that are strongly self-absorbed and blended, also making it difficult to use in our inflow analysis.

Line asymmetry can be measured by calculating a normalized velocity difference $\delta V$ as in the following equation \citep{Mardones97}: 
\[
     \delta V= \frac{V_{\rm thick} - V_{\rm thin}}{\Delta V_{\rm thin}}
\]
where $V_{\rm thick}$, $V_{\rm thin}$, and $\Delta$$V_{\rm thin}$ are the peak velocities of the
optically thick and thin lines, and the FWHM of the optically thin line, respectively.
The observed lines can be classified as either blue or red profile if the difference
between $V_{\rm thick}$ and $V_{\rm thin}$ is greater than a quarter of $\Delta V_{\rm thin}$.
\citet{Mardones97} established a criteria for assessing the line asymmetry as follows:
blue profile (B) for  $\delta V$ $<$ -0.25, red profile (R) for $\delta V$ $>$ 0.25, and 
neutral profile (N) for  -0.25 $ \le \delta V$ $\le$ 0.25.
We adopted these criteria to search for cores showing infall motions. 
The observed line parameters and derived $\delta V$ are listed in Table \ref{tab:10}.
The resulting mean values of $\delta V$ in each cloud are -0.09, -0.07, and -0.06
for the  \lam, Orion A, and B clouds, respectively.
The distribution of $\delta V$ shows the largest spread in the Orion A cloud (from -1.41 to +1.22), 
implying the diversity in the evolutionary stage of star-forming cores in the cloud.
For the 19 infall candidates, the infall rate ($\dot{M}_{\rm inf}$) is roughly estimated using $\dot{M}_{\rm inf} = 4\pi R^{2}V_{\rm inf}\rho$ \citep{Lopez10},
and their line profiles are presented in Figure \ref{fig:9}.
An infall velocity ($V_{\rm inf}$) is determined from $V_{\rm HCO^{+}}$ - $V_{\rm H^{13}CO^{+}}$,
$\rho = M/(4/3\pi R^{3})$ is the core volume density, and $R$ is the radius of the core. 
We have used $R$ and $M$ derived from the dust continuum emission at 850 $\micron$ of SCUBA-2 (Paper I).
The $\dot{M}_{\rm inf}$ is listed in the last column of Table \ref{tab:10}.
The mean $\dot{M}_{\rm inf}$ is the highest in the Orion B cloud (5.0 $\times$ 10$^{-5}$ $M_{\sun}$ yr$^{-1}$), which is two times higher than that of 
the \lam\ (2.5$\times$ 10$^{-5}$ $M_{\sun}$ yr$^{-1}$) and Orion A clouds (2.2 $\times$ 10$^{-5}$ $M_{\sun}$ yr$^{-1}$).

To quantify the dominance of the blue profile, 
the blue excess ($E$) was measured using the following relation \citep{Mardones97},
\[
     E= \frac{N_{\rm blue} - N_{\rm red}}{N_{\rm total}}
\]

\noindent where $N_{\rm blue}$, $N_{\rm red}$, and $N_{\rm total}$ represent
the numbers of cores having blue ($N_{\rm blue}$), red profiles ($N_{\rm red}$), and the total samples ($N_{\rm total}$), respectively.
We found the highest blue excess of $E$=0.25 in the Orion B cloud and a less significant blue excess of $E$=0.16 in the Orion A cloud. 
The estimated values in the Orion A and B clouds are comparable to the results in previous infall studies in active star-forming regions \citep[e.g.,][]{Fuller05, Wu07, He15, Yoo18}. 
On the other hand, we found no significant blue excess in the \lam\ cloud ($E$=0.00). 
The highest blue excess in the Orion B cloud indicates the dominance of the blue profile caused by infall motions, which is well consistent with the fact that 
the Orion B cloud is the most active star-forming region among the three clouds, while the \lam\ cloud is considered as a non-active star-forming region.
These results are summarized in Table \ref{tab:11}. 
In summary, our results support that UV radiation can heat the gas and suppress core formation in the \lam\ cloud.

\section{Discussion} \label{sec:Discussion}
Stars influence their local environment through various energetic processes, such as 
protostellar outflows, stellar winds, intense UV radiation, and supernova explosions.
These processes usually occur around massive OB-type stars, and thus, the stellar feedback by massive stars can trigger new star formation or 
otherwise destroy the potential sites of star formation.
Positive stellar feedback in the sense of triggered star formation has been steadily studied by simulations and observations
\citep{Bibas11,Koenig12,Thompson04,Lim14,Lim18}.
In contrast, in some cases, the stellar feedback can lower the integrated SFE, to a few percent, and eventually leads to the dissolution of molecular clouds after a dynamical time (10 - 30 Myr) 
\citep{Dale13, Kruijssen19, Lada03, Walch12}.
This process is usually referred to as negative feedback or suppressed star formation.

The supernova explosion plays an important role in shaping molecular clouds and regulating star formation \citep{PZ01,Smith04}. 
According to several observational and simulation studies, the large-scale expansion of the \lam\ cloud can be attributed to a supernova
 because of its high expansion velocity at 14 -- 25 km s$^{-1} $\citep{CS96,DM1999, DM02,Kounkel18, Lang2000, MM87, Mathieu08}.
The resulting shock wave created the low-density cavity and aborted star formation in the central region near the massive star $\lambda$ Ori \citep{Mathieu08}. 
A series of numerical simulations by \citet{Shima17} compared the effect solely by star formation, photoionization by massive stars, and the feedback by supernovae as well as photoionization. 
Their result argues that a sudden thermal injection from a supernovae is less effective than the continuous energy of UV radiation at preventing further fragmentation and reducing the accretion rates of forming protostars.
Hence, the environmental effect on ongoing star formation in the \lam\ cloud is likely dominated by the radiative feedback from $\lambda$ Ori.

The intense UV radiation fields of massive stars can increase the dissociation and ionization of cloud material and affect the physical and chemical properties of cores.
This study shows that FUV radiation modifies the relative molecular abundances in the \lam\ cloud via the photolysis of molecules, as presented in Section \ref{sec:Results}.
Furthermore, FUV absorption by dust grains can heat the grains themselves, or gas via photoelectric ejection \citep{Draine1978, HT99}.
We found that the dust temperature of clumps in the \lam\ cloud is higher on average than in the Orion A and B clouds.
A decreasing gradient of dust temperature with projected distance from $\lambda$ Ori was also found in the \lam\ cloud (Figure 11 of Paper I).
This suggests that the central star has a great influence on various scales, including cores, clumps, and even the entire molecular cloud.

The EUV photons (i.e., with energy greater than 13.6 eV) from $\lambda$ Ori ionize atomic hydrogen 
and contribute to the formation of a giant H \textsc{ii} region (Sh 2-264, a radius of 34.5 pc) in the \lam\ cloud \citep{MM87}.
The high-energy photons can photo-evaporate the dense material, leading to the lowest dense gas fraction (0.1) in the \lam\ cloud,
which can be interpreted as a low CFE.
Using the column-density power spectrum of 29 clouds, 
\citet{FK13} measured the SFEs and found that
the SFEs increase from zero in large scale H \textsc{i} clouds and non-star forming clouds to 1 - 10 \% in typical star-forming molecular clouds, and up to $>$ 10\% in dense cores.
This study suggests that the core-to-star formation efficiency eventually approaches to 0.3 -- 0.7.
The CFE was the lowest in the \lam\ cloud, and thereby the SFE should also be lowest in the cloud.

On the other hand, the Orion B cloud with the highest CFE (0.2) is expected to have the highest SFE.
The EUV photons can also hinder the further fragmentation of clumps.
This leads to a lower multiplicity (the number of cores in one clump, see Section 4.3. in Paper I) in the \lam\ cloud (1.8) than in the Orion A (3.2) and B (3.3) clouds. 
Among the three clouds, the cores in the \lam\ cloud have the lowest median values of mass (0.77 M$_{\sun}$), size
(0.09 pc), column density (8.2 $\times$ 10$^{22}$ cm$^{-2}$), and volume density (2.5 $\times$ 10$^{5}$ cm$^{-3}$).
These physical properties suggest that after the supernova explosion, the remnant of the cloud has been continuously affected by UV radiation; dense gas became deficient, and thus core formation was suppressed in the \lam\ cloud.

Morphological differences are also found between clouds or substructures with similar scales
(see Figures 1 to 5 in Paper I).
The entire molecular clouds of the Orion A and B have filamentary structures, while the \lam\ cloud has a ring-shaped structure. 
One difference among the observed substructures is the shape of clumps and the number of cores within those clumps.
The clumps in the Orion A and B clouds well follow the filamentary shapes of their natal clouds, while those in the \lam\ cloud are compact and spatially isolated.
The elongated and complex filaments can be explained by a cascade of turbulent motions
\citep[e.g.,][]{MK04,HF12}, which is supported by the large line widths of \hcop, \cth, HCN, and \htco\ (see Section \ref{subsec:widths}).
Furthermore, the cores within filaments appear to contain more gas at very high density (n(H$_{2}) >$ 10$^{6}$ cm$^{-3}$, see Table 5 in Paper I) and to be more highly clumped than those cores without filaments in the \lam\ cloud.
The fragmentation of clumps might be enhanced by turbulence in the Orion A and B clouds, which results in the high multiplicity in the two clouds.

We have examined the effects of stellar feedback on cores in the OMC, focusing on chemical processes.
This study is the first finding  of observational support for negative stellar feedback on core formation and evolution.
We found that a precedent massive star formation suppressed the further star formation in the \lam\ cloud due to the radiative feedback from $\lambda$ Ori.
In the majority of prior star formation studies, positive stellar feedback has mainly been considered.
As presented in this study, however, negative feedback could be more significant in the star formation process than positive feedback in some cases.
Therefore, not only the positive aspects but also the negative effects of stellar feedback should be carefully considered for a detailed understanding of star formation.

\section{Summary} \label{sec:Summary}
We performed single-point molecular line observations toward 80 cores in the \lam, Orion A, and B clouds
to investigate the chemical properties of cores in different star-forming environments.
Eleven molecular lines, including two maser lines,
were observed using the KVN 21-m telescopes.
The results from this work, combined with our previous study (Paper I), provide many observational features of suppressed star formation in the \lam\ cloud as follows:
\begin{enumerate}
\item    The detection rates of dense gas tracers such as $\rm N_{2}H^+$, $\rm HCO^+$, and $\rm H^{13}CO^+$ lines were the lowest in the \lam\ cloud among the three clouds.
This result supports our scenario that high energy photons 
from the massive star ($\lambda$ Ori) can photo-evaporate dense dust and gas, 
leading to the low detection rates toward cores in the $\lambda$ Orionis cloud. 

\item The significant variation in deuterium fractionation [HDCO]/[$\rm H_{2}CO$] ranging from 0.01 to 0.26 indicates that 
cores have formed in diverse conditions of temperature and density.
High deuterium fractionations in excess of 0.1 were found in the eight starless cores of the Orion A and B clouds, 
while none of the starless cores in the \lam \ cloud showed such a high ratio.
The low D/H ratio in the \lam\ cloud is likely due to the FUV radiation from the massive OB stars.
FUV radiation heats the entire cloud, hindering the deuteration of \htco\ on the grain surfaces.

\item  The fractional abundance of $N(\rm N_{2}H^{+})/ N(\rm HCO^+)$ can be used as a chemical clock. 
This ratio was the highest in the \lam\ cloud (0.26). 
 The Orion A and B clouds had moderate (0.16) and the lowest (0.10) values, respectively.
The highest ratio in the \lam\ cloud implies that the cores in the \lam\ cloud stayed longer in the prestellar stage than those in the Orion A and B clouds.
 The formation and evolution of cores seems to have been delayed by stellar feedback in the \lam\ cloud.

\item  The mean abundances of PDR tracers, \cth\ and HCN, were about three and two times higher in the \lam\ cloud (4.5$\times$10$^{-8}$ and 7.6$\times$10$^{-10}$ cm$^{-2}$) than 
in the Orion A (1.6$\times$10$^{-8}$ and 3.3$\times$10$^{-10}$ cm$^{-2}$) and B (1.5$\times$10$^{-8}$ and 4.0$\times$10$^{-10}$ cm$^{-2}$) clouds, respectively.
Photodissociation and dissociative recombination processes which were active due to the intense UV radiation from $\lambda$ Ori significantly impacted the chemistry of the cores.

\item The blue excess ($E$) in the \hcop \ (1 -- 0) line was the highest in cores in the Orion B cloud ($E$ = 0.2), which 
indicates the most active star formation is occurring in the Orion B cloud among the three clouds.
On the other hand, we could not find any sign of blue excess ($E$ = 0) in the \lam\ cloud, implying suppressed star-forming activity.

\end{enumerate}

Our findings well support the idea that feedback from massive stars can control star formation in a negative way.
A more detailed study of the kinematics and temperature distribution within the molecular cores with higher spatial resolution mapping observations is necessary to interpret 
negative feedback on star formation properly.

\section*{Acknowledgement}
This work was supported by the Basic Science Research Program
through the National Research Foundation of Korea
(grant No. NRF-2018R1A2B6003423) and the Korea Astronomy and Space Science Institute under the R\&D
program supervised by the Ministry of Science, ICT and Future Planning. 
We are grateful to the staff of the KVN who helped to operate the array and to correlate the data. 
The KVN is a facility operated by the KASI (Korea Astronomy and Space Science Institute). 
The KVN observations and correlations are supported through the high-speed network connections among the KVN sites provided by the KREONET (Korea Research Environment Open NETwork), 
which is managed and operated by the KISTI (Korea Institute of Science and Technology Information).
Tie Liu acknowledges the supports from the international partnership program of Chinese academy of sciences through grant No.114231KYSB20200009, National Natural Science Foundation of China (NSFC) through grant NSFC No.12073061, and Shanghai Pujiang Program 20PJ1415500.
BL acknowledges the support of the NRF grant funded by the Korea government (MSIT) (grant No. NRF-2019R1C1C1005224).
K.T. was supported by JSPS KAKENHI Grant Number 20H05645.

\clearpage

\appendix 
\section{Column densities of cores} \label{sec:CD}
 We calculated column densities of seven molecules 
 (\hcop, \htcop, \htco, HDCO, \nthp, \cth, and HCN). 
 These lines are assumed to be optically thin, except for the HCO$^+$ line. 
Since the core sizes are similar to the main-beam sizes, beam filling factor ($\eta_{f}$) was treated as 1.
 Under the consideration that the gas and dust are well coupled in cores 
 with high densities (n $>$ 10$^{5}$ cm$^{-3}$) via collisions 
 \citep{GK74}, the dust temperatures ($T_{\rm d}$) are almost the same as  
the excitation temperatures ($T_{\rm ex}$) \citep{Galli02}. 
The resulting column densities of seven molecules are given in Table \ref{tab:12}.
Figure \ref{fig:9} shows the spectra of nine molecular emission lines including SiO and \httco\ for 80 cores.
 
\subsection{Optically thick case: HCO$^+$ column density}
To estimate the total column density ($N_{\rm tot}$) of a given molecule from its observed line, 
we use the previously derived equation of \citet{Scoville86} as below:
\begin{equation} \label{eqn:CD}
N_{\rm tot} = \frac{3k} {8\pi^{3}B\mu^{2}} \frac{(T_{\rm ex}+hB/3k)}{J+1} \frac{{\rm exp}[hBJ(J+1)/kT_{\rm ex}]}{1-{\rm exp}(-h\nu/kT_{\rm ex})}
\int{\tau_{\nu}}dV 
\end{equation}
where $h$, $k$, $\nu$, $B$, $J$, and $\mu$ represent  
the Planck's constant, the Boltzmann's constant, 
the frequency of the transition in GHz, the rotational constant, the quantum number for total rotational angular momentum,
and the dipole moment, respectively.
$T_{\rm ex}$ is the excitation temperature.
The total column density of HCO$^+$ is calculated with dust temperatures ($T_{\rm d}$) 
from the PGCC catalog.
The HCO$^+$ line is optically thick, and therefore Eq \ref{eqn:CD} becomes:

\begin{equation}
N_{\rm tot} = \frac{3k} {8\pi^{3}B\mu^{2}} \frac{(T_{\rm d}+hB/3k)}{J+1} \frac{{\rm exp}[hBJ(J+1)/kT_{\rm d}]}{1-{\rm exp}(-h\nu/kT_{\rm d})}
\frac{\tau_{\nu}}{1-e^{-\tau_{\nu}}} \frac{\int{T_{B}dV}} {J(T_{\rm d}) - J(T_{\rm bg})} 
\end{equation}
where $J(T)$ and $\tau_{\nu}$ denote a effective source radiation temperature ($\frac{h\nu} {k} \frac{1}{(e^{h\nu/kT} -1)}$) and
optical depth ($-{\rm ln} \left[1 - \frac{T_{B}}{J(T_{\rm d}) - J(T_{\rm bg})} \right]$), respectively. 
$T_{\rm bg}$ is the background radiation temperature of 2.73 K. 

\subsection{Optically thin cases}
\paragraph{H$_{2}$CO and HDCO column density}
\htco \ molecules exist in the ortho or para form, 
according to the spin configurations of two hydrogen atoms.
In this study, we only observed the transitions of  ortho-\htco, 
and therefore the \htco\ column densities should be corrected by multiplying the ortho to para ratio. 
The statistical ortho to para ratio of 3:1 \citep{Minh95,Guzman13} was adopted.
The column density in a lower state 
(defined as energy level u $\rightarrow$ l = ((J+1) $\rightarrow$ J ) 
can be obtained by following equation \citep{Kang15}:
\begin{equation}
N_l = 93.5\frac{g_l}{g_u} \frac{\nu^3}{A_{ul}}\frac{1}{T_{\rm ex}[1-{\rm exp}(-h\nu/kT_{ex})]} \int T_{\rm mb} \ dV,
\end{equation}
where $T_{\rm ex}$, $g$, $A_{ul}$, $\int T_{\rm mb} \ dv$, and $V$ represent  
the excitation temperature, the statistical weight, the Einstein coefficient in s$^{-1}$, the integrated line intensity, 
and the velocity in km s$^{-1}$, respectively.

The total column density of ortho-\htco, $N_{\rm ortho}$, is related to
the column density, ${N_l} =$ $N(J_{Ka,Kc})$ in the lower state $J_{Ka,Kc}$ as below:
\begin{equation}
\frac{N_{\rm ortho}}{N(J_{K\textsubscript a,K\textsubscript c})} = 
\frac{Z}{g(J_{K\textsubscript a,K\textsubscript c})} {\rm exp} 
 \left(\frac{E(J_{K\textsubscript a,K\textsubscript c})}{kT_{\rm ex}}\right)
\end{equation}
where g(J$_{Ka,Kc}$), E(J$_{KaKc}$), and Z
are the statistical weight for the ortho-\htco\  (3(2J+1)), 
the energy of a J$_{KaKc}$ level, and the partition function, respectively. 
If the molecules follow the Boltzmann distribution of a single $T_{\rm ex}$, 
the partition function is expressed as follow:
\begin{equation}
Z = \sum z = \sum g(J_{K\textsubscript a,K\textsubscript c}) {\rm exp} \left(\frac{-E(J_{K\textsubscript a,K\textsubscript c})}{kT_{\rm ex}} \right).
\end{equation}

In the case of HDCO molecules, 
it is not necessary to correct the ortho to para ratio because there is no H-pair.
Deuterated species are, in general, 10 -- 100 times less abundant than the main species
\citep{RM00}, 
and thus HDCO transitions are optically thin.
The statistical weight $g$ depends only on the rotational angular momentum as 
$g(J_{K\textsubscript a,K\textsubscript c}) = 2J + 1$.
The total column density ($N_{\rm tot}$) is related to the upper state of column density 
($N_{\rm u}$) as below: 
\begin{equation}
N_{\rm tot} = \frac{N_{\rm u}} {f_{\rm u}}
\end{equation}
where $f_{\rm u}$ is fractional population of an upper state defined as 
\begin{equation}
f_{\rm u}= \frac{g_{\rm u}} {Q(T_{\rm ex})}  \ e^{-E_{\rm u} / kT}
\end{equation}
where $g_{\rm u}$, $E_{\rm u}$, and $Q(T_{\rm ex})$ are
the degeneracy of the upper state, the energy of the upper state, and the partition function.
The partition function can be approximated as 
$Q(T_{\rm ex}) = \frac{kT_{\rm}}{hB} + \frac{1}{3}$ \citep{McDowell88}.
$N_{u}$ is the column density in the upper level of the transition
($N_{u} = \frac{8\pi k\nu^2}{A_{ul}hc^3} \int T_{\rm mb}dV$).

\paragraph{N$_2$H$^+$ column density}
The column density of N$_2$H$^+$ molecule was calculated following 
the equation of \citet{Furuya06}:
\begin{equation}
N({\rm N_{2}H^{+}}) = 3.30 \times 10^{11} \frac{(T_{\rm ex} + 0.75)}{1 - e^{(-4.47/T_{\rm ex})}} \left(\frac{\tau_{tot}}{1.0}\right) 
\left( \frac{\Delta V}{1.0 \ \rm km \ s^{-1}}\right) \rm cm^{-2}.
\end{equation}
The optical depth $\tau_{tot}$ and line width $\Delta V$ 
are obtained from the hyperfine structure fitting to the N$_2$H$^+$ lines.
Excitation temperature are substituted with dust temperature as described above.

\paragraph{$H^{13}CO^+$, HCN, and $C_{2}H$ column density}
Column densities for HCN and \cth\ molecules were calculated 
using the equation of \citet{Caselli02}: 
\begin{equation}
N_{\rm tot} = \frac{8\pi\nu_{ul}^3}{c^3} \frac{Q(T_{\rm ex})}{A_{ul}g_u} 
\frac{e^{E_{u}/kT_{\rm ex}}}{ e^{h\nu_{ul}/kT_{\rm ex}}-1} 
\frac{\int{T_{\rm mb}dV}} {J(T_{\rm ex}) - J(T_{\rm bg})},
\end{equation}
where $\nu_{ul}$, c, $A_{ul}$, $g_{u}$, $E_{u}$, $T_{\rm ex}$, and $\int T_{\rm mb}dV$    
are the transition frequency, the speed of light,
the Einstein coefficient, the statistical weight, 
the energy of the upper level, 
the excitation temperature,
and the integrated intensity.
$Q(T_{\rm ex})$ is the partition function ($\sum_{j=0}^{\infty} g_{u}e^{-E_{u}/kT_{\rm ex}}$).

\section{Sources with SiO thermal lines and masers} \label{sec:maser}
The SiO thermal line is considered to be the classical tracer of protostellar jets because
the emission line is less contaminated by infalling envelopes or swept-up cavities,
and traces regions close to protostars \citep{Codella13}.
The SiO thermal lines were detected in the two cores G191.90-11.21 N (G191 N) 
and G192.32-11.88 N (G192 N) of the \lam \ cloud.
The 22 GHz \hto\ and 44 GHz class I \chtoh\ masers are usually associated with inner jets and outflow lobes, respectively \citep[e.g.,][]{Kurtz04,Kim18}. 
Water masers were detected in three out of 80 cores, whereas 
a methanol maser was detected only in G205.46-14.56 S1 that was one of the water maser-detected cores. 
Figure \ref{fig:8} presents the observed maser spectra in flux density (Jy) by multiplying 13.3 Jy K$^{-1}$ \citep{Kim18}. 
We address the overall properties of the five cores and their neighbor cores 
in the following subsections.

\subsection{G191.90-11.21 N and G192.32-11.88 N  } \label{subsec:SiO}
The two cores G191N and G192N are of particular interest because they were
 classified as starless cores in Paper I. 
In our previous SCUBA-2 survey, the 850 $\micron$ dust continuum data revealed 
very weak emissions from G191N.
The IR source associated with G191N was classified as 
a candidate of star-forming galaxy (SFG) 
due to its low brightness in the 3.36 $\micron$ band of the
$\it WISE$ data (Paper I).
The associated source is not a SFG, but a YSO, with a low $L_{\rm bol} $ (0.1 $L_{\sun}$) and $T_{\rm bol}$ (13 K), and a high mass-to-luminosity ratio (3.4).
These values suggest that the G191N could still be in the earliest stage of star formation.  
A more detailed investigation using interferometric observations is required to clarify the origin of the SiO thermal emission. 
Within a radius of 1$\arcmin$, we could not identify any known YSO. 
The nearest YSO candidate is J053130.03+130016.9, separated  83$\arcsec$ from the G191N, which makes it hard to consider it as the driving source of the thermal emission.

\citet{Liu16} identified G192N as a youngest class 0 source, more evolved than the first hydrostatic core (FSHC). 
This source was not detected in either the Spitzer/Multiband Imaging Photometer (MIPS) 
24 $\micron$ band or  Infrared Array Camera (IRAC) bands, 
but it has a large envelope seen in the Spitzer/MIPS 70 $\micron$ 
and SMA 1.3 mm continuum observations \citep{Liu16}. 
The $^{12}$CO (2--1) and  $^{13}$CO (2--1) observations with SMA  also revealed a well-collimated outflow. 
In the AllWISE catalog \citep{Wright10},
no source was found within a radius of 10$\arcsec$, which
implies G192N is still deeply embedded in its dense envelope.
The N$_{2}$H$^+$ to HCO$^+$ abundance ratio of G192 N is higher than that of other protostellar cores 
in the \lam\ cloud, indicating G192N is chemically younger.
 
\subsection{G196.92-10.37} \label{subsec:G196}
We detected the 22 GHz \hto \ maser towards the core G196.92-10.37 (G196) 
in the bright-rimmed cloud (BRC) 18 within the \lam \ cloud. 
BRCs are the illuminated part of molecular clouds usually found at the border of H \textsc{ii}  regions. 
A number of observational studies have found signposts of triggered star formation in BRCs.
YSOs distributed in BRCs exhibit an age gradient, where younger sources are deeply embedded in
the BRC, and older ones are located in the outer region of the BRCs or the H \textsc{ii}  regions \citep{Lee05, Ogura07, Fukuda13, Lim14}.

G196 is the only core detected in the BRC 18 (Paper I).
In Figure 1 of paper I, 
G196 was found to be an isolated and compact source in the 850 \micron\ continuum image.
This core has the largest mass (5.4 $M_{\sun}$) and averaged column density (1.8$\times$10$^{23}$ cm$^{-2}$) among the cores in the \lam \ cloud,
harboring a Class 0 source with  
$L_{\rm bol} =$ 14.9 $L_{\sun}$ and $T_{\rm bol} =$ 68 K (Paper I).

\citet{Qin03} conducted $^{12}$CO (J = 1) observations towards BRC 18 and 
found a bipolar outflow near the IRAS 05417+0907. 
The Class 0 YSO and IRAS 05417+0907 are considered to be the same source with an angular separation of 4\arcsec,
and thus the YSO is regarded as the driving source of the \hto\ maser.

The luminosity of maser emission was estimated from the integrated flux density equations (1) and (2) of \citet{Kim18}.
The luminosity of water maser $L \textsubscript{H$_{2}$O}$ is as low as 7.7$\times10^{-9}$ $L_{\sun}$. 
The low $L \textsubscript{H$_{2}$O}$ and 
the $T_{\rm bol}$ in the border of the criteria between Class 0 and Class I (70 K)
imply that the maser driving source is entering a Class I evolutionary stage 
rather than staying in an embedded Class 0 stage.

Except for the G196, we could not find any other cores in the vicinity of BRC 18.
Thus, this BRC does not exhibit clear evidence that star formation
has been triggered, even considering the higher mass and column density of G196.
A critical ionizing photon flux, which can indicate the possibility of triggered star formation, can be calculated using
$\Phi_{\rm CRIT} =$ 6$\times$ 10$^{13}$ cm$^{-2}$ s$^{-1}$ $\times $($M_{\rm CLOUD}$/M$_{\sun}$)$^{-3}$  \citep{Bibas11} .
When an ionizing photon flux exceeds the $\Phi_{\rm CRIT}$, the cloud is dispersed without any star formation being triggered. 
We measured the ionizing photon flux illuminating the BRC 18 using the equation (1) of \citet{Morgan04} and obtained $\Phi \sim$ 2.3$\times$10$^{10}$ cm$^{-2}$ s$^{-1}$.
This value is two orders of magnitude higher than the $\Phi_{\rm CRIT}$ of BRC 18 
(3.9$\times$10$^{8}$ cm$^{-2}$ s$^{-1}$) if we adopt the minimum mass of BRC 18 as 115 $M_{\sun}$ (Neha et al. in preparation).
These results support that a new generation of stars would be hard to find in BRC 18.

\subsection{G210.49-19.79} \label{subsec:G210}
G210.49-10.79 is located in the Orion A cloud, and is divided into three cores, G210.49-19.79E1, G210.49-19.79E2, and G210.49-19.79W (hereafter G210W). 
We detected the H$_{2}$O maser emission in G210W.
The luminosity $L \textsubscript{H$_{2}$O}$ was about 4.43$\times10^{-8}$ $L_{\sun}$ derived from equation (1) of \citet{Kim18}. 
G210W contains two millimeter sources, HH 1-2 MMS 2 and MMS 3 \citep{Chini97,Chini01}, associated with HOPS 168 and 167, respectively. 
The position of HOPS 168 is consistent with the position of G210W, 
while HOPS 167 is 40$\arcsec$ away to the southeast of HOPS 168.
HOPS 168 was classified as a Class 0 YSO with $L_{\rm bol}$ =41.6$L_{\sun}$ and $T_{\rm bol} =$ 51 K 
(Paper I), while HOPS 167 was classified as a flat-spectrum YSO with $L_{\rm bol}$ = 0.2 $L_{\sun}$ and $T_{\rm bol} =$ 568 K \citep{Furlan16}. 
The former is supposed to be in an active state of mass infall 
(3$\times10^{-5}$ $M_{\sun}$ yr$^{-1}$) and accretion \citep{Fischer10}. 
 Thus, HOPS 168 is probably the youngest and the most luminous YSO in G210.49-19.79, 
which drives the water maser emission.

\subsection{G205.46-14.56} \label{subsec:G205}
G205.46-14.56 (G205.46) is located in the Orion B cloud. 
This clump shows a filamentary structure in the 850 $\micron$ dust continuum emission map. 
We identified a total of nine cores in the northern (N1, N2, and N3), middle (M1, M2, and M3), 
and southern (S1, S2, and S3) parts of this clump, respectively (Paper I).
Both water (22 GHz) and methanol (44 GHz) masers were detected in G205.46S1 core for the first time in this work.
A \chtoh\ class I maser (36, 44, 84, and 95 GHz transitions) is, in general, emitted from the interface 
between outflows and the surrounding protostellar environment 
\citep{Plambeck88,Plambeck90,Voronkov14,Cyga09}, and therefore this maser emission traces star-forming activities.
       
In the southern part, G205.46S1 (also known as HH25) is the brightest core in this field.
Its peak intensity is about 1.8 Jy beam$^{-1}$ at 850 $\micron$ dust continuum. 
The core mass is 11.4 $M_{\sun}$ which is the largest in the Orion B cloud (Paper I). 
Our observations found the first detection of water maser toward G205.46 S1.
In Paper I, we identified one YSO (Class 0) within G205.46S1 using the AllWISE catalog. 
\citet{Furlan16} found two YSOs (HOPS 316 and HOPS 358) in this core, and 
these two YSOs are separated from each other by 7$\arcsec$.
This separation can not be resolved with the KVN single-dish observations. 

HOPS 358 was identified as a PACS Bright Red Source (PBRS) with 
an observed 70 $\micron$/24 $\micron$ color ratio greater than 1.65 \citep{Stutz13}. 
A protostellar jet (HH 25) associated with a collimated molecular outflow was found in this field \citep{GD98,Gibb04}, 
and HOPS 358 is spatially coincident with the position of HH 25. 
This source is also included in the high velocity outflow catalog \citep{Wu04}.
HOPS 358 is thus the most probable driving source of the water and methanol masers. 
The estimated water and methanol maser luminosities are 
$L \textsubscript{H$_{2}$O}=$ 4.4$\times10^{-8}$ and 
$L \textsubscript{CH$_{3}$OH}=$ $4.2\times10^{-8}$, respectively. 
The $L_{\rm bol}$ of HOPS 358 is 25.0 $L_{\sun}$ \citep{Furlan16}.
The other association ($14\farcs1$ away from the peak position of 850 \micron\ continuum, HOPS 316) is classified as a Class 0 YSO \citep{Furlan16}.

\section{Correlation of H$_{2}$CO line with star-forming activity}    \label{sec:H2CO}
The relationship between \htco\ emission and star-forming activity has been suggested in several studies \citep{Bernstein95,Bachiller97}, 
but an observational signature to ascertain it is still unclear. 
Here we indirectly suggest a relation between \htco\ line widths and star-forming activities.

In our sample of 80 cores, ten cores showed broad \htco\ line widths ($\Delta$V $>$ 2.0 km s$^{-1}$) in the three clouds.
In the \lam\ cloud, 
three out of the 14 detected cores showed broad line widths (G191N, G192N, and G196). 
We found three cores (G210W, G209.55-19.68S2, and G211.16-19.33S) having large $\Delta$V in the Orion A cloud and  
those of four cores (G205.46S1, G205.46S2, G206.12-15.76, and G206.93-16.61W3) in the OrionB cloud.
Either an SiO emission or water/methanol maser was detected in five cores out of ten cores (see Section \ref{sec:maser}), 
and therefore the large line widths are likely related to jets/outflows.
The mean $\Delta$V of the ten cores is 2.6 km s$^{-1}$, which is almost three times larger than 
that  of the rest 70 cores (0.9 km s$^{-1}$).

Among the ten cores with the broad $\Delta$V, the remaining five cores did not show SiO emission or water/methanol maser.
The five cores are G209.55-19.68S2, G211.16-19.33S, G205.46S2, G206.12-15.76, and G206.93-16.61W3.
The former two cores and the latter three cores are located in the Orion A and B clouds, respectively.
G209.55-19.68S2 was classified as a starless core in Paper I, and no YSO candidate was found in the HOPS catalog within a radius of 10\arcsec.
However, \citet{Stutz13} classified this core as a protostar candidate due to a strong 870 \micron\ detection and the {\it Spitzer}/IRAC 24 \micron\ emission which traces material accreting from the core onto the central protostar.
This protostar candidate may be a PBRS with high envelope densities and infall rates given its red spectral energy distribution (SED) and high luminosity at sub-mm.
This deeply embedded source may be attributed to the observed line width,
but further observations are required to constrain the broadening source.
A Class I YSO (HOPS 130) was identified in G211.16-19.33 S (Paper I), but
neither outflow nor jet emissions were detected toward this core.
Further observations are needed to confirm the origin of line width broadening.

The bipolar outflow driven by HOPS 385 
in G205.46 S2 has been reported in previous studies \citep{Gibb93,Wu04}.
G206.12-15.76 incubates a Class 0 YSO (HOPS 400) classified as a PBRS \citep{Stutz13}.
The IRAC image at 4.5 \micron\ reveals an extended emission region around the YSO, 
which seems to be associated with its outflow.
G206.93-16.61 W3 contains a Class 0 YSO (HOPS 399) showing a molecular outflow \citep{Sandell99}. 
Recently, a collimated outflow on a 0.2 pc scale was found towards the YSO \citep{Tobin16}.
The CO line emission from the strong outflow is highly redshifted up to +40 km s$^{-1}$ from the systemic velocity of HOPS 399.
A multi-epoch survey of 22 GHz H$_{2}$O and 
44 GHz Class I CH$_{3}$OH maser lines was carried out with the KVN telescope,
but no emission was detected in HOPS 399 \citep{Bae11}.
The non-detection of maser emission is consistent with our observation results.

According to the above studies, eight out of ten cores with large $\Delta$V are associated with outflows.
We thus propose that the large line widths of \htco\ could indicate the presence of jet or molecular outflow in dense cores.
The relation between \htco\ line widths and star-forming activities needs to be further examined in different star-forming regions.


\clearpage

\begin{figure}[b]
\centering
\resizebox{4.5cm}{5.5cm}{\includegraphics{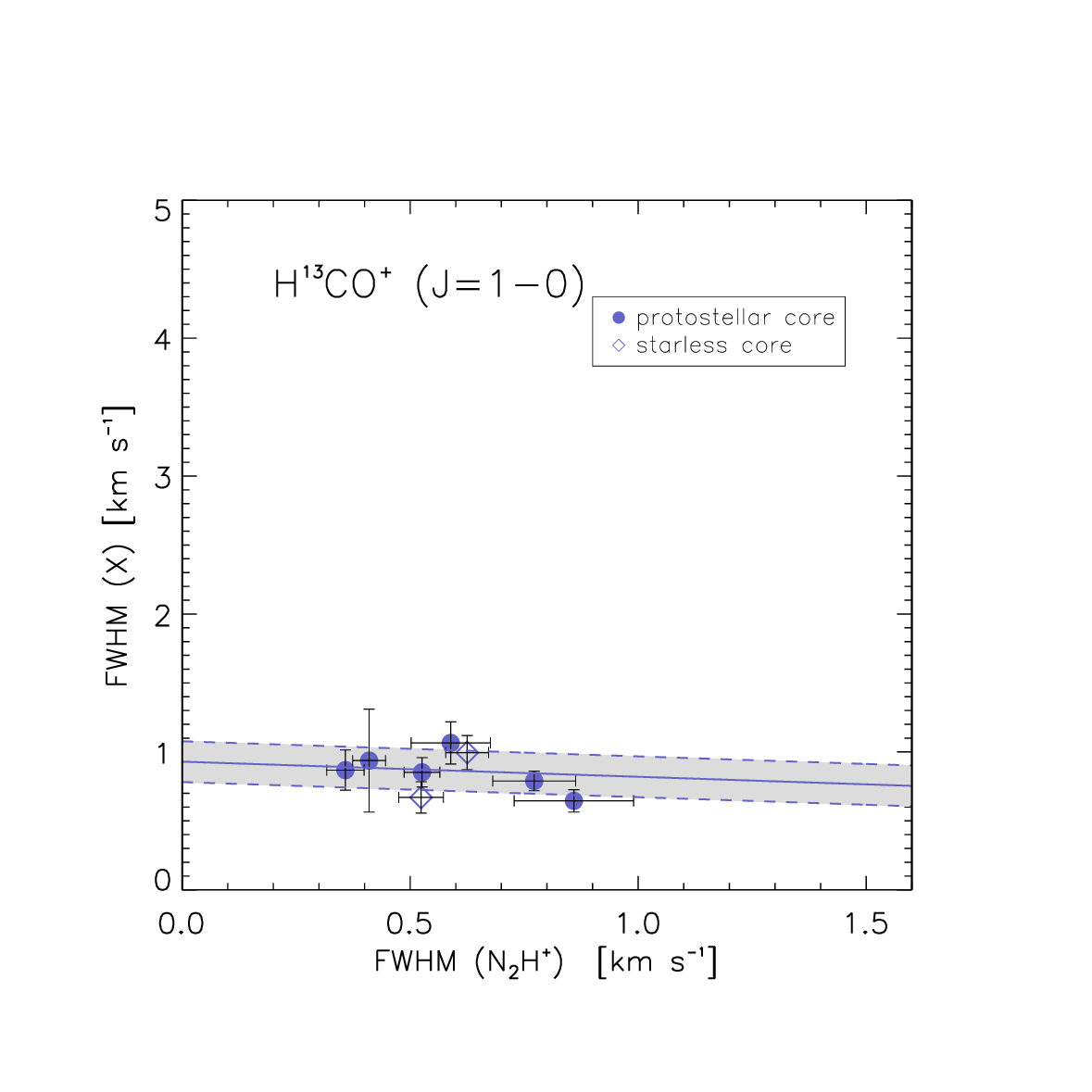}\hspace*{-0.25\textwidth}\vspace*{-0.3\textwidth}}
\resizebox{4.5cm}{5.5cm}{\includegraphics{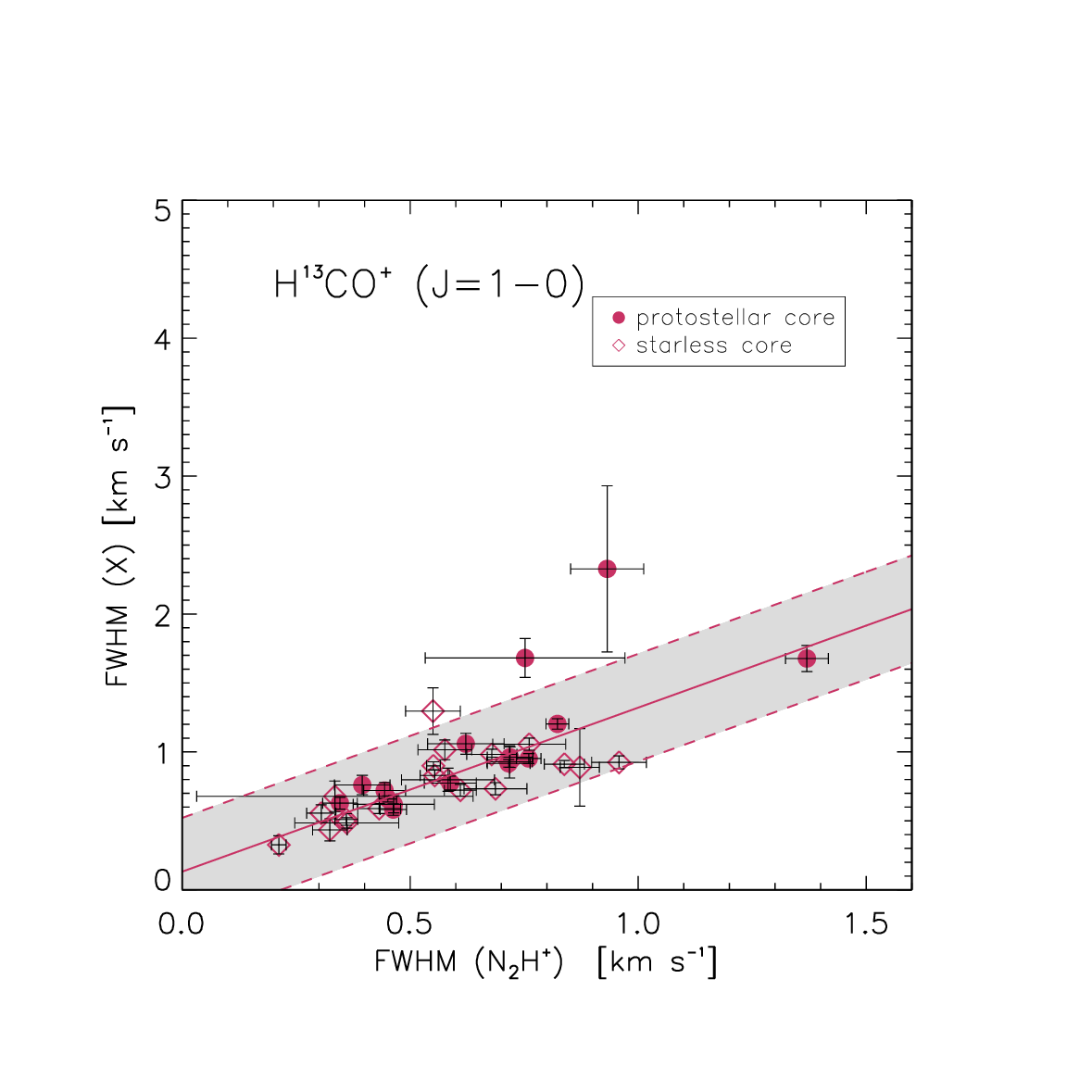}\hspace*{-0.25\textwidth}\vspace*{-0.3\textwidth}}
\resizebox{4.5cm}{5.5cm}{\includegraphics{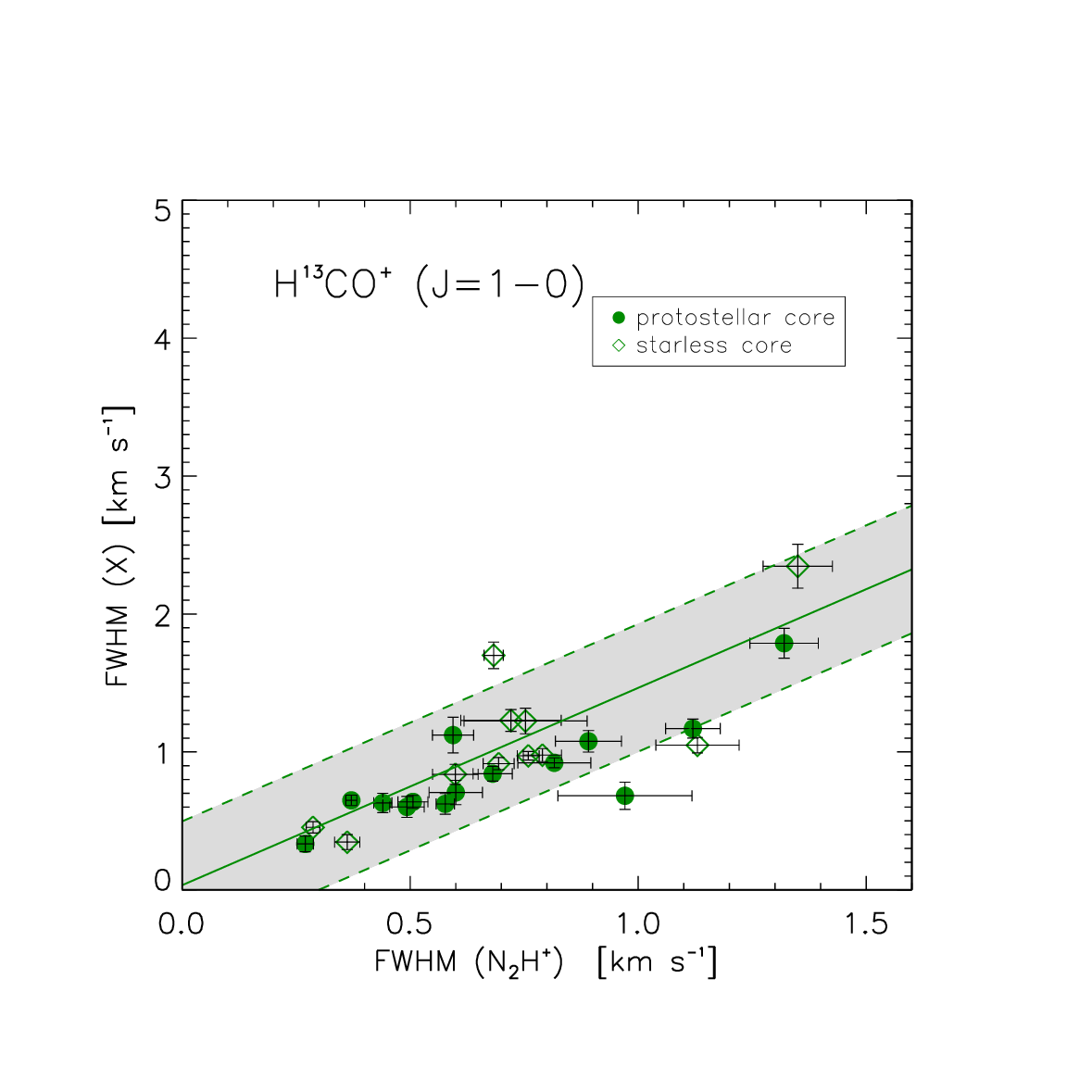}\hspace*{-0.25\textwidth}\vspace*{-0.3\textwidth}}
\resizebox{4.5cm}{5.5cm}{\includegraphics{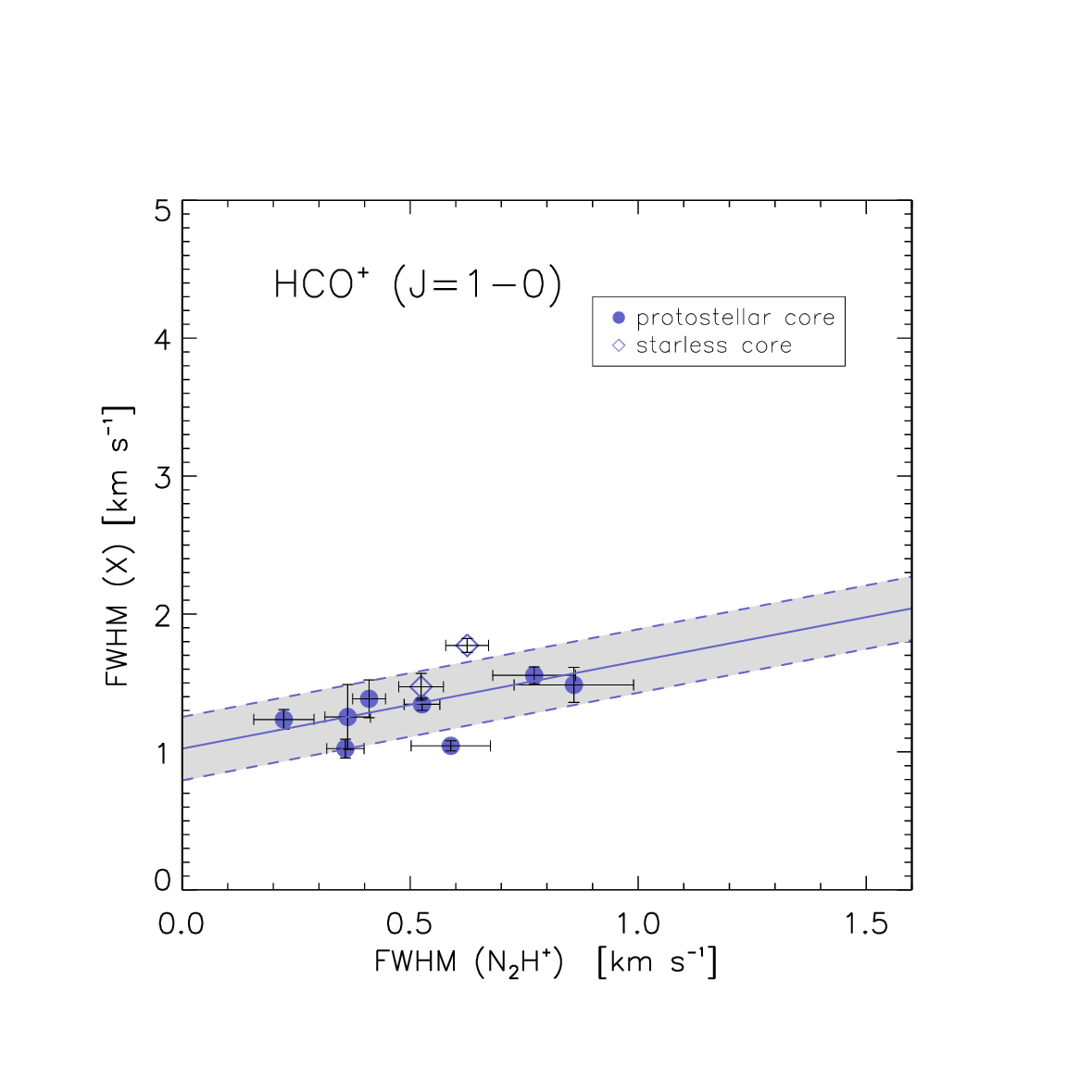}\hspace*{-0.25\textwidth}\vspace*{-0.3\textwidth}}
\resizebox{4.5cm}{5.5cm}{\includegraphics{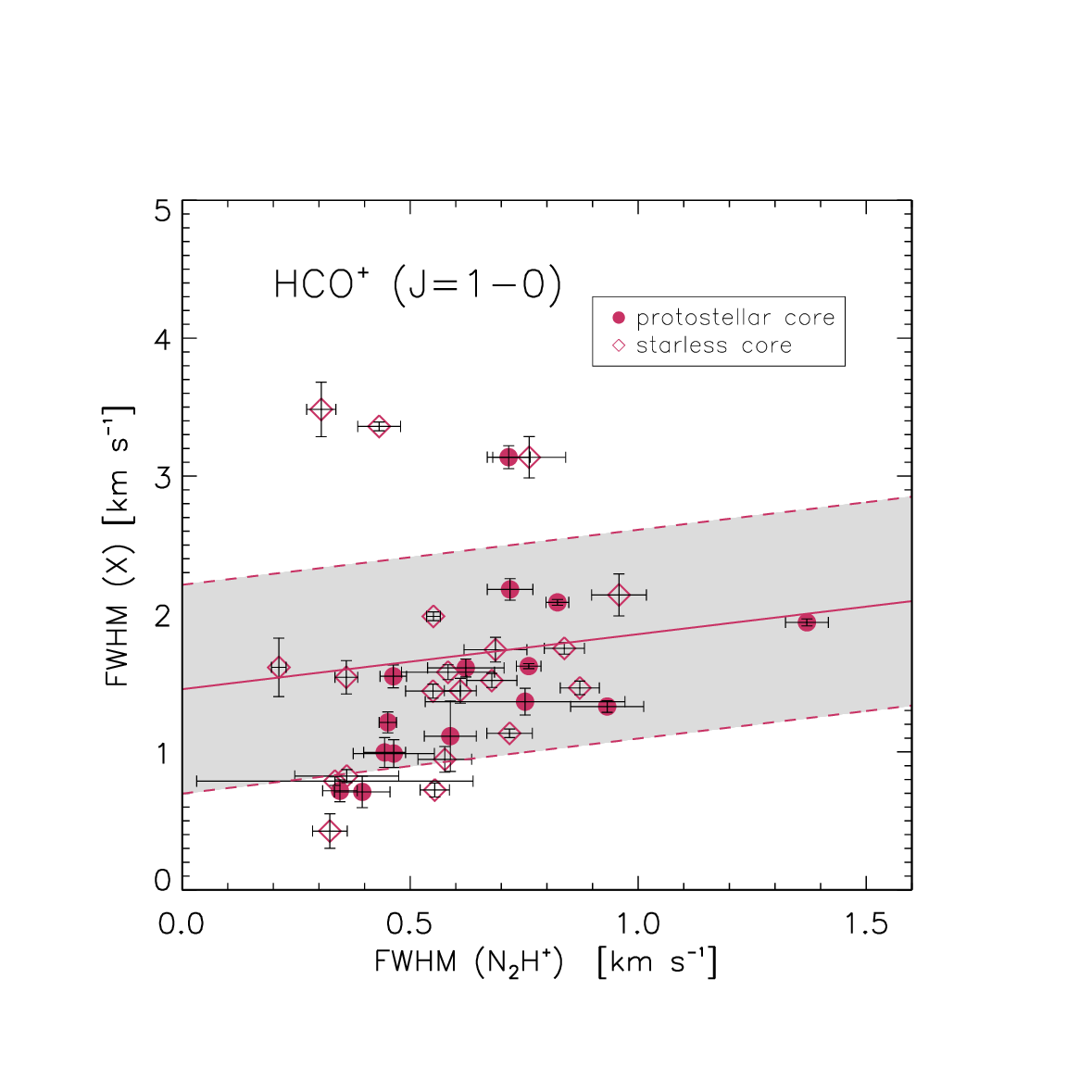}\hspace*{-0.25\textwidth}\vspace*{-0.3\textwidth}}
\resizebox{4.5cm}{5.5cm}{\includegraphics{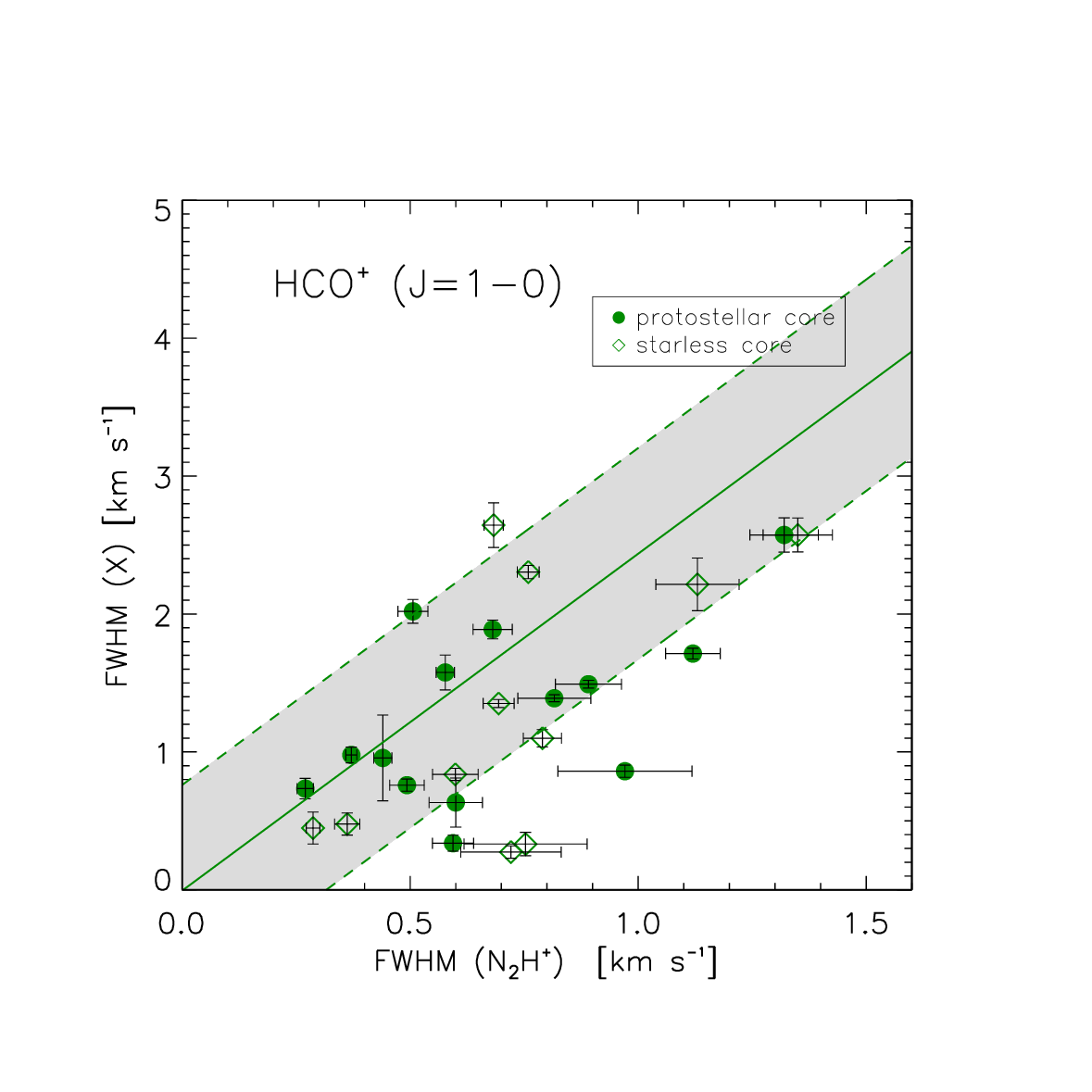}\hspace*{-0.25\textwidth}\vspace*{-0.3\textwidth}}
\resizebox{4.5cm}{5.5cm}{\includegraphics{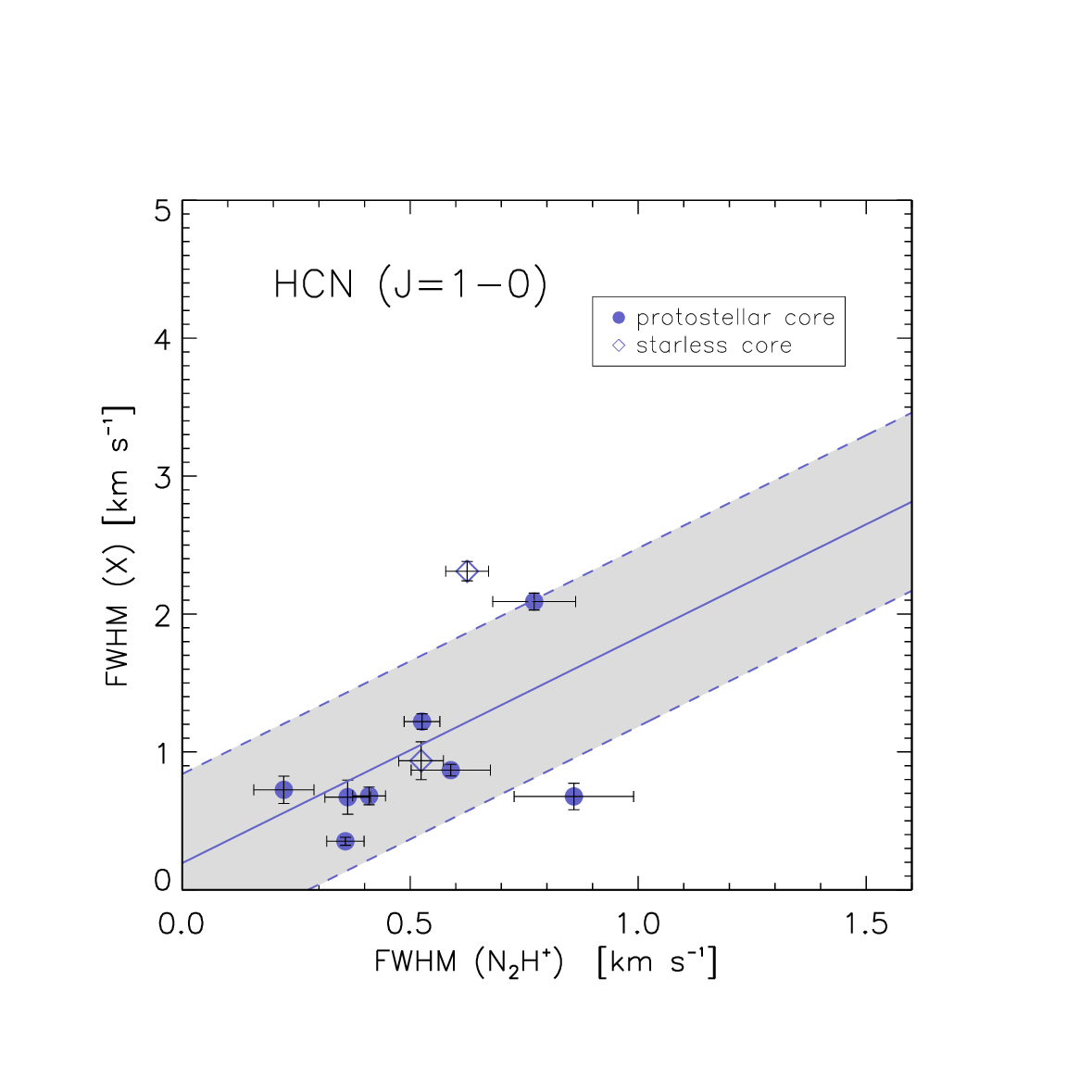}\hspace*{-0.25\textwidth}\vspace*{-0.3\textwidth}}
\resizebox{4.5cm}{5.5cm}{\includegraphics{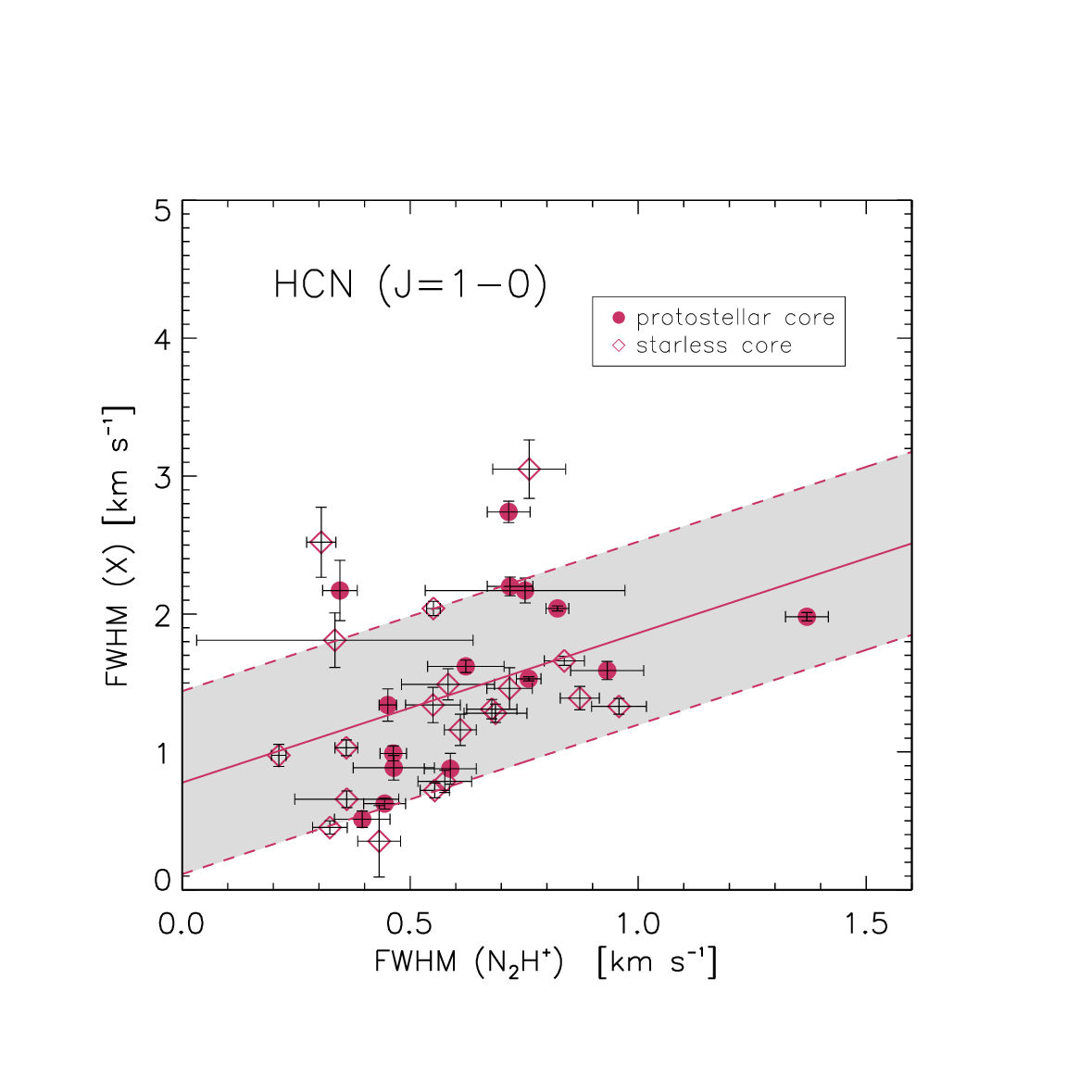}\hspace*{-0.25\textwidth}\vspace*{-0.3\textwidth}}
\resizebox{4.5cm}{5.5cm}{\includegraphics{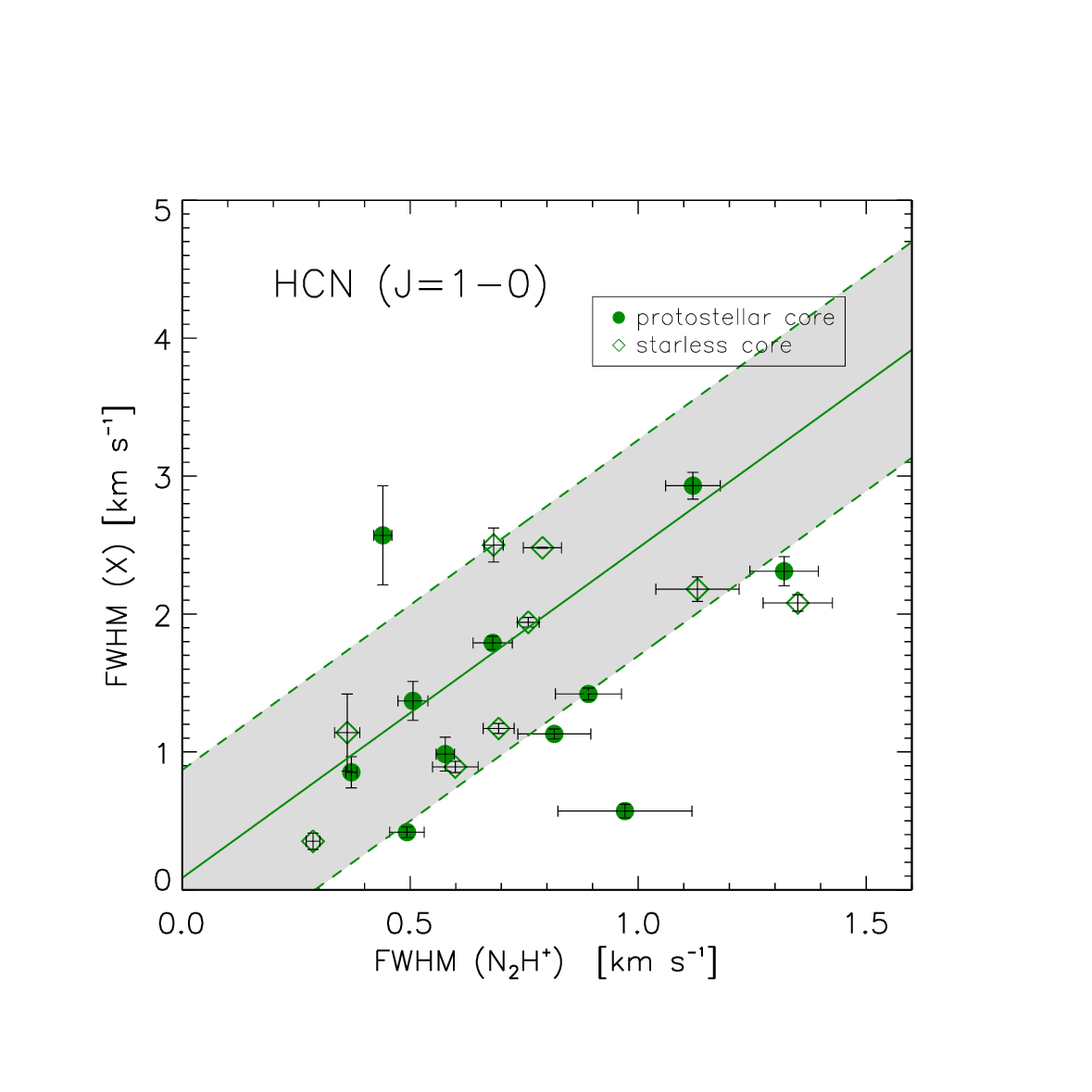}\hspace*{-0.25\textwidth}\vspace*{-0.3\textwidth}}
\resizebox{4.5cm}{5.5cm}{\includegraphics{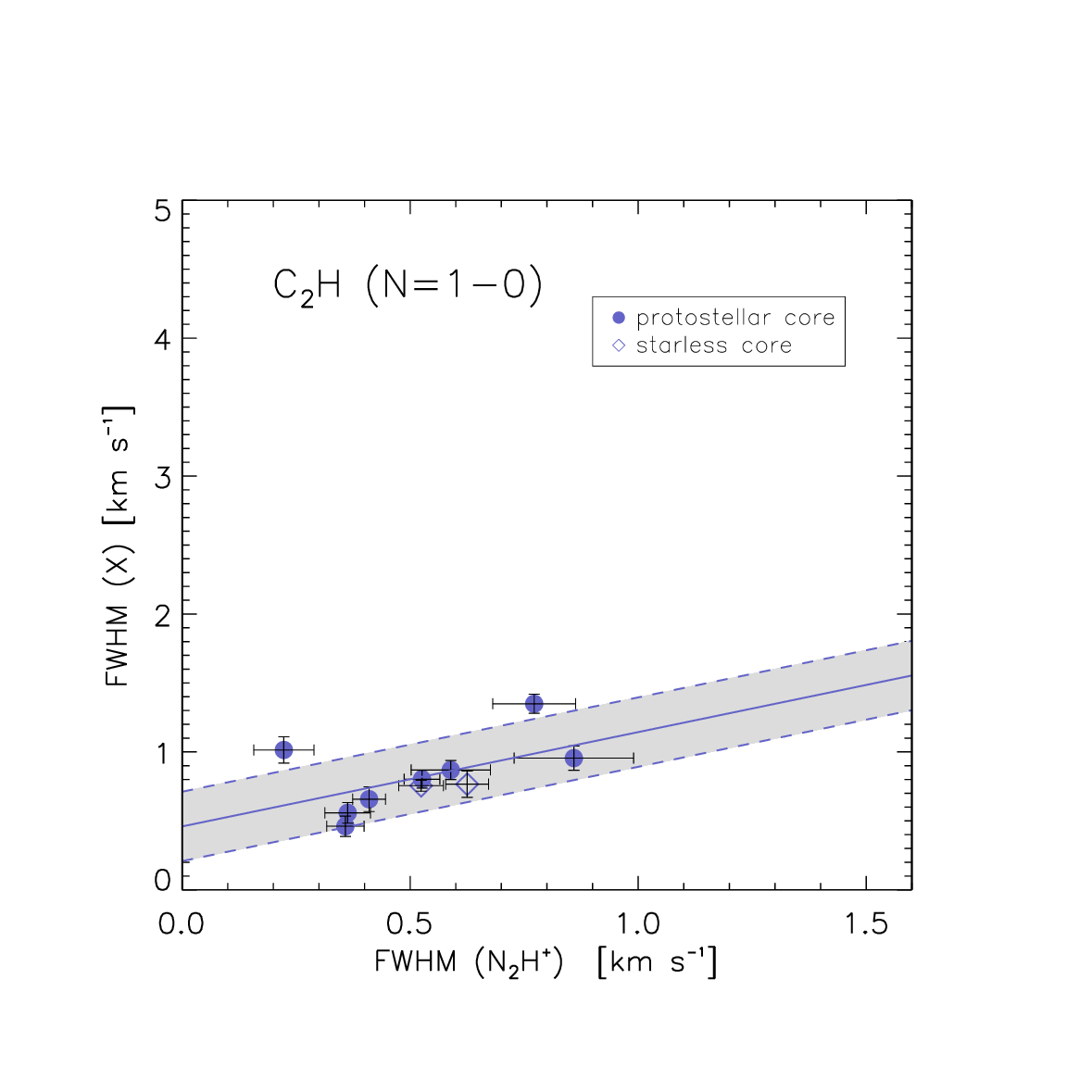}\hspace*{-0.25\textwidth}\vspace*{-0.3\textwidth}}
\resizebox{4.5cm}{5.5cm}{\includegraphics{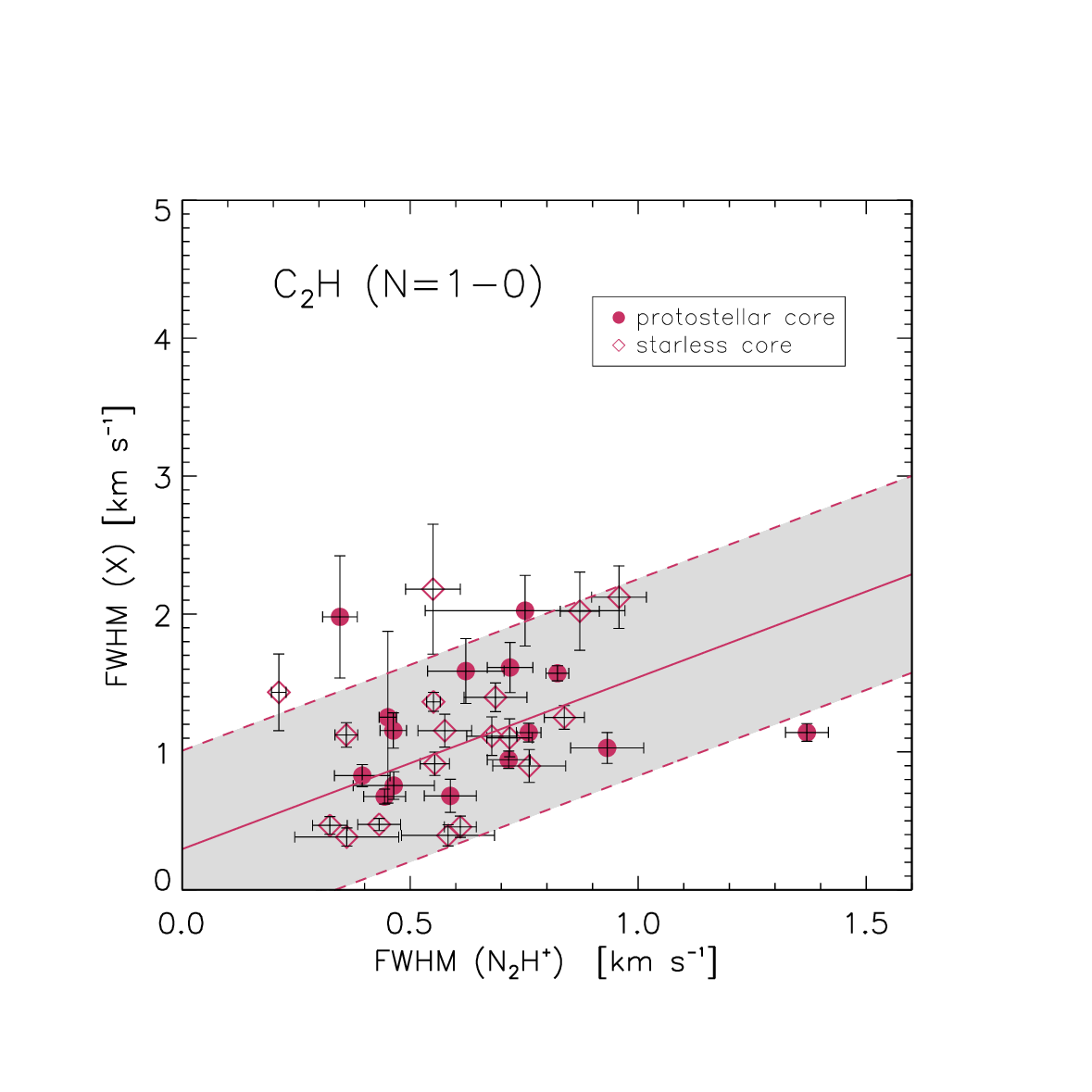}\hspace*{-0.25\textwidth}\vspace*{-0.3\textwidth}}
\resizebox{4.5cm}{5.5cm}{\includegraphics{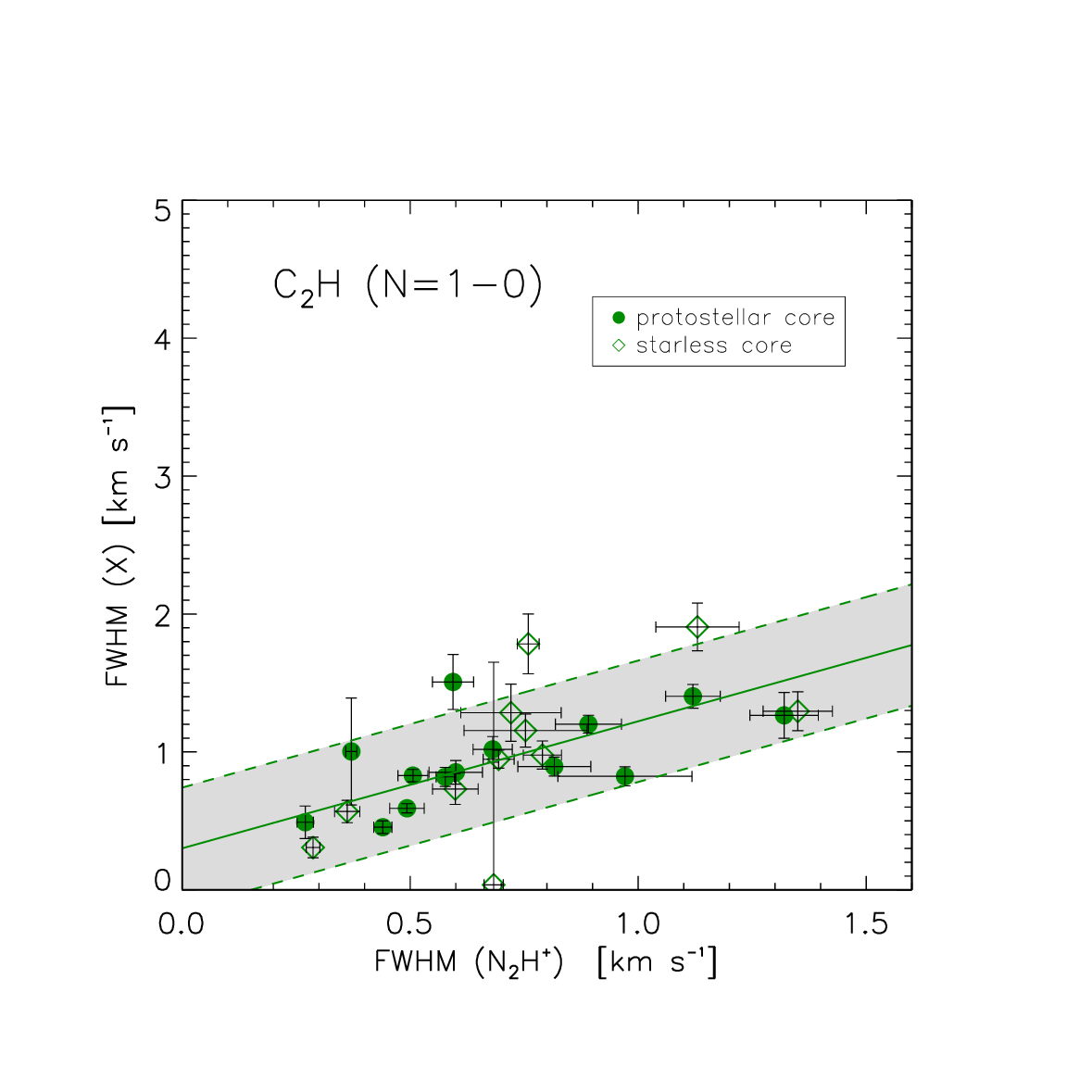}\hspace*{-0.25\textwidth}\vspace*{-0.3\textwidth}}
\end{figure}

\clearpage

\begin{figure}[b]
\centering
\resizebox{4.5cm}{5.5cm}{\includegraphics{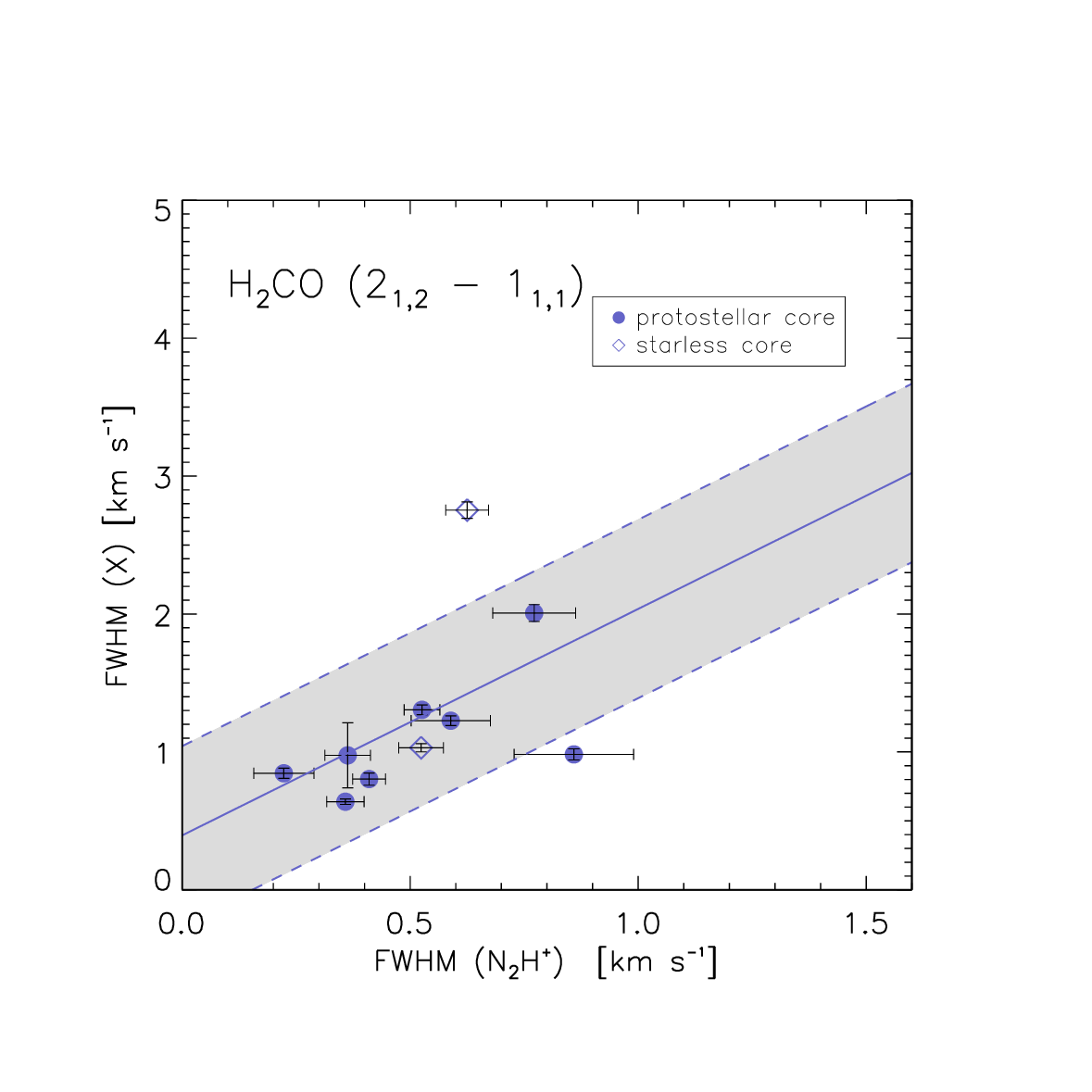}\hspace*{-0.25\textwidth}\vspace*{-0.3\textwidth}}
\resizebox{4.5cm}{5.5cm}{\includegraphics{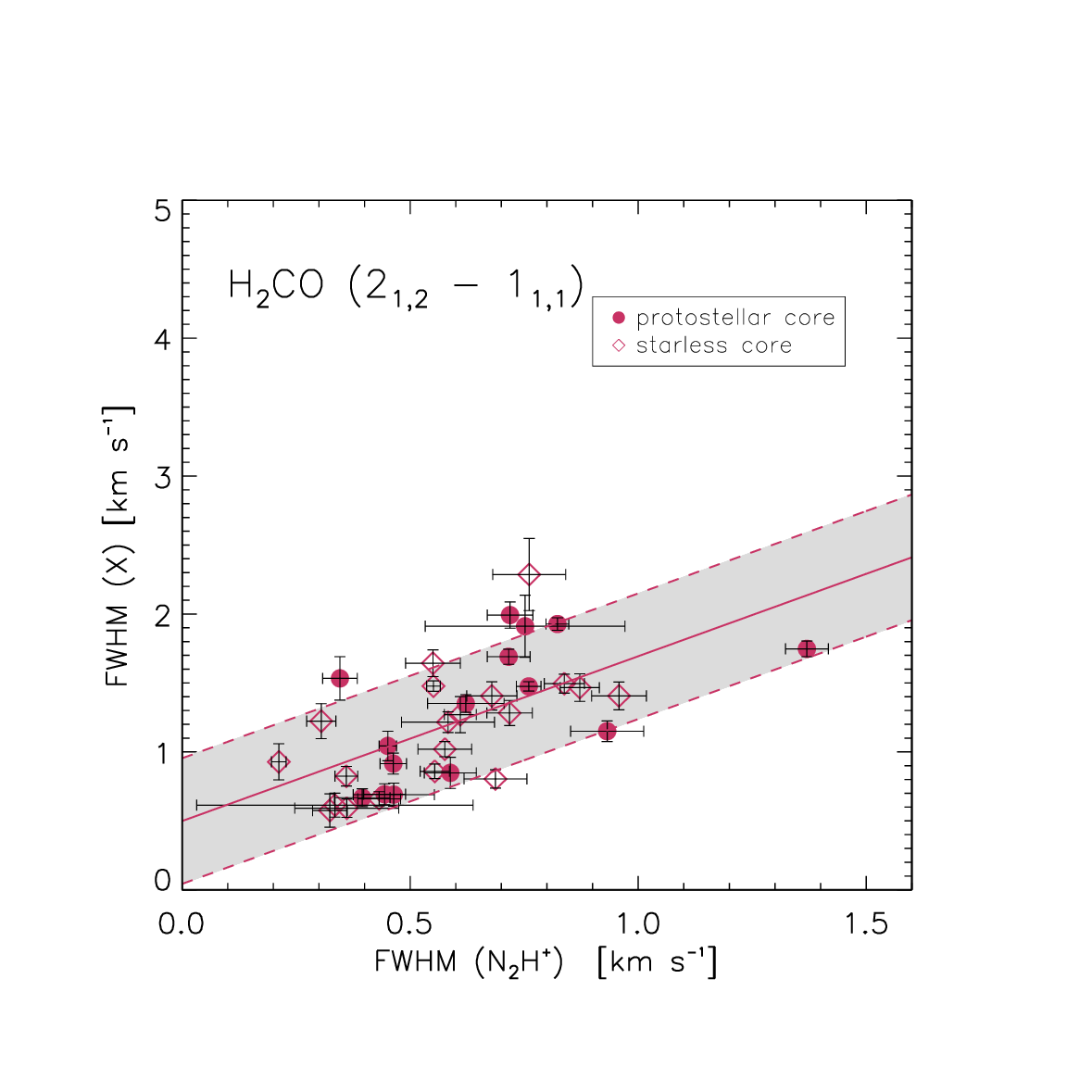}\hspace*{-0.25\textwidth}\vspace*{-0.3\textwidth}}
\resizebox{4.5cm}{5.5cm}{\includegraphics{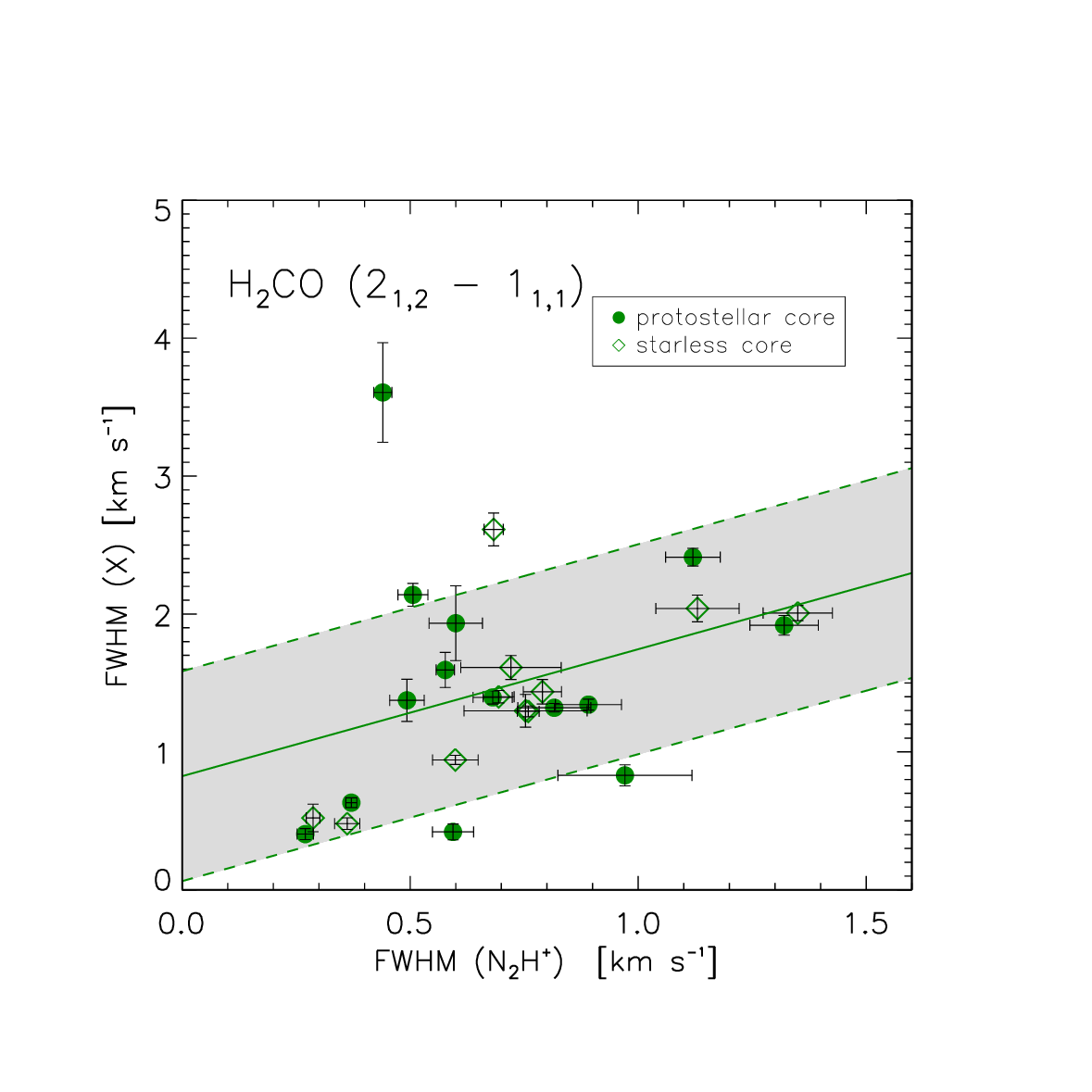}\hspace*{-0.25\textwidth}\vspace*{-0.3\textwidth}}
\resizebox{4.5cm}{5.5cm}{\includegraphics{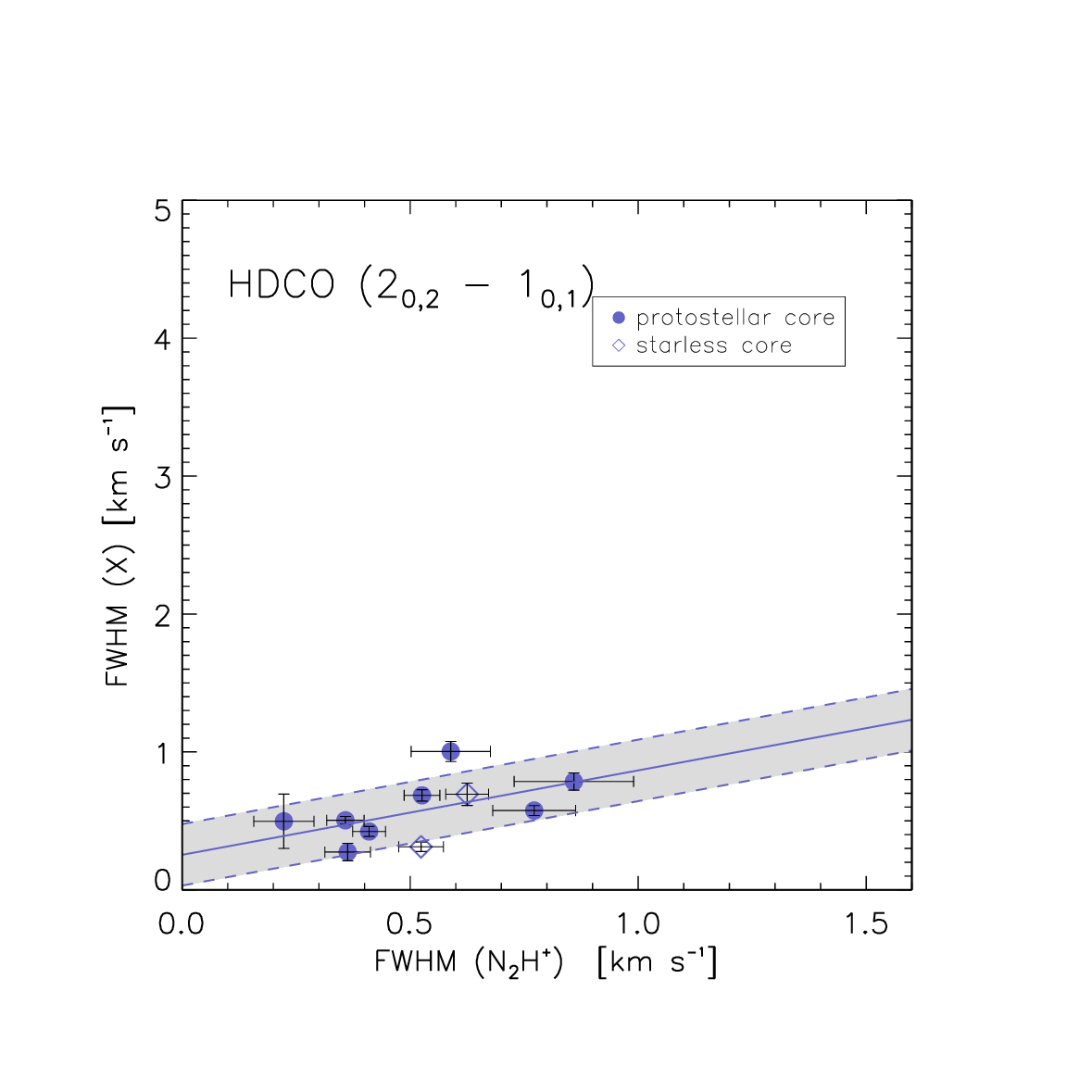}\hspace*{-0.25\textwidth}\vspace*{-0.3\textwidth}}
\resizebox{4.5cm}{5.5cm}{\includegraphics{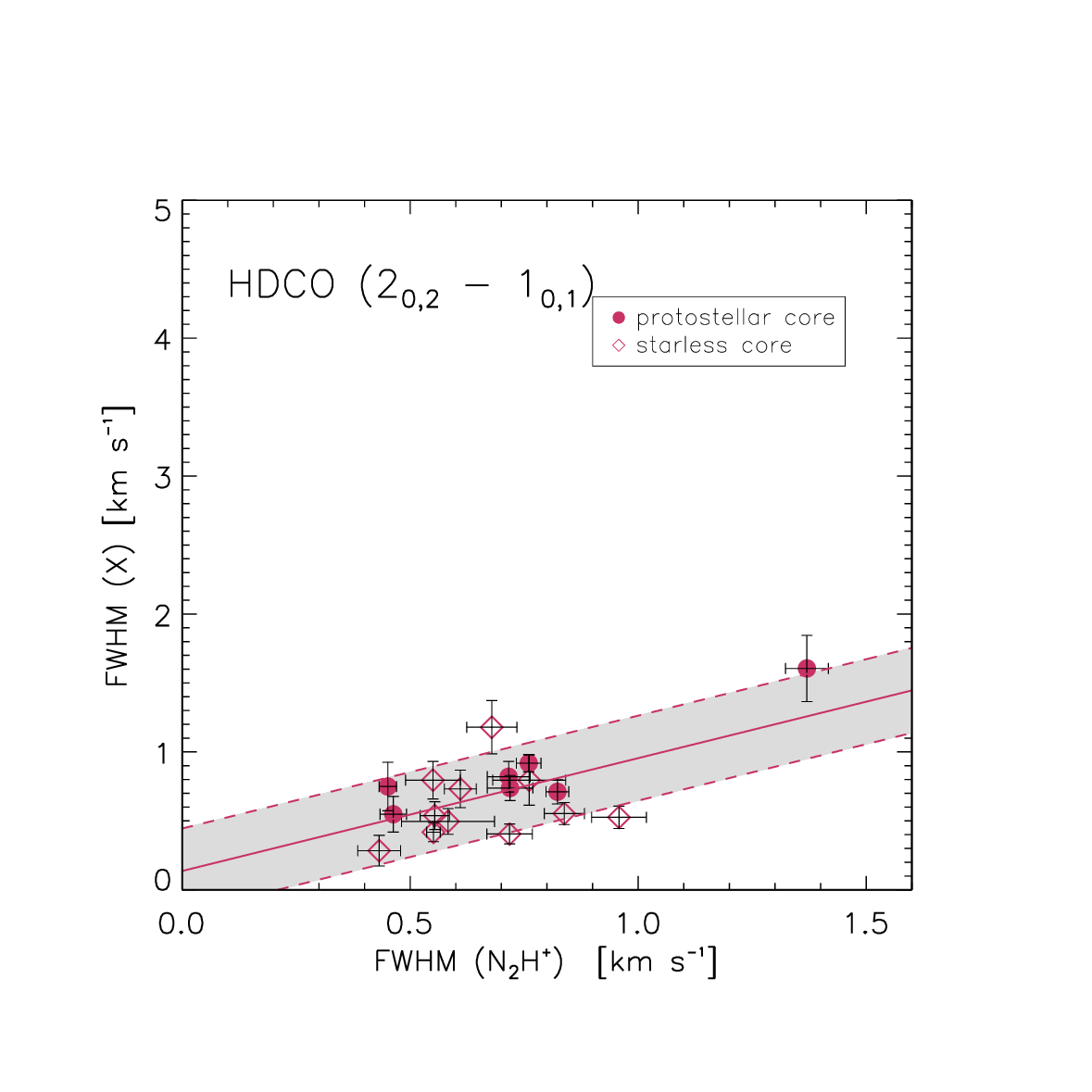}\hspace*{-0.25\textwidth}\vspace*{-0.3\textwidth}}
\resizebox{4.5cm}{5.5cm}{\includegraphics{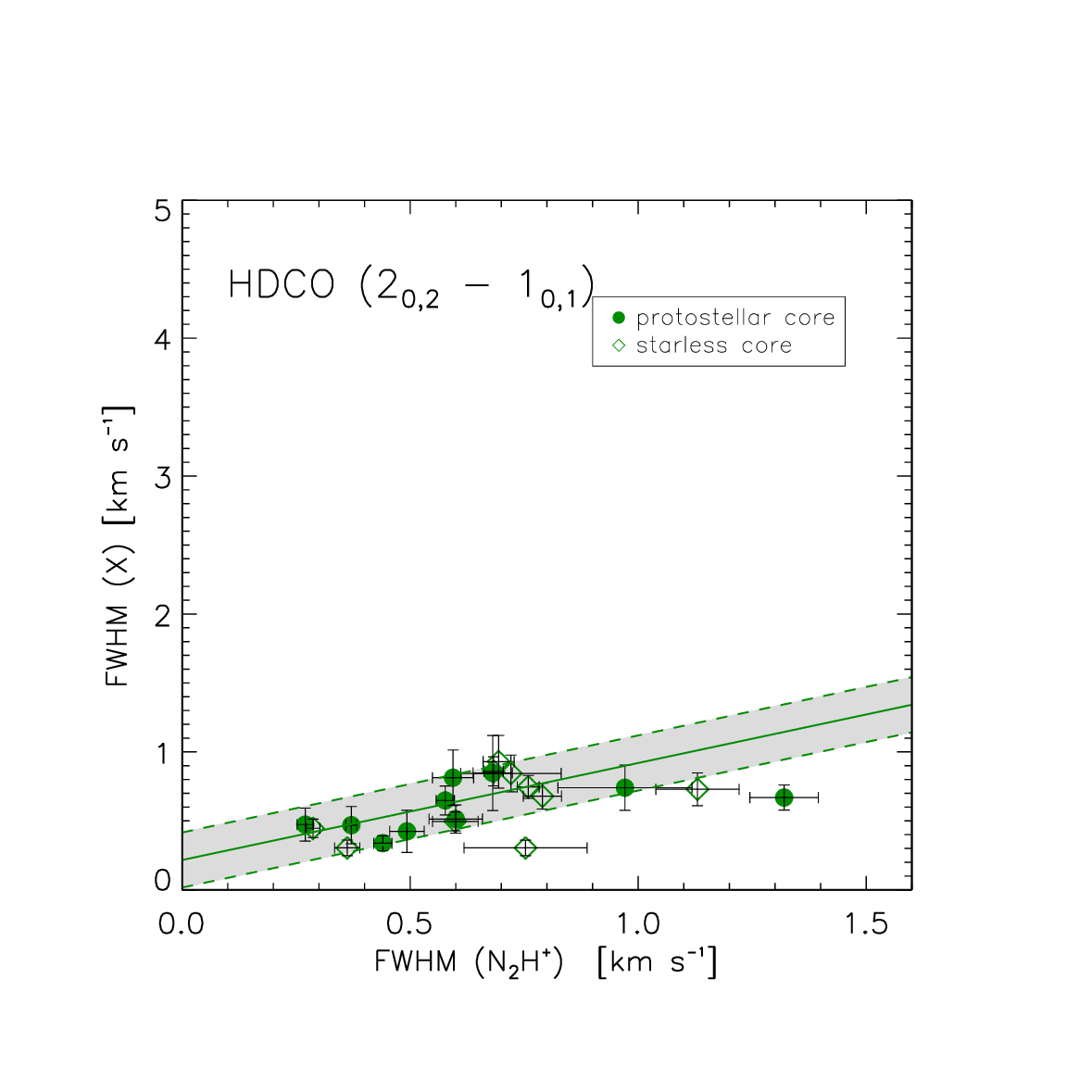}\hspace*{-0.25\textwidth}\vspace*{-0.3\textwidth}}
\caption{Relations between velocity dispersions of \nthp\ and six different molecular lines. The relations obtained from cores in the \lam, Orion A, and Orion B clouds are displayed from the left to right. Filled circles and open diamonds represent protostellar cores and starless cores, respectively. The grey shaded region in each panel presents the standard deviations ($\pm$ $\sigma_{\rm width}$) of $\Delta V_{\rm core}$ for the molecular line.} \label{fig:fig2}
\end{figure}

\clearpage

\begin{figure*}
\gridline{\fig{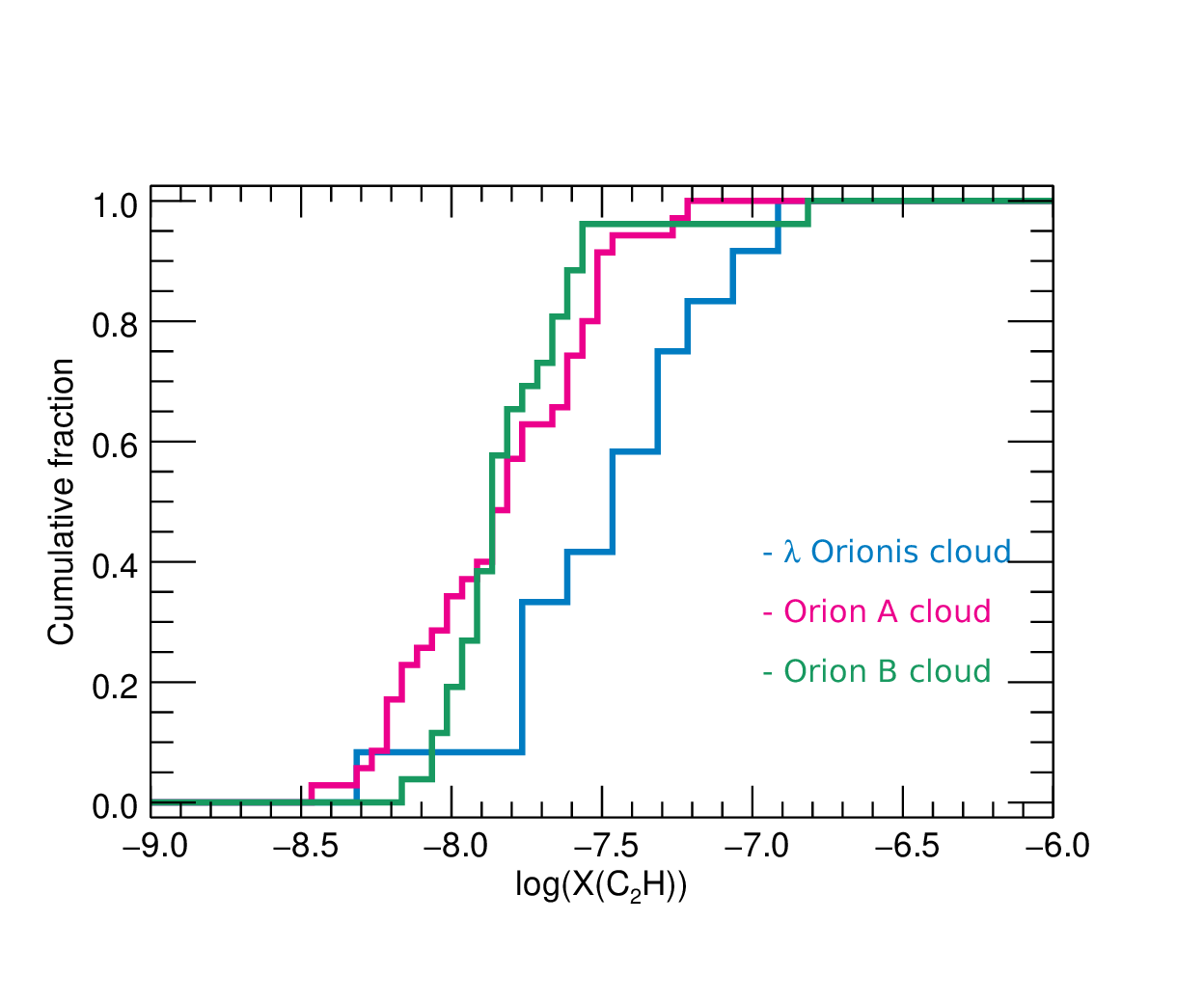}{0.5\textwidth}{(a) \cth}
          \fig{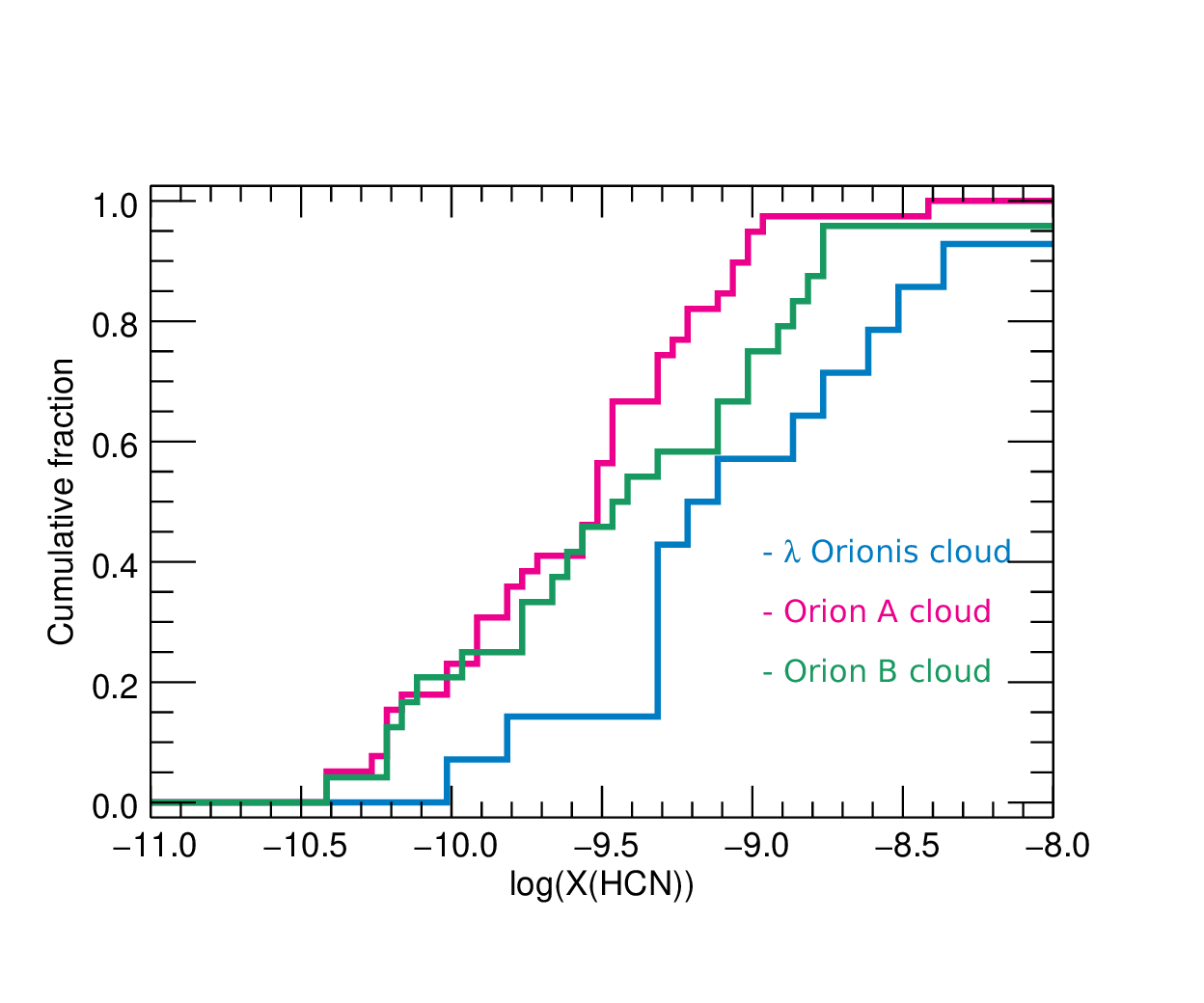}{0.5\textwidth}{(b) \hcn}
          }
\caption{Cumulative distribution of the C$_{2}$H (left panel) and HCN (right panel) abundance. 
Different colors denote different clouds.
The abundances of \cth\ and HCN are higher in the \lam\ cloud than in the Orion A and B clouds.
 \label{fig:3}} 
\end{figure*}

\begin{figure}[ht!]
 \epsscale{0.7}
\plotone{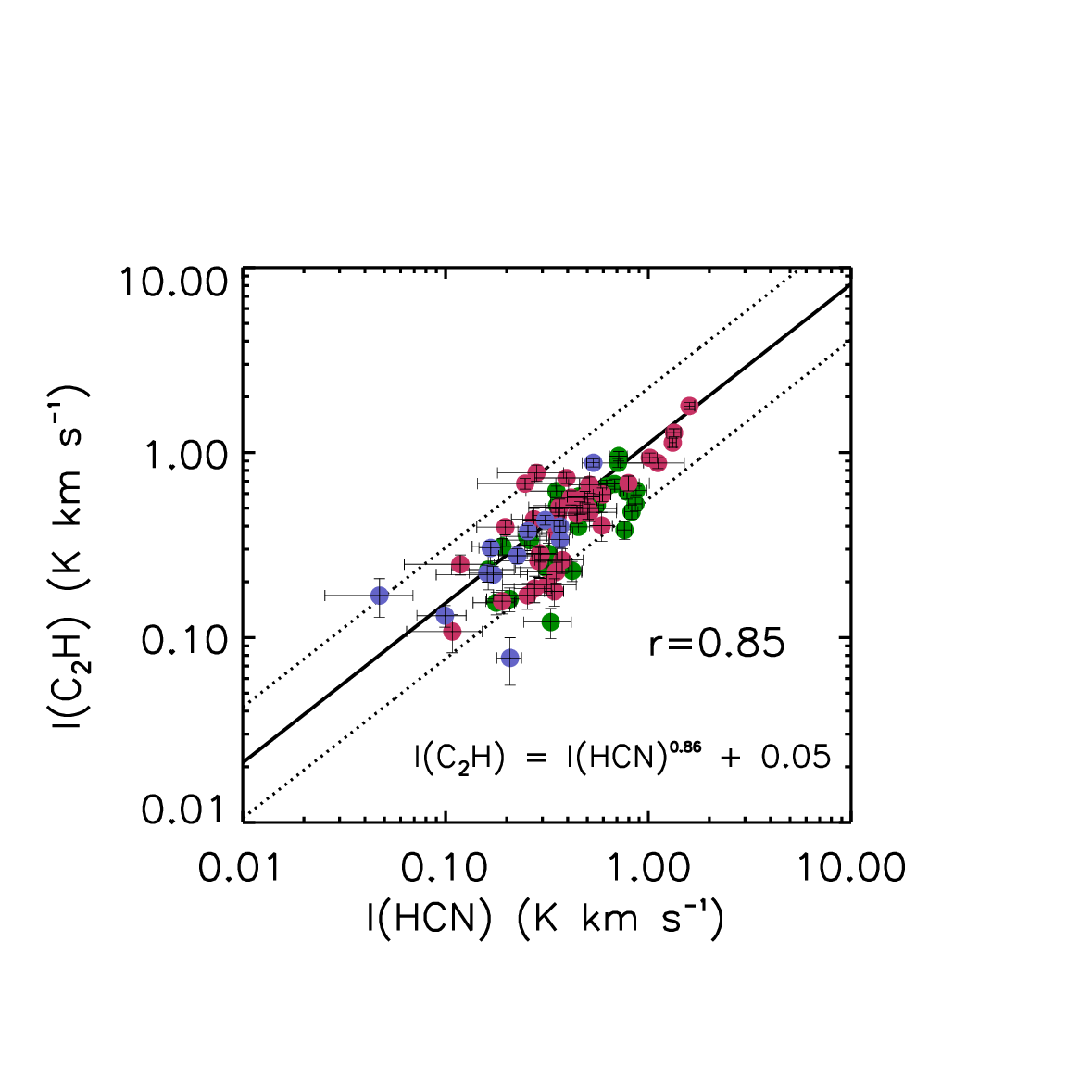}
\caption{Integrated intensities of HCN versus those of C$_{2}$H for cores in the three molecular clouds.
The black solid line indicates a least-square fit to the all data points and the dotted lines indicate 1$\sigma$ from the line.  
Different colors denote data points of different clouds ($\lambda$: blue, Orion A: magenta, Orion B: green).
The Pearson correlation coefficient and relation (log value) are given in the lower right corner.
 \label{fig:4}}
\end{figure}

\clearpage

\begin{figure*}
\gridline{\fig{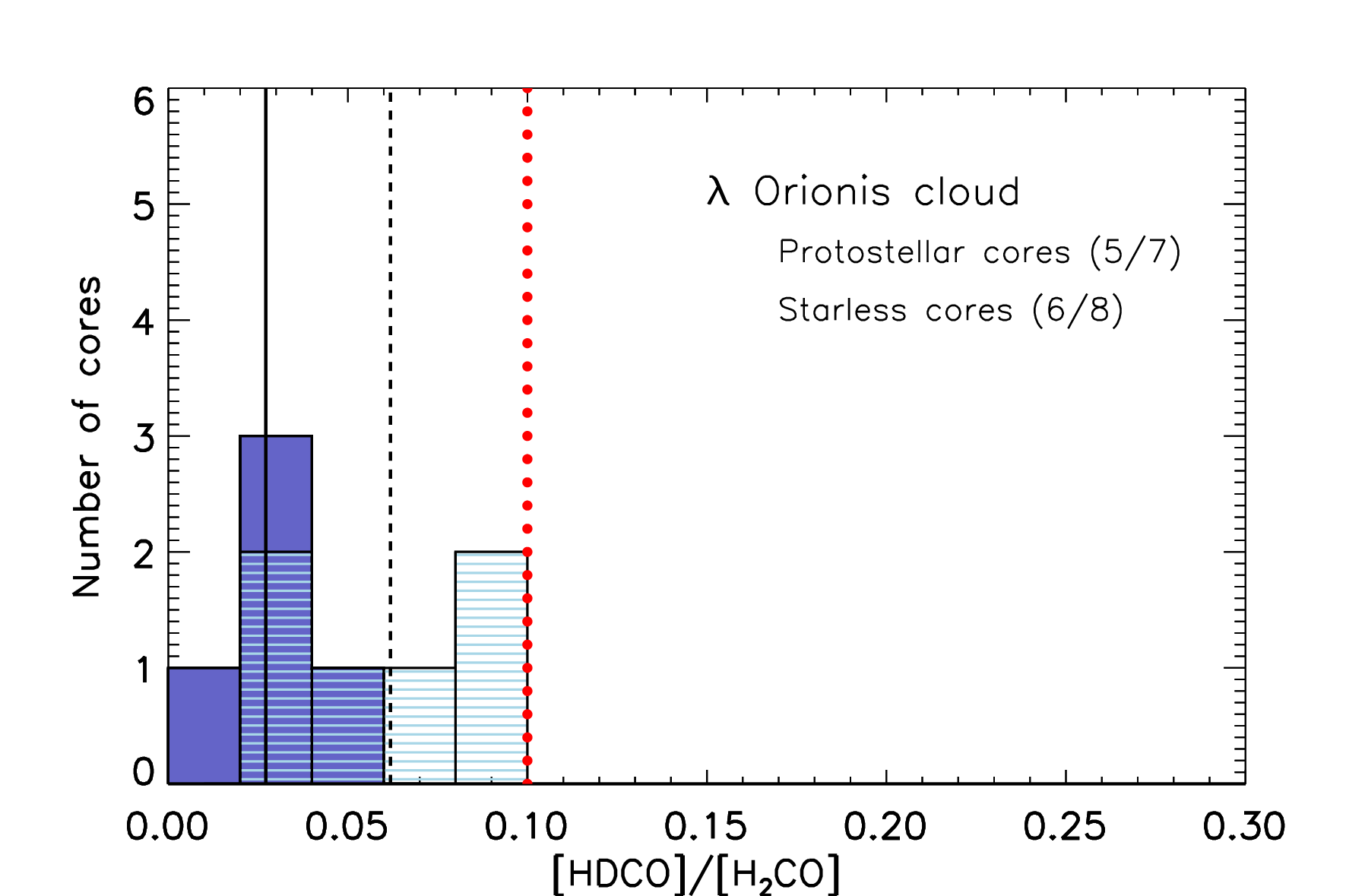}{0.35\textwidth}{(a) $\lambda$ Orionis cloud}
          \fig{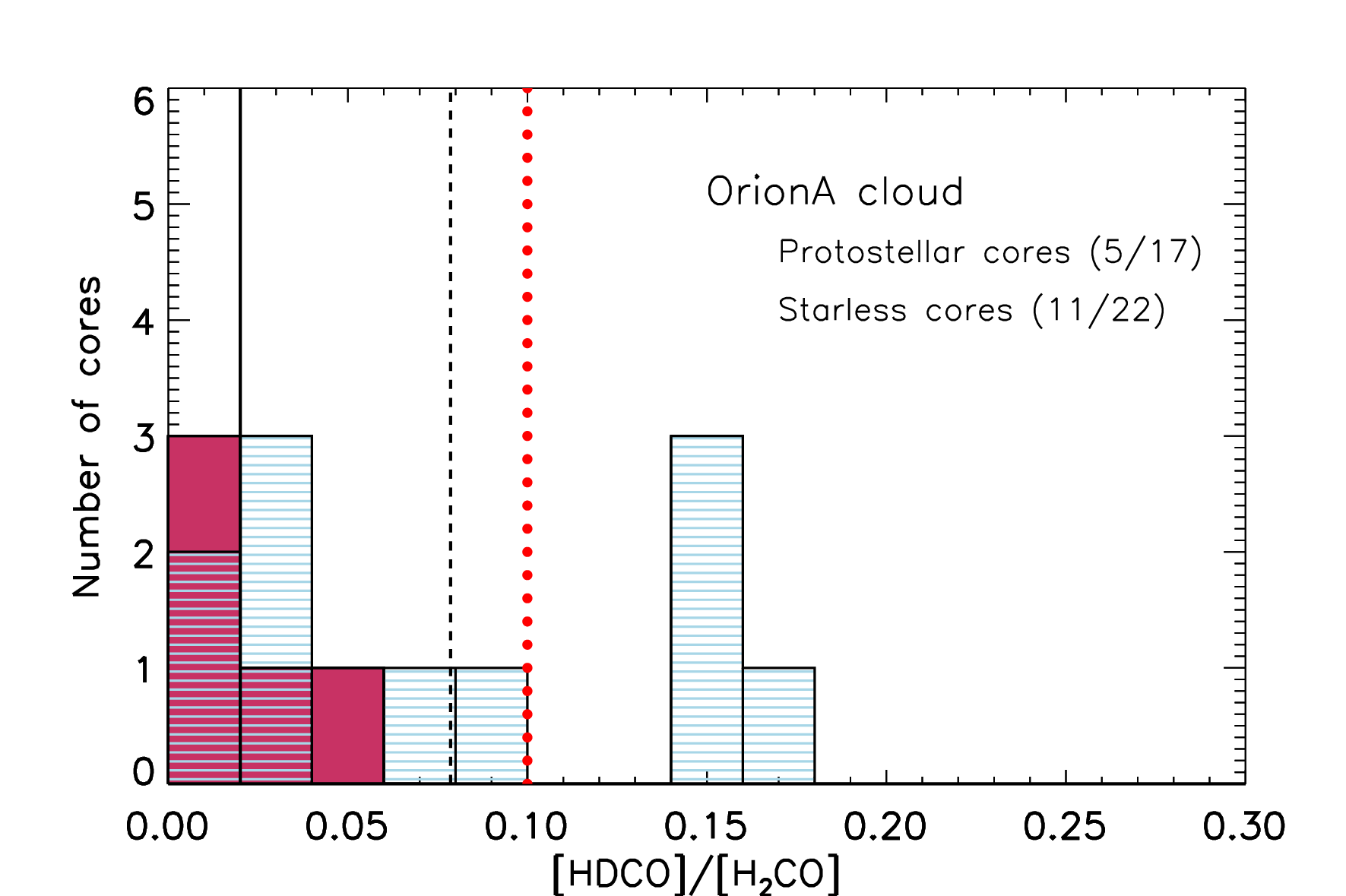}{0.35\textwidth}{(b) Orion A cloud}
          \fig{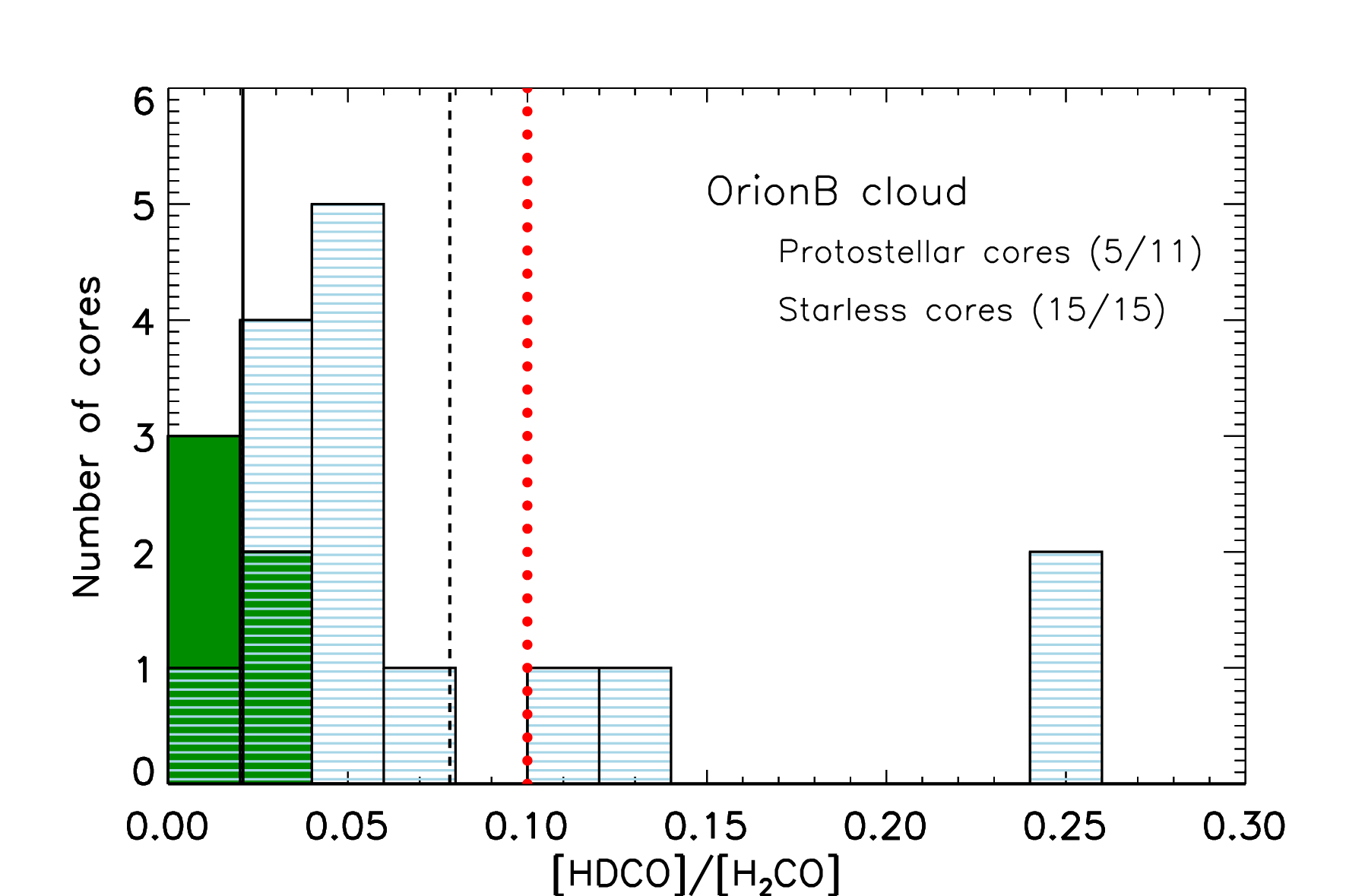}{0.35\textwidth}{(c) Orion B cloud}
          }
\caption{Distributions of [HDCO]/[\htco] in three different clouds. 
 The [HDCO]/[\htco] distributions of protostellar and starless cores are presented by filled and shaded histograms, respectively. 
 The numbers of detected protostellar and starless cores are presented in each panel.
 The black solid and dashed lines represent mean values of protostellar and starless cores, respectively.
 The red dotted line indicates the [HDCO]/[\htco] ratio of 0.1.
\label{fig:5}}
\end{figure*}

\begin{figure*}
\gridline{\fig{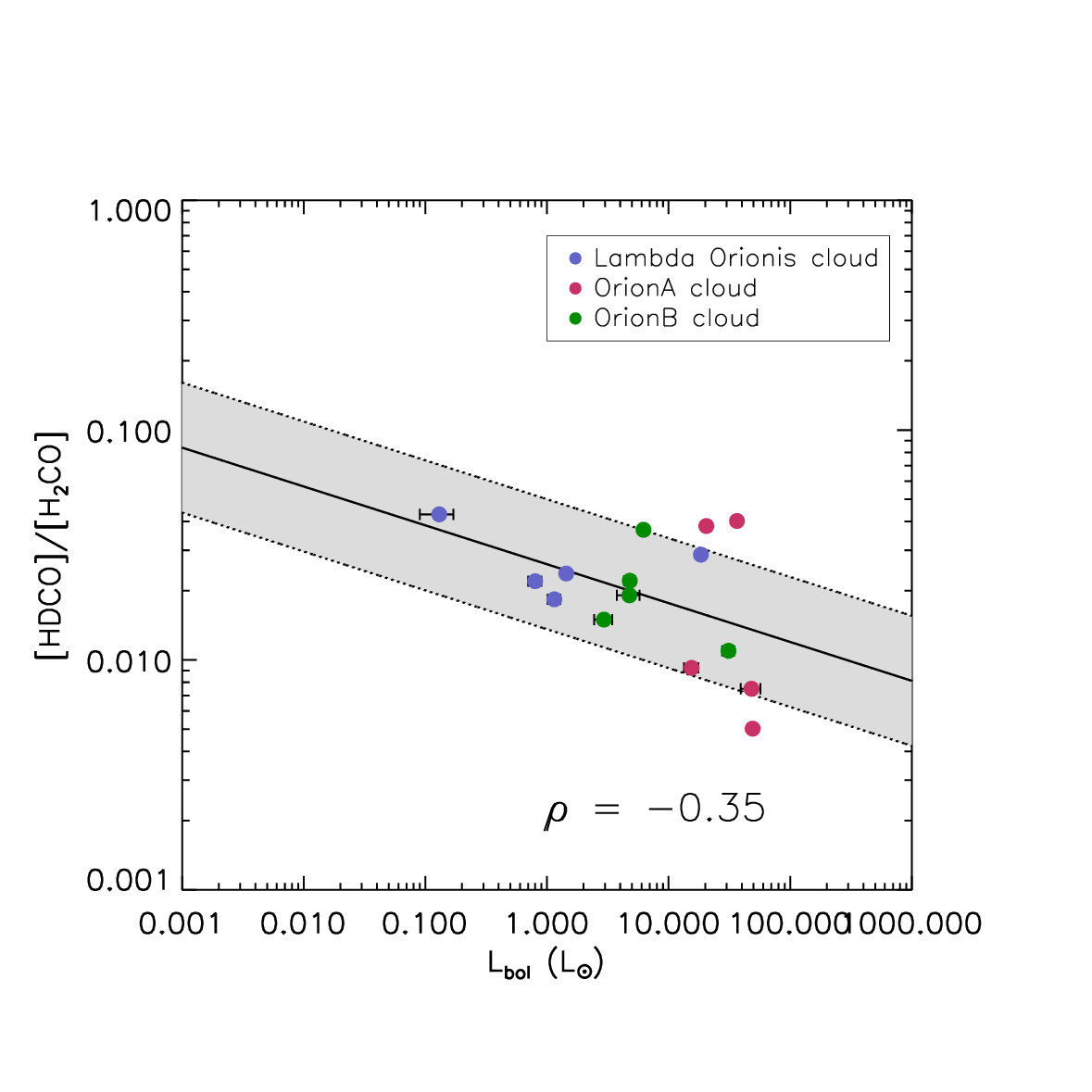}{0.38\textwidth}{}\hspace*{-0.07\textwidth}  
          \fig{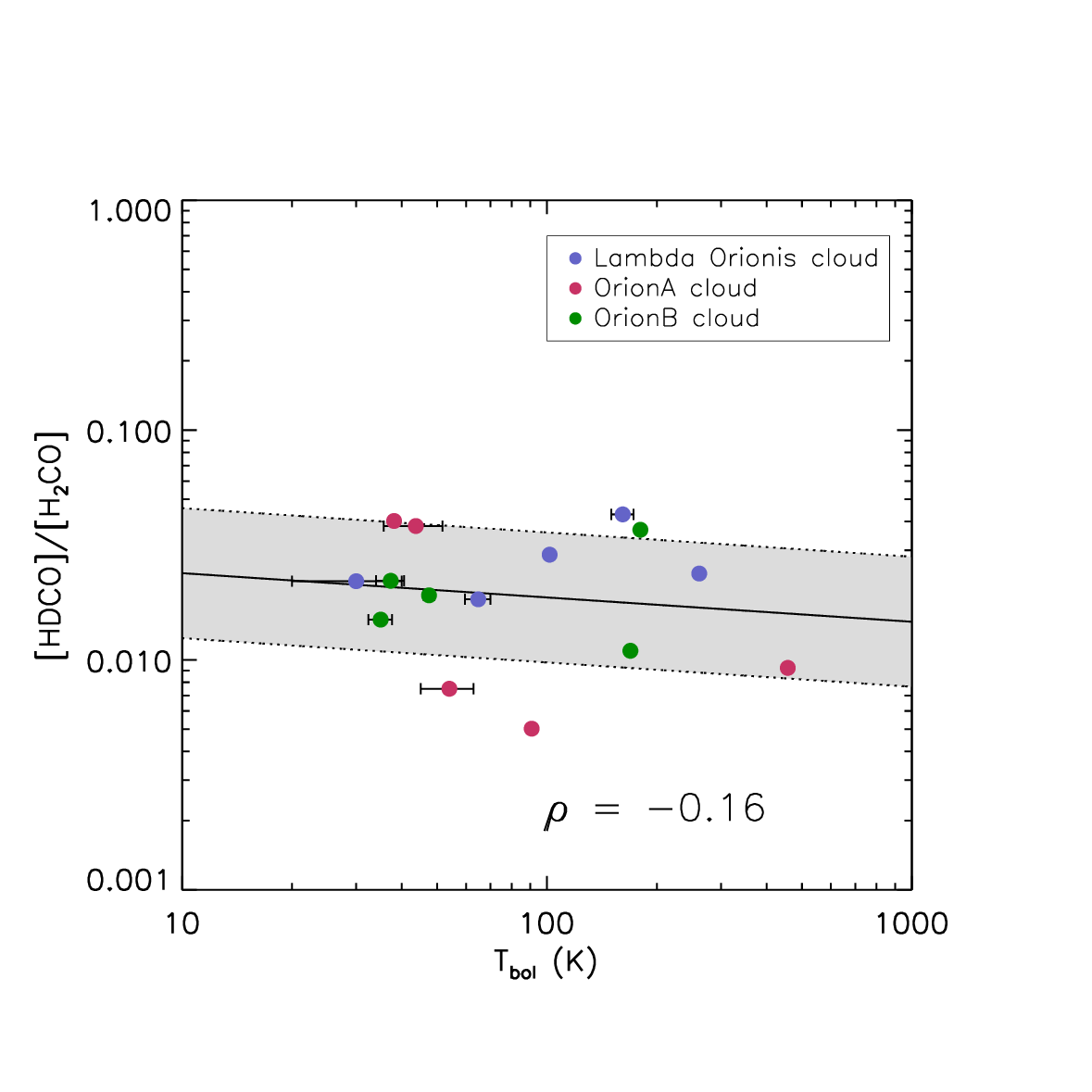}{0.38\textwidth}{}\hspace*{-0.07\textwidth}
          \fig{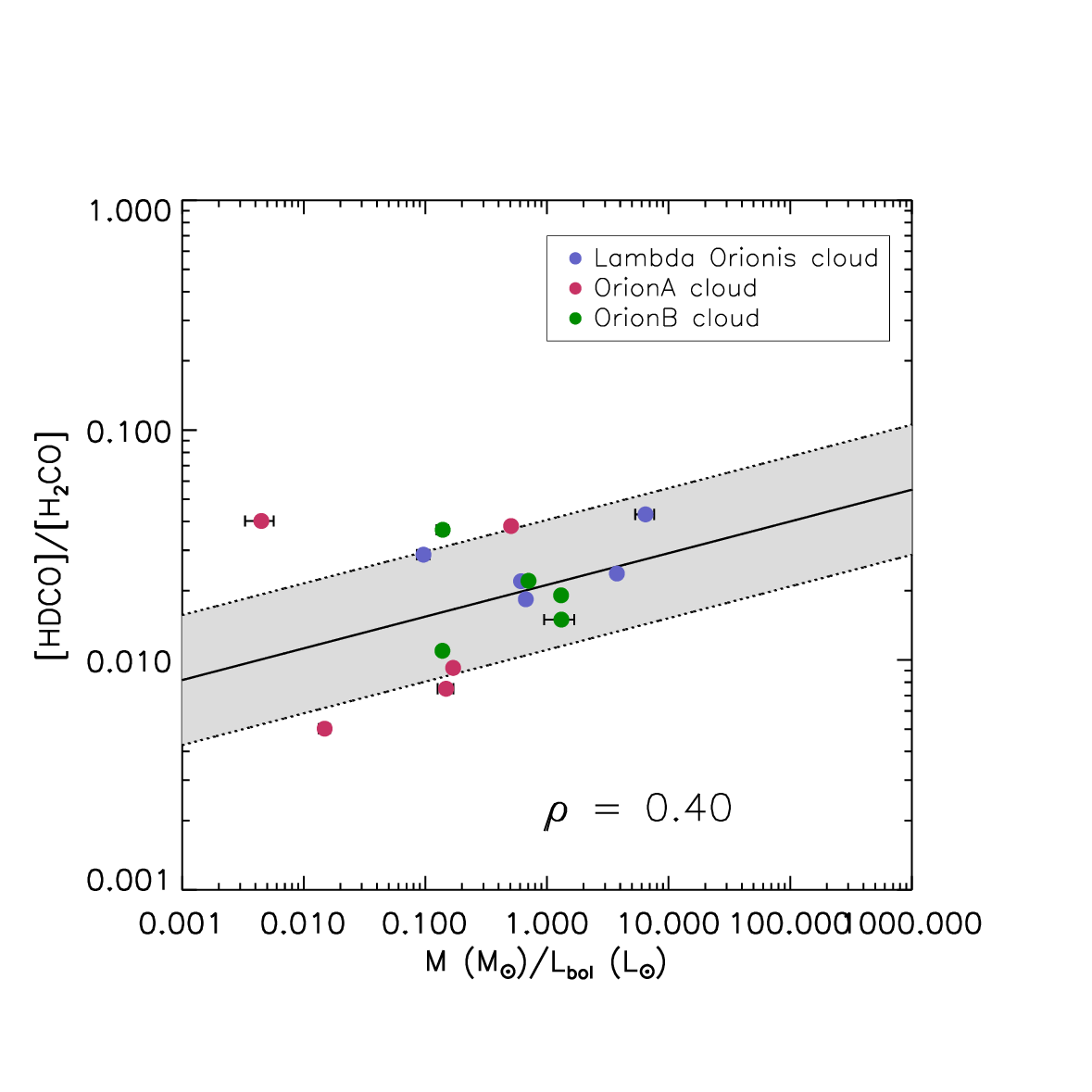}{0.38\textwidth}{}\hspace*{-0.07\textwidth}
         }
\caption{[HDCO]/[$\rm H_{2}CO$] ratios for protostellar cores in each cloud with respect to the bolometric luminosity (left), the temparature (middle), and the core mass-to-bolometric luminosity ratios (right). 
Different colors indicate different regions as used in Figure \ref{fig:1}. 
The gray shaded region corresponds to the standard deviation (0.01) of the D/H ratios from the linear least-square fitting result.
The Pearson correlation coefficient is shown in the bottom right corner of each panel.
\label{fig:6}}
\end{figure*}          
                    
\clearpage        

\begin{figure*}
\gridline{\fig{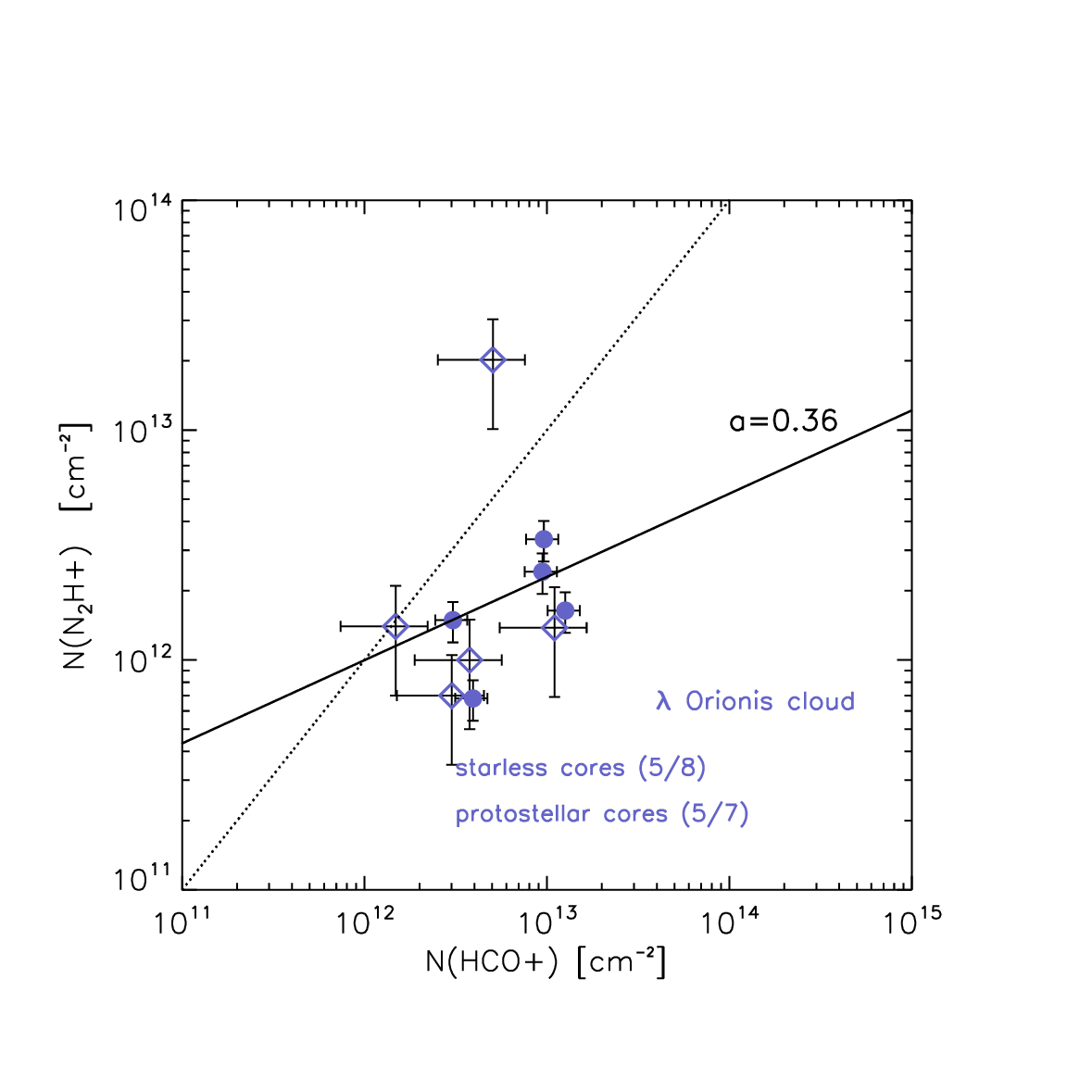}{0.37\textwidth}{(a) $\lambda$ Orionis cloud}\hspace*{-0.05\textwidth}
          \fig{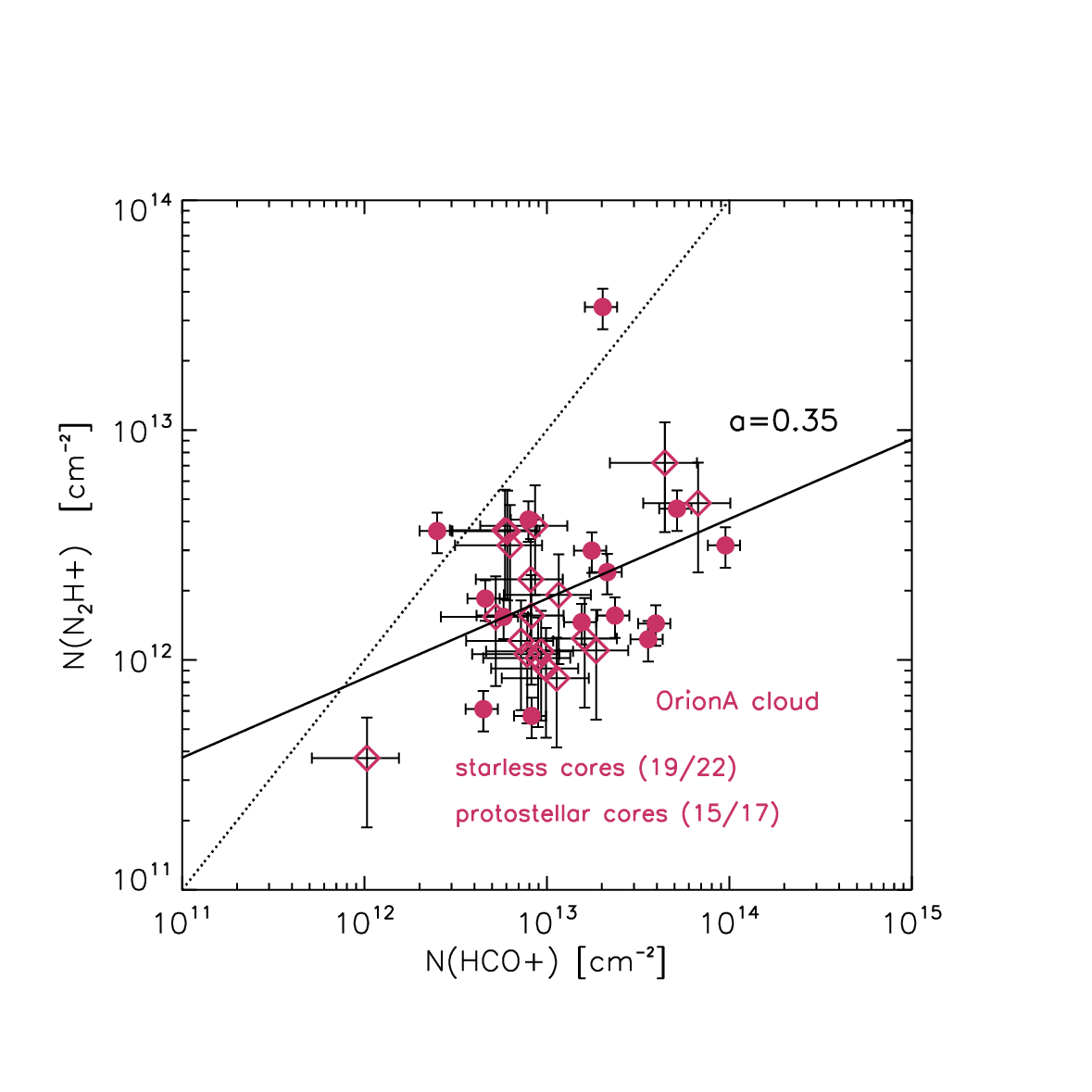}{0.37\textwidth}{(b) Orion A cloud}\hspace*{-0.05\textwidth}
          \fig{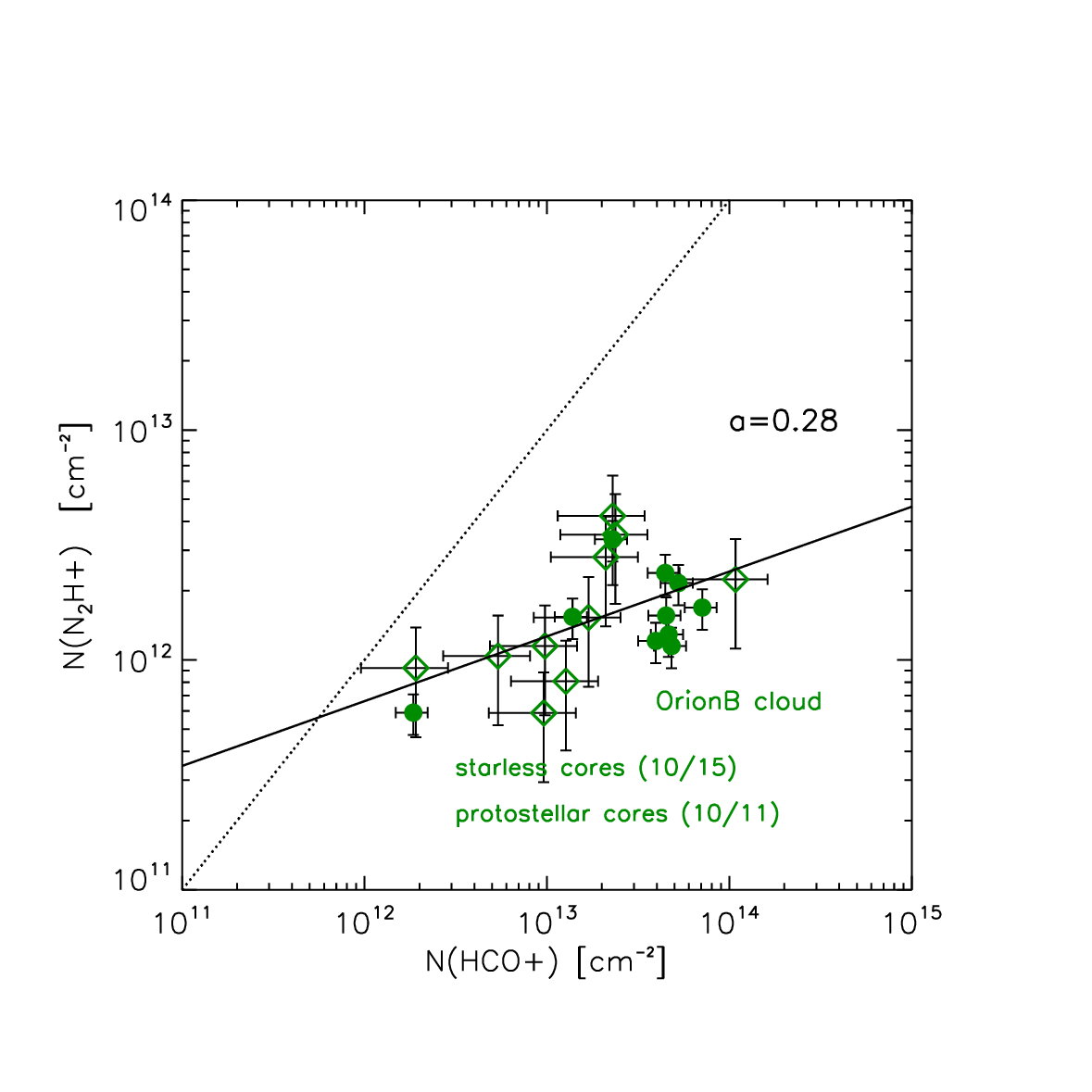}{0.37\textwidth}{(c) Orion B cloud}\hspace*{-0.05\textwidth}
          }
\caption{Column density of $\rm N_{2}H^+$ with respect to the column density of $\rm HCO^+$.
Filled circles and open diamonds represent protostellar and starless cores, respectively. 
The solid and dotted lines indicate least-square fits to the total cores and the ratio of 1.0, respectively.
The slope of the fitted relation is given in the upper right corner of each panel.
\label{fig:7}}
\end{figure*}

\clearpage

\begin{figure}[ht!]
\plotone{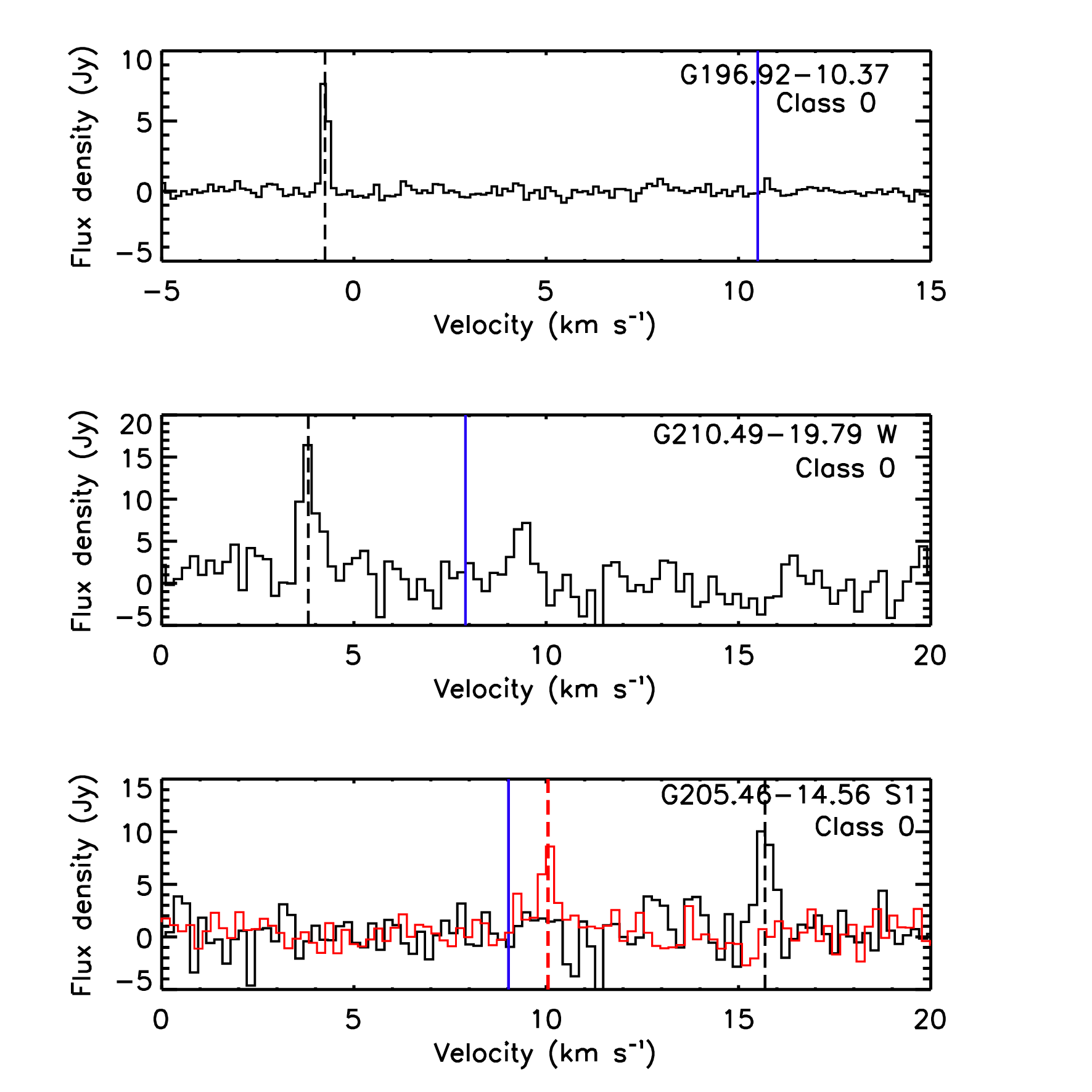}
\caption{Detected \hto\ maser spectra.  
The name and evolutionary stage of a given central object are shown at the upper right corner.
The vertical lines indicate the systemic velocity (blue solid line) and the maser peak velocity (black dashed line). 
For G205.46-14.56 S1, the detected 44 GHz methanol maser spectrum is also presented in red color and its peak velocity is marked by the red vertical line. 
 \label{fig:8}}
\end{figure}

\begin{figure}
\figurenum{9}
\plotone{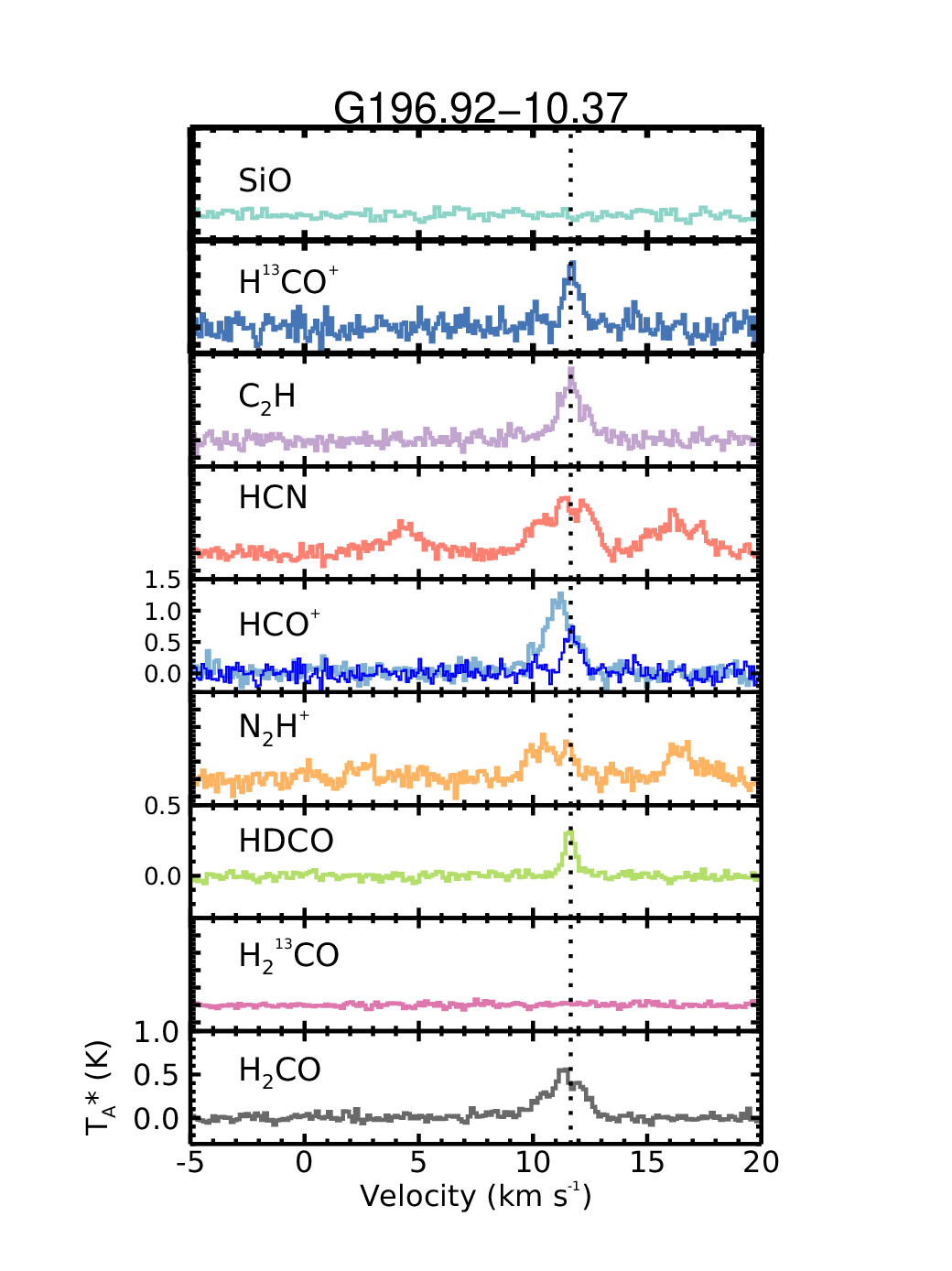}
\caption{Spectra of 80 cores observed with the KVN telescopes in nine molecular lines. The vertical dotted line indicates the systemic velocity of the core. The y-axis is 0 K to 1 K in  $T_{A}^*$
scale as shown in panel of \htco\ except for some molecules such as \hcop\ and HDCO. The blue thin line in the \hcop\ panel represent the \htcop\ line emission.
\textbf{The complete figure set (80 images) is available in the online journal.}}
\label{fig:9}
\end{figure}

\begin{deluxetable*}{lclCrlc}
\tablecaption{Observed Transitions and Telescope Parameters \label{tab:1}}
\tablecolumns{7}
\tablenum{1}
\tablewidth{0pt}
\tablehead{
\colhead{Molecule \tablenotemark{}} &
\colhead{Rest frequency\tablenotemark{}} &
\colhead{Transition \tablenotemark{}} &
\colhead{HPBW \tablenotemark{a}} & \colhead{$\eta_{\rm mb}$} & \colhead{$T_{\rm sys}$} \\
\colhead{} & \colhead{(GHz)} &  \colhead{} &
\colhead{(\arcsec)} & \colhead{YS, US, TN} & \colhead{}
}
\startdata
H$_2$O   & 22.23508 & 6$_{16} - 5_{23}$ &  126 & 0.45, 0.49, 0.48 & 130 \\
SiO  & 43.42379 & $v=0, \, J= 1-0$ & 63 & 0.48, 0.46, 0.48 & 150 \\
CH$_3$OH   & 44.06943 & 7$_{0} - 6_{1} \, ,\,A^+$ & 60 & 0.47, 0.45, 0.47 & 240\\
H$^{13}$CO$^{+}$ & 86.75428 &  $J=1-0$ & 32 & 0.41, 0.44, 0.44 & 260 \\
C$_{2}$H & 87.31689 &  $N=1-0, J=5/2-3/2, F=2-1$ & 31 & 0.40, 0.43, 0.43 & 190 \\
HCN & 88.63184 &  $J=1-0$ & 30 & 0.39, 0.41, 0.41 & 220 \\
HCO$^+$ & 89.18852 &  $J=1-0$ & 30 & 0.38, 0.41, 0.41 & 230 \\
N$_2$H$^+$ & 93.17339 &  $J=1-0$ & 29 & 0.35, 0.37, 0.37 & 190 \\
HDCO & 128.81286 &  2$_{0,2} - 1_{0,1}$ & 21 & 0.39, 0.36, 0.36 & 210 \\
H$_{2}$$^{13}$CO & 137.44995 & 2$_{1,2} - 1_{1,1}$ & 20 & 0.34, 0.32, 0.32 & 200 \\
H$_2$CO & 140.83952 &  2$_{1,2} - 1_{1,1}$ & 19 & 0.33, 0.30, 0.30 & 220 \\
\enddata
\tablenotetext{a}{ The Half Power Beam Width of three single-dish in Yonsei, Ulsan, and Tamna site.}
\tablecomments{\hto\ and \chtoh\ are maser emission lines.}
\end{deluxetable*}

\clearpage

\startlongtable
\begin{deluxetable*}{ll  ccc c ccc c ccc c ccc} 
\tablecaption{Properties of \htcop, \cth, HCN, and \hcop\ Lines} 
\tabletypesize{\scriptsize}\tablecolumns{17}
\tablewidth{0pc} \setlength{\tabcolsep}{0.05in}
\tablenum{2}
 \tablehead{\\
 Cloud & Core &\multicolumn{3}{c}{\htcop} && \multicolumn{3}{c}{\cth} &&
\multicolumn{3}{c}{HCN} && \multicolumn{3}{c}{\hcop}   \\
 \cline{3-5} \cline{7-9} \cline{11-13} \cline{15-17}
           &    & $T_{\rm peak}$ & $V_{\rm LSR}$  & FWHM & &$T_{\rm peak}$ & $V_{\rm LSR}$  & FWHM  & &$T_{\rm peak}$ & $V_{\rm LSR}$  & FWHM  & &$T_{\rm peak}$ & $V_{\rm LSR}$  & FWHM \\
    &  & (K) &    \multicolumn{2}{c}{(km s$^{-1}$)} &&    (K) &    \multicolumn{2}{c}{(km s$^{-1}$)} &&  (K) &    \multicolumn{2}{c}{(km s$^{-1}$)} &&  (K) &    \multicolumn{2}{c}{(km s$^{-1}$)} 
}\startdata
$\lambda$ Orionis	&	G190.15-13.75N	&	\nodata	&	\nodata	&	\nodata	&&	\nodata	&	\nodata	&	\nodata	&&	\nodata	&	\nodata	&	\nodata	&&	\nodata	&	\nodata	&	\nodata	\\
	&	G190.15-13.75S	&	0.23	&	1.32	&	0.51	&&	\nodata	&	\nodata	&	\nodata	&&	0.10	&	1.61	&	1.52	&&	0.88	&	1.27	&	0.97	\\
	&	G191.90-11.21N	&	\nodata	&	\nodata	&	\nodata	&&	\nodata	&	\nodata	&	\nodata	&&	0.49	&	10.60	&	2.26	&&	0.42	&	10.87	&	3.91	\\
	&	G191.90-11.21S	&	0.20	&	10.55	&	0.87	&&	0.27	&	10.57	&	0.46	&&	0.19	&	10.50	&	0.35	&&	0.65	&	10.51	&	1.03	\\
	&	G192.12-10.90N	&	0.43	&	9.96	&	0.67	&&	0.54	&	9.85	&	0.76	&&	0.52	&	9.87	&	0.94	&&	0.66	&	9.61	&	1.48	\\
	&	G192.12-10.90S	&	\nodata	&	\nodata	&	\nodata	&&	0.09	&	10.11	&	1.76	&&	0.16	&	9.83	&	0.64	&&	\nodata	&	\nodata	&	\nodata	\\
	&	G192.12-11.10	&	0.47	&	10.22	&	0.65	&&	0.27	&	10.12	&	0.96	&&	0.25	&	10.10	&	0.68	&&	0.26	&	10.29	&	1.49	\\
	&	G192.32-11.88N	&	0.46	&	12.16	&	1.00	&&	0.40	&	12.09	&	0.77	&&	1.49	&	12.40	&	2.31	&&	1.38	&	12.41	&	1.76	\\
	&	G192.32-11.88S	&	0.48	&	12.10	&	0.85	&&	0.46	&	12.02	&	0.80	&&	1.33	&	12.10	&	1.22	&&	1.34	&	12.23	&	1.36	\\
	&	G196.92-10.37	&	0.65	&	11.66	&	0.79	&&	0.61	&	11.61	&	1.35	&&	0.58	&	11.40	&	2.09	&&	1.08	&	11.06	&	1.57	\\
	&	G198.69-09.12N1	&	0.42	&	11.07	&	0.73	&&	0.10	&	10.90	&	0.72	&&	0.67	&	11.10	&	1.17	&&	0.99	&	10.91	&	1.25	\\
	&	G198.69-09.12N2	&	\nodata	&	\nodata	&	\nodata	&&	0.28	&	10.72	&	1.01	&&	0.29	&	10.80	&	0.73	&&	0.84	&	10.81	&	1.24	\\
	&	G198.69-09.12S	&	0.40	&	11.10	&	1.07	&&	0.41	&	11.04	&	0.87	&&	2.05	&	10.80	&	0.87	&&	1.43	&	11.03	&	1.04	\\
	&	G200.34-10.97N	&	0.30	&	13.42	&	0.94	&&	0.31	&	13.51	&	0.66	&&	0.35	&	13.50	&	0.68	&&	0.54	&	13.68	&	1.40	\\
	&	G200.34-10.97S	&	\nodata	&	\nodata	&	\nodata	&&	0.38	&	13.68	&	0.56	&&	0.15	&	13.80	&	0.67	&&	0.37	&	14.31	&	1.25	\\
Orion A	&	G207.36-19.82N1	&	0.31	&	10.68	&	2.31	&&	0.53	&	10.80	&	1.03	&&	2.01	&	10.70	&	1.59	&&	1.76	&	10.68	&	1.35	\\
	&	G207.36-19.82N2	&	0.59	&	11.06	&	0.98	&&	0.40	&	10.93	&	1.10	&&	1.33	&	10.80	&	1.46	&&	2.23	&	10.75	&	1.16	\\
	&	G207.36-19.82N3	&	0.48	&	11.19	&	0.80	&&	0.42	&	11.09	&	0.40	&&	0.52	&	11.10	&	1.49	&&	1.21	&	11.09	&	1.59	\\
	&	G207.36-19.82N4	&	0.38	&	11.17	&	1.21	&&	0.19	&	10.99	&	2.18	&&	0.47	&	11.20	&	1.34	&&	1.33	&	11.18	&	1.46	\\
	&	G208.68-19.20N1	&	1.31	&	11.14	&	0.93	&&	0.93	&	11.19	&	1.14	&&	7.67	&	11.10	&	1.53	&&	4.83	&	11.16	&	1.63	\\
	&	G208.68-19.20N2	&	1.63	&	11.15	&	0.92	&&	0.88	&	11.03	&	1.36	&&	3.95	&	11.00	&	2.04	&&	2.92	&	11.13	&	1.99	\\
	&	G208.68-19.20N3	&	1.33	&	11.07	&	1.20	&&	1.07	&	10.83	&	1.57	&&	5.53	&	10.90	&	2.04	&&	3.83	&	10.99	&	2.09	\\
	&	G208.68-19.20S	&	0.60	&	10.61	&	1.59	&&	0.77	&	10.77	&	1.14	&&	4.60	&	10.80	&	1.98	&&	3.51	&	10.82	&	1.95	\\
	&	G209.55-19.68N1	&	0.47	&	7.32	&	0.97	&&	0.23	&	7.72	&	2.02	&&	2.12	&	6.97	&	1.39	&&	2.08	&	7.03	&	1.48	\\
	&	G209.55-19.68N2	&	0.29	&	8.15	&	0.55	&&	\nodata	&	\nodata	&	\nodata	&&	0.40	&	8.36	&	2.52	&&	0.54	&	8.04	&	3.49	\\
	&	G209.55-19.68N3	&	0.53	&	8.16	&	0.73	&&	0.38	&	8.06	&	0.46	&&	0.24	&	7.94	&	1.16	&&	0.80	&	7.75	&	1.45	\\
	&	G209.55-19.68S1	&	0.38	&	7.49	&	1.08	&&	0.24	&	7.56	&	0.90	&&	0.37	&	8.08	&	3.05	&&	0.97	&	7.42	&	3.15	\\
	&	G209.55-19.68S2	&	0.62	&	8.03	&	0.66	&&	0.21	&	8.19	&	1.25	&&	0.24	&	8.32	&	1.34	&&	1.26	&	7.89	&	1.26	\\
	&	G209.77-19.40E1	&	0.31	&	8.02	&	0.50	&&	0.48	&	8.19	&	1.12	&&	0.43	&	8.16	&	1.03	&&	0.77	&	8.18	&	1.57	\\
	&	G209.77-19.40E2	&	0.35	&	8.07	&	0.56	&&	0.46	&	8.18	&	1.16	&&	0.43	&	8.29	&	0.99	&&	0.76	&	8.29	&	1.56	\\
	&	G209.77-19.40E3	&	0.30	&	7.94	&	0.71	&&	0.49	&	8.32	&	1.40	&&	0.55	&	8.18	&	1.28	&&	0.79	&	8.16	&	1.75	\\
	&	G209.77-19.40W	&	0.18	&	8.31	&	0.33	&&	0.29	&	8.28	&	1.43	&&	0.29	&	8.16	&	0.98	&&	0.45	&	8.36	&	1.63	\\
	&	G209.77-19.61E	&	0.35	&	8.06	&	0.62	&&	0.12	&	8.40	&	1.98	&&	0.22	&	8.31	&	2.17	&&	0.79	&	7.95	&	0.75	\\
	&	G209.77-19.61W	&	0.15	&	7.19	&	0.59	&&	\nodata	&	\nodata	&	\nodata	&&	1.06	&	7.49	&	1.81	&&	1.73	&	7.20	&	0.80	\\
	&	G209.79-19.80E	&	0.43	&	5.45	&	0.49	&&	0.41	&	5.50	&	0.38	&&	0.93	&	5.43	&	0.66	&&	1.41	&	5.52	&	0.85	\\
	&	G209.79-19.80W	&	0.56	&	5.89	&	0.99	&&	0.43	&	5.98	&	1.11	&&	1.52	&	5.91	&	1.31	&&	1.56	&	5.86	&	1.53	\\
	&	G209.94-19.52N	&	0.45	&	7.95	&	0.83	&&	0.52	&	8.01	&	0.91	&&	1.31	&	7.73	&	0.72	&&	1.51	&	7.77	&	0.75	\\
	&	G209.94-19.52S1	&	0.43	&	8.09	&	0.93	&&	0.30	&	8.14	&	2.12	&&	0.38	&	8.20	&	1.33	&&	0.75	&	8.29	&	2.14	\\
	&	G209.94-19.52S2	&	0.27	&	7.70	&	1.05	&&	0.28	&	7.69	&	1.59	&&	2.63	&	7.73	&	1.62	&&	1.64	&	7.76	&	1.62	\\
	&	G210.49-19.79E1	&	0.73	&	8.82	&	0.91	&&	0.66	&	8.77	&	1.25	&&	4.50	&	8.78	&	1.66	&&	2.42	&	8.86	&	1.76	\\
	&	G210.49-19.79E2	&	0.95	&	10.22	&	0.91	&&	0.68	&	10.23	&	0.94	&&	0.89	&	9.45	&	2.74	&&	1.98	&	9.52	&	3.14	\\
	&	G210.49-19.79W	&	0.60	&	9.11	&	0.93	&&	0.40	&	9.21	&	1.61	&&	2.08	&	9.08	&	2.20	&&	1.73	&	9.19	&	2.18	\\
	&	G210.97-19.33N	&	0.15	&	3.91	&	0.84	&&	0.26	&	3.76	&	0.63	&&	0.16	&	5.48	&	4.04	&&	0.87	&	3.59	&	1.59	\\
	&	G210.97-19.33S2	&	0.34	&	3.47	&	1.68	&&	0.36	&	3.79	&	2.02	&&	1.18	&	3.68	&	2.17	&&	1.90	&	3.25	&	1.39	\\
	&	G211.16-19.33N1	&	0.48	&	3.27	&	0.80	&&	0.45	&	3.33	&	0.83	&&	0.29	&	3.16	&	0.51	&&	0.97	&	3.04	&	0.76	\\
	&	G211.16-19.33N2	&	0.52	&	3.55	&	0.74	&&	0.64	&	3.51	&	0.68	&&	0.57	&	3.39	&	0.63	&&	0.89	&	3.37	&	1.00	\\
	&	G211.16-19.33N3	&	0.61	&	3.30	&	0.59	&&	0.56	&	3.35	&	0.47	&&	0.39	&	3.33	&	0.35	&&	1.31	&	3.36	&	0.89	\\
	&	G211.16-19.33N4	&	0.43	&	4.55	&	0.81	&&	0.36	&	4.60	&	0.68	&&	1.48	&	4.26	&	0.88	&&	0.72	&	4.19	&	1.21	\\
	&	G211.16-19.33N5	&	0.34	&	4.48	&	1.02	&&	0.30	&	4.49	&	1.15	&&	0.27	&	4.49	&	0.79	&&	1.36	&	3.38	&	0.96	\\
	&	G211.16-19.33S	&	0.18	&	3.31	&	0.71	&&	\nodata	&	\nodata	&	\nodata	&&	0.15	&	4.42	&	2.23	&&	0.46	&	4.09	&	3.30	\\
	&	G215.44-16.38	&	0.19	&	11.38	&	0.44	&&	0.32	&	11.39	&	0.47	&&	1.59	&	11.20	&	0.45	&&	0.70	&	11.31	&	0.43	\\
	&	G215.87-17.62N	&	0.31	&	9.21	&	0.63	&&	0.31	&	9.19	&	0.76	&&	0.94	&	9.34	&	0.89	&&	1.22	&	9.34	&	1.02	\\
	&	G215.87-17.62M	&	0.25	&	9.02	&	0.78	&&	0.41	&	8.95	&	0.45	&&	0.22	&	9.03	&	0.74	&&	0.28	&	9.00	&	1.35	\\
	&	G215.87-17.62S	&	\nodata	&	\nodata	&	\nodata	&&	\nodata	&	\nodata	&	\nodata	&&	0.80	&	9.57	&	0.48	&&	0.30	&	9.61	&	1.02	\\
Orion B	&	G201.52-11.08	&	\nodata	&	\nodata	&	\nodata	&&	0.20	&	9.26	&	0.51	&&	0.56	&	9.07	&	1.01	&&	0.76	&	9.37	&	0.46	\\
	&	G201.72-11.22	&	\nodata	&	\nodata	&	\nodata	&&	0.49	&	9.47	&	0.31	&&	0.30	&	9.55	&	0.35	&&	0.57	&	9.45	&	0.45	\\
	&	G203.21-11.20E1	&	0.48	&	10.22	&	1.22	&&	0.29	&	10.25	&	1.16	&&	0.20	&	9.88	&	0.37	&&	0.46	&	9.62	&	0.19	\\
	&	G203.21-11.20E2	&	0.63	&	10.29	&	1.23	&&	0.21	&	10.54	&	1.28	&&	0.19	&	10.20	&	0.68	&&	0.63	&	9.65	&	0.32	\\
	&	G203.21-11.20W1	&	0.47	&	10.46	&	1.13	&&	0.21	&	10.41	&	1.51	&&	\nodata	&	\nodata	&	\nodata	&&	0.59	&	9.64	&	0.38	\\
	&	G203.21-11.20W2	&	0.61	&	10.21	&	0.74	&&	0.34	&	10.10	&	0.85	&&	\nodata	&	\nodata	&	\nodata	&&	0.34	&	9.63	&	0.63	\\
	&	G204.4-11.3A2E	&	0.64	&	1.56	&	0.63	&&	0.91	&	1.60	&	0.59	&&	0.72	&	1.67	&	0.42	&&	1.43	&	1.77	&	0.79	\\
	&	G204.4-11.3A2W	&	0.58	&	1.67	&	0.71	&&	0.59	&	1.59	&	0.82	&&	0.88	&	1.76	&	0.57	&&	1.51	&	1.75	&	0.88	\\
	&	G205.46-14.56N2	&	0.73	&	9.84	&	1.04	&&	0.69	&	9.76	&	1.20	&&	4.20	&	9.92	&	1.42	&&	4.28	&	9.86	&	1.51	\\
	&	G205.46-14.56N3	&	0.78	&	9.90	&	0.94	&&	0.67	&	9.88	&	0.95	&&	4.72	&	9.93	&	1.17	&&	4.66	&	9.94	&	1.36	\\
	&	G205.46-14.56M1	&	0.73	&	10.02	&	0.93	&&	0.42	&	9.97	&	0.89	&&	4.23	&	10.00	&	1.13	&&	4.40	&	10.06	&	1.40	\\
	&	G205.46-14.56S1	&	0.56	&	10.12	&	1.18	&&	0.41	&	10.05	&	1.40	&&	2.62	&	9.87	&	2.93	&&	3.55	&	9.99	&	1.72	\\
	&	G205.46-14.56S2	&	0.67	&	10.42	&	0.65	&&	0.71	&	10.47	&	0.83	&&	1.05	&	10.40	&	1.37	&&	2.16	&	10.54	&	2.04	\\
	&	G205.46-14.56S3	&	0.41	&	10.41	&	0.61	&&	0.44	&	10.29	&	0.82	&&	0.84	&	10.40	&	0.98	&&	1.26	&	10.36	&	1.60	\\
	&	G206.12-15.76	&	0.41	&	8.51	&	0.64	&&	0.49	&	8.39	&	0.45	&&	0.39	&	7.95	&	2.57	&&	0.58	&	8.51	&	0.96	\\
	&	G206.21-16.17N	&	0.62	&	9.87	&	0.65	&&	0.22	&	9.90	&	1.00	&&	2.19	&	9.87	&	0.85	&&	2.20	&	9.84	&	1.00	\\
	&	G206.21-16.17S	&	0.43	&	9.53	&	0.33	&&	0.23	&	9.51	&	0.49	&&	0.32	&	9.62	&	0.35	&&	1.33	&	9.55	&	0.77	\\
	&	G206.69-16.60N	&	0.28	&	11.40	&	0.36	&&	0.26	&	11.32	&	0.57	&&	0.14	&	11.80	&	1.14	&&	0.80	&	11.42	&	0.48	\\
	&	G206.69-16.60S	&	0.51	&	12.03	&	0.80	&&	0.29	&	11.99	&	0.73	&&	2.37	&	12.10	&	0.89	&&	2.62	&	12.12	&	0.94	\\
	&	G206.93-16.61E1	&	1.38	&	9.83	&	0.98	&&	0.32	&	10.02	&	1.78	&&	3.71	&	10.00	&	1.94	&&	4.16	&	10.00	&	2.31	\\
	&	G206.93-16.61E2	&	0.79	&	9.71	&	0.85	&&	0.48	&	9.84	&	1.02	&&	2.69	&	9.77	&	1.79	&&	3.68	&	9.67	&	1.90	\\
	&	G206.93-16.61W1	&	0.95	&	9.29	&	0.96	&&	0.51	&	9.37	&	0.98	&&	1.13	&	9.69	&	2.48	&&	3.81	&	9.02	&	1.13	\\
	&	G206.93-16.61W3	&	0.54	&	9.64	&	1.70	&&	0.56	&	9.57	&	0.04	&&	0.60	&	9.80	&	2.50	&&	1.49	&	9.39	&	2.65	\\
	&	G206.93-16.61W4	&	0.78	&	9.89	&	1.03	&&	0.32	&	10.09	&	1.91	&&	0.39	&	9.73	&	2.18	&&	1.35	&	9.39	&	2.21	\\
	&	G206.93-16.61W5	&	0.55	&	9.40	&	1.79	&&	0.28	&	9.85	&	1.27	&&	0.58	&	9.54	&	2.31	&&	1.89	&	9.33	&	2.58	\\
	&	G206.93-16.61W6	&	0.44	&	9.39	&	2.35	&&	0.35	&	10.34	&	1.30	&&	2.57	&	10.00	&	2.08	&&	1.89	&	9.33	&	2.58	\\
\enddata
\label{tab:2}
\end{deluxetable*}

\clearpage

\startlongtable
\begin{deluxetable*}{ll  ccc c ccc c ccc} 
\tablecaption{Properties of \nthp, HDCO, and \htco\ Lines} 
\tabletypesize{\scriptsize}\tablecolumns{13}
\tablewidth{0pc} \setlength{\tabcolsep}{0.05in}
\tablenum{3}
 \tablehead{\\
 Cloud & Core &\multicolumn{3}{c}{\nthp} && \multicolumn{3}{c}{HDCO} && \multicolumn{3}{c}{\htco}   \\
 \cline{3-5} \cline{7-9} \cline{11-13} 
    &  & $T_{\rm peak}$ & $V_{\rm LSR}$  & FWHM & &$T_{\rm peak}$ & $V_{\rm LSR}$  & FWHM  & &$T_{\rm peak}$ & $V_{\rm LSR}$  & FWHM \\
    &  & (K) &    \multicolumn{2}{c}{(km s$^{-1}$)} &&    (K) &    \multicolumn{2}{c}{(km s$^{-1}$)} &&  (K) &    \multicolumn{2}{c}{(km s$^{-1}$)} 
}
\startdata
$\lambda$ Orionis&G190.15-13.75N	&\nodata	&	\nodata	&	\nodata	&&	\nodata	&	\nodata	&	\nodata	&&	\nodata	&	\nodata	&	\nodata	\\
	&	G190.15-13.75S	&	\nodata	&	\nodata	&	\nodata	&&	0.08	&	1.42	&	0.59	&&0.41	&	1.30	&	0.75	\\
	&	G191.90-11.21N	&	\nodata	&	\nodata	&	\nodata	&&	\nodata	&	\nodata	&	\nodata	&&	0.74	&	11.00	&	2.68	\\
	&	G191.90-11.21S	&	0.60	&	9.32	&	0.36	&&	0.33	&	10.54	&	0.50	&&		0.68	&	10.50	&	0.64	\\
	&	G192.12-10.90N	&	0.46	&	8.68	&	0.52	&&	0.23	&	10.06	&	0.31	&&		0.61	&	9.85	&	1.03	\\
	&	G192.12-10.90S	&	\nodata	&	\nodata	&	\nodata	&&	\nodata	&	\nodata	&	\nodata	&&		0.13	&	10.05	&	0.57	\\
	&	G192.12-11.10	&	0.28	&	8.97	&	0.86	&&	0.21	&	10.19	&	0.79	&&	0.49	&	10.18	&	0.98	\\
	&	G192.32-11.88N	&	0.79	&	10.90	&	0.63	&&	0.16	&	12.11	&	0.69	&&		0.64	&	12.67	&	2.75	\\
	&	G192.32-11.88S	&	0.77	&	10.80	&	0.53	&&	0.23	&	12.61	&	0.69	&&	0.68	&	12.08	&	1.31	\\
	&	G196.92-10.37	&	0.47	&	10.50	&	0.77	&&	0.31	&	11.69	&	0.58	&&	0.47	&	11.44	&	2.01	\\
	&	G198.69-09.12N1	&	\nodata	&	\nodata	&	\nodata	&&	0.16	&	11.10	&	0.52	&&		0.65	&	11.15	&	1.04	\\
	&	G198.69-09.12N2	&	0.11	&	9.37	&	0.22	&&	\nodata	&	\nodata	&	\nodata	&&		0.51	&	10.80	&	0.84	\\
	&	G198.69-09.12S	&	0.27	&	9.76	&	0.59	&&	0.13	&	11.08	&	1.00	&&		0.67	&	11.03	&	1.23	\\
	&	G200.34-10.97N	&	0.83	&	12.20	&	0.41	&&	0.23	&	13.41	&	0.42	&&		0.54	&	13.43	&	0.80	\\
	&	G200.34-10.97S	&	0.37	&	12.50	&	0.36	&&	0.13	&	13.72	&	0.27	&&		0.22	&	13.82	&	0.73	\\
Orion A&	G207.36-19.82N1	&	0.95	&	9.48	&	0.93	&&	\nodata	&	\nodata	&	\nodata	&&	0.84	&	10.85	&	1.15	\\
	&	G207.36-19.82N2	&	1.34	&	9.84	&	0.72	&&	0.35	&	11.22	&	0.41	&&		0.63	&	10.97	&	1.28	\\
	&	G207.36-19.82N3	&	0.49	&	9.94	&	0.58	&&	0.37	&	11.32	&	0.50	&&		0.71	&	11.30	&	1.22	\\
	&	G207.36-19.82N4	&	0.69	&	9.86	&	0.55	&&	0.26	&	11.15	&	0.78	&&		0.63	&	11.07	&	1.64	\\
	&	G208.68-19.20N1	&	2.34	&	9.92	&	0.76	&&	0.59	&	11.18	&	0.92	&&		2.02	&	11.20	&	1.48	\\
	&	G208.68-19.20N2	&	1.96	&	9.94	&	0.55	&&	0.47	&	11.20	&	0.42	&&		1.63	&	11.09	&	1.48	\\
	&	G208.68-19.20N3	&	2.91	&	9.86	&	0.82	&&	0.36	&	11.17	&	0.71	&&		1.59	&	10.96	&	1.93	\\
	&	G208.68-19.20S	&	2.24	&	9.45	&	1.37	&&	0.21	&	10.32	&	1.44	&&		1.29	&	10.74	&	1.75	\\
	&	G209.55-19.68N1	&	1.78	&	6.09	&	0.87	&&	\nodata	&	\nodata	&	\nodata	&&		0.74	&	7.28	&	1.47	\\
	&	G209.55-19.68N2	&	0.89	&	6.98	&	0.31	&&	\nodata	&	\nodata	&	\nodata	&&		0.37	&	8.20	&	1.22	\\
	&	G209.55-19.68N3	&	0.81	&	6.85	&	0.61	&&	0.17	&	8.07	&	0.73	&&		0.44	&	8.03	&	1.27	\\
	&	G209.55-19.68S1	&	0.81	&	6.09	&	0.76	&&	\nodata	&	\nodata	&	0.80	&&		0.31	&	7.59	&	2.29	\\
	&	G209.55-19.68S2	&	1.51	&	6.85	&	0.45	&&	0.23	&	8.18	&	0.75	&&		0.50	&	7.95	&	1.04	\\
	&	G209.77-19.40E1	&	1.14	&	6.80	&	0.36	&&	\nodata	&	\nodata	&	\nodata	&&		0.54	&	8.09	&	0.82	\\
	&	G209.77-19.40E2	&	1.15	&	6.82	&	0.46	&&	0.16	&	7.99	&	0.55	&&		0.60	&	8.15	&	0.91	\\
	&	G209.77-19.40E3	&	0.65	&	6.76	&	0.69	&&	\nodata	&	\nodata	&	\nodata	&&		0.72	&	7.90	&	0.80	\\
	&	G209.77-19.40W	&	1.72	&	7.13	&	0.21	&&	\nodata	&	\nodata	&	\nodata	&&		0.40	&	8.36	&	0.93	\\
	&	G209.77-19.61E	&	0.77	&	6.78	&	0.35	&&	\nodata	&	\nodata	&	\nodata	&&		0.27	&	8.20	&	1.53	\\
	&	G209.77-19.61W	&	0.41	&	6.00	&	0.34	&&	\nodata	&	\nodata	&	\nodata	&&		0.44	&	7.18	&	0.61	\\
	&	G209.79-19.80E	&	0.21	&	4.24	&	0.36	&&	\nodata	&	\nodata	&	\nodata	&&		0.54	&	5.50	&	0.59	\\
	&	G209.79-19.80W	&	0.70	&	4.56	&	0.68	&&	0.18	&	5.96	&	1.18	&&		0.64	&	6.00	&	1.41	\\
	&	G209.94-19.52N	&	0.80	&	6.76	&	0.55	&&	0.30	&	7.97	&	0.54	&&		0.74	&	7.88	&	0.86	\\
	&	G209.94-19.52S1	&	1.28	&	6.78	&	0.96	&&	0.29	&	8.14	&	0.53	&&		0.60	&	8.21	&	1.41	\\
	&	G209.94-19.52S2	&	0.57	&	6.15	&	0.62	&&	\nodata	&	\nodata	&	\nodata	&&		0.66	&	7.73	&	1.35	\\
	&	G210.49-19.79E1	&	1.56	&	7.54	&	0.84	&&	0.28	&	8.83	&	0.55	&&		1.01	&	8.89	&	1.50	\\
	&	G210.49-19.79E2	&	1.14	&	8.96	&	0.72	&&	0.35	&	10.30	&	0.69	&&		1.10	&	10.11	&	1.69	\\
	&	G210.49-19.79W	&	0.77	&	7.90	&	0.72	&&	0.25	&	9.11	&	0.57	&&		0.70	&	9.32	&	1.99	\\
	&	G210.97-19.33N	&	\nodata	&	\nodata	&	\nodata	&&	\nodata	&	\nodata	&	\nodata	&&	0.19	&	3.72	&	1.48	\\
	&	G210.97-19.33S2	&	0.18	&	2.07	&	0.75	&&	\nodata	&	\nodata	&	\nodata	&&	0.42	&	3.44	&	1.79	\\
	&	G211.16-19.33N1	&	0.90	&	2.13	&	0.40	&&	\nodata	&	\nodata	&	\nodata	&&		0.49	&	3.18	&	0.67	\\
	&	G211.16-19.33N2	&	0.67	&	2.32	&	0.44	&&	\nodata	&	\nodata	&	\nodata	&&		0.60	&	3.58	&	0.69	\\
	&	G211.16-19.33N3	&	0.91	&	2.16	&	0.43	&&	0.37	&	14.39	&	0.28	&&		0.61	&	3.33	&	0.66	\\
	&	G211.16-19.33N4	&	0.35	&	3.40	&	0.59	&&	\nodata	&	\nodata	&	\nodata	&&		0.61	&	4.51	&	0.85	\\
	&	G211.16-19.33N5	&	0.93	&	3.17	&	0.58	&&	\nodata	&	\nodata	&	\nodata	&&		0.63	&	4.55	&	1.03	\\
	&	G211.16-19.33S	&	\nodata	&	\nodata	&	\nodata	&&	\nodata	&	\nodata	&	\nodata	&&	0.17	&	4.92	&	2.92	\\
	&	G215.44-16.38		&	0.37	&	10.30	&	0.32	&&	\nodata	&	\nodata	&	\nodata	&&	0.22	&	11.38	&	0.58	\\
	&	G215.87-17.62N	&	0.14	&	8.05	&	0.46	&&	\nodata	&	\nodata	&	\nodata	&&	0.38	&	9.34	&	0.72	\\
	&	G215.87-17.62M	&	\nodata	&	\nodata	&	\nodata	&&	\nodata	&	\nodata	&	\nodata	&&		0.44	&	9.03	&	0.85	\\
	&	G215.87-17.62S	&	\nodata	&	\nodata	&	\nodata	&&	\nodata	&	\nodata	&	\nodata	&&		0.19	&	9.96	&	0.45	\\
Orion B&	G201.52-11.08		&	\nodata	&	\nodata	&	\nodata	&&	\nodata	&	\nodata	&	\nodata	&&	0.29	&	9.16	&	0.56	\\
	&	G201.72-11.22		&	1.42	&	8.25	&	0.29	&&	0.15	&	9.46	&	0.45	&&	0.34	&	9.59	&	0.52	\\
	&	G203.21-11.20E1	&	0.35	&	9.20	&	0.75	&&	0.29	&	10.66	&	0.30	&&		0.37	&	10.17	&	1.30	\\
	&	G203.21-11.20E2	&	0.72	&	9.02	&	0.72	&&	0.24	&	10.40	&	0.84	&&		0.45	&	10.29	&	1.61	\\
	&	G203.21-11.20W1	&	0.72	&	9.38	&	0.59	&&	0.25	&	10.67	&	0.81	&&		0.52	&	9.82	&	0.42	\\
	&	G203.21-11.20W2	&	0.55	&	8.97	&	0.60	&&	0.24	&	10.16	&	0.51	&&		0.19	&	10.19	&	1.79	\\
	&	G204.4-11.3A2E	&	1.09	&	0.31	&	0.49	&&	0.31	&	1.75	&	0.42	&&		0.56	&	1.60	&	1.37	\\
	&	G204.4-11.3A2W	&	0.60	&	0.33	&	0.97	&&	0.25	&	1.78	&	0.74	&&		0.59	&	1.77	&	0.83	\\
	&	G205.46-14.56N2	&	1.02	&	8.79	&	0.89	&&	\nodata	&	\nodata	&	\nodata	&&		1.46	&	9.94	&	1.34	\\
	&	G205.46-14.56N3	&	1.59	&	8.79	&	0.69	&&	0.26	&	10.10	&	0.93	&&	1.28	&	9.94	&	1.40	\\
	&	G205.46-14.56M1	&	1.32	&	8.88	&	0.82	&&	\nodata	&	\nodata	&	\nodata	&&		1.01	&	10.02	&	1.32	\\
	&	G205.46-14.56S1	&	1.88	&	9.02	&	1.12	&&	\nodata	&	\nodata	&	\nodata	&&		0.86	&	10.02	&	2.41	\\
	&	G205.46-14.56S2	&	1.82	&	9.21	&	0.51	&&	\nodata	&	\nodata	&	\nodata	&&		0.53	&	10.60	&	2.14	\\
	&	G205.46-14.56S3	&	1.79	&	9.18	&	0.58	&&	0.18	&	10.53	&	0.65	&&		0.47	&	10.29	&	1.59	\\
	&	G206.12-15.76		&	1.67	&	7.20	&	0.44	&&	0.35	&	8.39	&	0.34	&&		0.54	&	8.37	&	0.61	\\
	&	G206.21-16.17N	&	1.94	&	8.59	&	0.37	&&	0.14	&	9.88	&	0.47	&&	0.63	&	9.83	&	0.63	\\
	&	G206.21-16.17S	&	1.04	&	8.33	&	0.27	&&	0.14	&	9.60	&	0.47	&&		0.44	&	9.59	&	0.40	\\
	&	G206.69-16.60N	&	0.74	&	10.10&	0.36	&&	0.23	&	11.36	&	0.30	&&		0.41	&	11.37	&	0.48	\\
	&	G206.69-16.60S	&	0.65	&	10.90&	0.60	&&	0.19	&	12.01	&	0.50	&&		0.84	&	12.05	&	0.94	\\
	&	G206.93-16.61E1	&	2.00	&	8.62	&	0.76	&&	0.22	&	9.84	&	0.75	&&		1.16	&	9.98	&	1.29	\\
	&	G206.93-16.61E2	&	0.85	&	8.49	&	0.68	&&	0.10	&	9.61	&	0.85	&&		0.91	&	9.75	&	1.40	\\
	&	G206.93-16.61W1	&	1.26	&	8.14	&	0.79	&&	0.22	&	9.38	&	0.68	&&		0.71	&	9.37	&	1.44	\\
	&	G206.93-16.61W3	&	1.35	&	8.09	&	0.68	&&	0.22	&	9.16	&	0.86	&&		0.58	&	9.88	&	2.61	\\
	&	G206.93-16.61W4	&	0.63	&	8.72	&	1.13	&&	0.23	&	10.07	&	0.73	&&		0.49	&	9.66	&	2.04	\\
	&	G206.93-16.61W5	&	0.64	&	8.10	&	1.32	&&	0.23	&	9.09	&	0.67	&&		0.63	&	9.48	&	1.92	\\
	&	G206.93-16.61W6	&	0.72	&	8.80	&	1.35	&&	\nodata	&	\nodata	&	\nodata	&&		0.89	&	9.63	&	2.01	\\
\enddata
\tablecomments{\httco\ was detected only in G207.36-19.82N3.  The $T_{\rm peak}$,  $V_{\rm LSR}$, and FWHM of the detected \httco\ line are 0.15 K, 11.25 km s$^{-1}$, and 0.92 km s$^{-1}$, respectively.}
\label{tab:3}
\end{deluxetable*}

\clearpage

\begin{deluxetable*}{lrrrrr}
\tablecaption{Statistics of the observed cores \label{tab:4}}
\tablecolumns{4}
\tablenum{4}
\tablewidth{0pt}
\tablehead{
\colhead{Cloud \tablenotemark{}} &
\colhead{Protostellar cores} & \colhead{} &
\colhead{Starless cores} &\colhead{} &
\colhead{Total}  \\
}
\startdata
$\lambda$ Orionis  &  7 & & 8 & & 15  \\
OrionA & 17& & 22 & & 39  \\
OrionB  & 11& & 15 & & 26  \\
\enddata
\end{deluxetable*}

\begin{deluxetable*}{lrrrr}
\tablecaption{Detection rates of the Observed lines \label{tab:5}}
\tablecolumns{5}
\tablenum{5}
\tablewidth{0pt}
\tablehead{
\colhead{Molecule \tablenotemark{}} &
\colhead{Rest frequency\tablenotemark{}} &
\colhead{$\lambda$ Orionis \tablenotemark{}} &
\colhead{Orion A \tablenotemark{}} &
\colhead{Orion B \tablenotemark{}}  \\
\colhead{} & \colhead{(GHz)} & 
 \colhead{(15)} &\colhead{(39)} & \colhead{(26)} 
}
\startdata
H$_2$O maser  & 22.23508 & 1 (7\%) & 1 (3\%) & 1 (4\%)   \\
SiO  & 43.42379 & 2 (13\%) & 0 (0\%) & 0 (0\%)  \\
CH$_3$OH maser  & 44.06943 & 0 (0\%)  &  0 (0\%) & 1 (4\%) \\
H$^{13}$CO$^{+}$ & 86.75428 & 10 (67\%)  & 38 (97\%) & 24 (92\%)  \\
C$_{2}$H & 87.31689 &  12 (80\%) & 35 (90\%) & 26 (100\%)  \\
HCN & 88.63184 & 14 (93\%) & 39 (100\%) & 24 (92\%) \\
HCO$^+$ & 89.18852 &  13 (87\%) & 39 (100\%) & 26 (100\%)  \\
N$_2$H$^+$ & 93.17339 &  10 (67\%) & 35 (90\%) & 25 (96\%) \\
HDCO & 128.81286 &  11 (73\%) & 17 (44\%) & 20 (77\%)  \\
H$_{2}$$^{13}$CO & 137.44995 & 0 (0\%) & 1 (3\%) & 0 (0\%)  \\
H$_2$CO & 140.83952 &  14 (93\%) & 39 (100\%) & 26 (100\%) \\
\enddata
\tablecomments{Detection rates toward 80 cores in the OMC.
The numbers indicate the number of detected cores in each cloud and the percentage is in parentheses.}
\end{deluxetable*}

\begin{deluxetable*}{l cc c cc c cc}
\tablecaption{Median values of the non-thermal and thermal velocity dispersions derived for seven molecules. \label{tab:6}} 
\tabletypesize{\scriptsize}\tablecolumns{9}
\tablewidth{0pc} 
\tablenum{6}
 \tablehead{
 \\
Molecule &\multicolumn{2}{c}{$\lambda$ Orionis} && \multicolumn{2}{c}{Orion A} && \multicolumn{2}{c}{Orion B}\\
\cline{2-3} \cline{5-6} \cline{8-9}&    $\sigma_{\rm NT}$ & $\sigma_{\rm T}$ &&  $\sigma_{\rm NT}$ & $\sigma_{\rm T}$ &&   $\sigma_{\rm NT}$ & $\sigma_{\rm T}$\\
&  (km s$^{-1}$)   & (km s$^{-1}$) & &  (km s$^{-1}$) & (km s$^{-1}$)&   & (km s$^{-1}$)   &  (km s$^{-1}$)  
}\startdata
H$^{13}$CO$^{+}$   & 0.36 & 0.06 & & 0.36 & 0.06 & & 0.37 & 0.07  \\
C$_{2}$H  		&  0.32 & 0.07 & & 0.47 & 0.07 & & 0.40 & 0.07  \\
HCN   			& 0.38 & 0.06  & & 0.57 & 0.06 & & 0.59 & 0.07 \\
HCO$^+$  		&  0.60 & 0.06 & & 0.65 & 0.06 & & 0.61 & 0.07   \\
N$_2$H$^+$  		&  0.23 & 0.06 & &0.25 & 0.06 & & 0.29 & 0.07 \\
HDCO 			&  0.22 & 0.06 & & 0.30 & 0.06 & & 0.27 & 0.06 \\
H$_2$CO  		& 0.47 & 0.06 & & 0.53 & 0.06 & & 0.59 & 0.07 \\
\enddata
\end{deluxetable*}

\clearpage

\begin{deluxetable*}{lrrrr}
\tablecaption{Median line widths of seven molecules toward cores within the three clouds \label{tab:7}}
\tablecolumns{4}
\tablenum{7}
\tablewidth{0pt}
\tablehead{
\colhead{Molecule \tablenotemark{}} &
\colhead{$\lambda$ Orionis \tablenotemark{}} &
\colhead{Orion A \tablenotemark{}} &
\colhead{Orion B \tablenotemark{}}  \\
\colhead{} &   \colhead{(km s$^{-1}$)} &\colhead{(km s$^{-1}$)} & \colhead{(km s$^{-1}$)} 
}
\startdata
H$^{13}$CO$^{+}$  & 0.8 & 0.8 & 0.9  \\
C$_{2}$H  &  0.8 & 1.1 & 1.0 \\
HCN  & 0.9 & 1.3 & 1.4 \\
HCO$^+$  &  1.1 & 1.5 & 1.3  \\
N$_2$H$^+$  &  0.5 & 0.6 & 0.7 \\
HDCO  &  0.5 & 0.7 & 0.6 \\
H$_2$CO  &  1.0 &1.2 & 1.4 \\
\enddata
\tablecomments{Median line widths of seven molecules which are frequently detected in the three clouds.
The velocity resolution is 0.1 km s$^{-1}$.}
\end{deluxetable*}

\begin{deluxetable*}{l rrrrrrrr }
\tablecaption{Statistics of abundance ratios of seven molecules toward the three clouds \label{tab:8}}
\tablecolumns{9}
\tablenum{8}
\tablewidth{0pt}
\tablehead{
\colhead{Cloud \tablenotemark{}} &
  X(\nthp)  & X(\hcop)& X(\htcop) & X(HCN)& X(\cth)  & X(\htco)& X(HDCO) & X(H$_{2}$$^{13}$CO)\\
 & (10$^{-10}$) & (10$^{-10}$) & (10$^{-10}$) &(10$^{-10}$) & (10$^{-8}$) & (10$^{-10}$) & (10$^{-10}$)  & (10$^{-10}$) \\
 \hline
&  \multicolumn{8}{c}{median} 
}
\startdata
\lam      &  0.96& 4.31& 0.87& 7.64& 4.51& 7.89& 0.25 & 0.13\\
OrionA  &  0.37& 2.80& 0.32& 3.27& 1.60& 7.05& 0.23 & 0.11\\
OrionB  &  0.27& 3.81& 0.39& 3.97& 1.48& 5.21& 0.24 & 0.09\\
\hline
 &  \multicolumn{8}{c}{mean} \\
\hline
\lam      &  2.88& 3.78& 0.88&21.34& 6.13& 14.82& 0.48&	0.24\\   
OrionA  &  0.73& 3.45& 0.36&  4.91& 1.98&  7.41&  0.44& 0.12\\
OrionB  &  0.75& 4.12& 0.45& 28.47& 2.23& 6.17& 0.24& 0.10\\
\enddata
\end{deluxetable*}

\begin{deluxetable*}{lrrrr}[b!]
\tablecaption{The mean f$_{\rm D}$ [HDCO]/[\htco] of cores within the three clouds \label{tab:9}}
\tablecolumns{5}
\tablenum{9}
\tablewidth{0pt}
\tablehead{
\colhead{cores \tablenotemark{}} &
\colhead{$\lambda$ Orionis \tablenotemark{}} &
\colhead{Orion A \tablenotemark{}} &
\colhead{Orion B \tablenotemark{}}  
}
\startdata
total cores &  0.046 $\pm$  0.001 & 0.060 $\pm$  0.004 & 0.064 $\pm$  0.005\\
protostellar cores &  0.028 $\pm$  0.005 & 0.020 $\pm$  0.007 & 0.021 $\pm$  0.004\\
starless cores & 0.056 $\pm$ 0.005& 0.079 $\pm$  0.006 & 0.081 $\pm$  0.006  \\
\enddata
\end{deluxetable*}

\clearpage

\startlongtable
\begin{deluxetable*}{llrrrrrr} 
\tablecaption{Spectral line velocities, line asymmetry, and infall rate \label{tab:10}}
\tablecolumns{8}
\tablenum{10}
\tablewidth{20pt}
\tablehead{
\colhead{Cloud \tablenotemark{}} &
\colhead{Source name \tablenotemark{}} &
\colhead{$V_{\rm thick}$ \tablenotemark{}} &
\colhead{$V_{\rm thin}$ \tablenotemark{}} & \colhead{$\Delta$$V_{\rm thin}$} & \colhead{$\delta$$v$} &
\colhead{profile \tablenotemark{}}  &
\colhead{$\dot{M}_{\rm inf}$}\\
\colhead{} & \colhead{} &  \colhead{(km s$^{-1}$)} &
\colhead{(km s$^{-1}$)} & \colhead{(km s$^{-1}$)} & \colhead{} & \colhead{} &\colhead{(10$^{-6}$ $M_{\sun}$ yr$^{-1}$)}
}
\startdata
$\lambda$ Orionis   & G190.15-13.75 N & \nodata &  \nodata  &  \nodata &  \nodata &  \nodata & \nodata	\\
	& G190.15-13.75S & 1.27  &  1.32 &  0.45 & -0.11  &  neutral	& \nodata	 \\
	& G191.90-11.21N & 10.87 &  \nodata &  \nodata &   \nodata&  \nodata & \nodata	\\
	& G191.90-11.21S &  10.51 & 10.55 & 0.87 &  -0.05 & neutral & \nodata	\\
	& G192.12-10.90N &  9.61 & 9.96  &0.67 &  -0.52 & blue & 5.78\\
	& G192.12-10.90S & \nodata &   \nodata&   \nodata&  \nodata &  \nodata & \nodata	\\
	& G192.12-11.10     &  10.29 & 10.22 & 0.65 & 0.12 & neutral & \nodata	\\
	& G192.32-11.88N &  12.41 & 12.15 & 0.88 & 0.29 & red & \nodata	\\
	& G192.32-11.88S &  12.23 & 12.10 & 0.85 & 0.15 & neutral & \nodata	\\
	& G196.92-10.37    & 11.06 & 11.66 & 0.79 & -0.76 & blue & 45.23\\
	& G198.69-09.12N1&  10.91 & 11.07 & 0.73 & -0.22 & neutral & \nodata	\\
	& G198.69-09.12N2&  10.81 &  \nodata &  \nodata &  \nodata &  \nodata & \nodata	\\
	& G198.69-09.12S & 11.03 & 11.10 & 1.07  & -0.07 & neutral	& \nodata	\\
	& G200.34-10.97N &  13.68 & 13.42 & 0.91 & 0.28 &  red	& \nodata	\\
	& G200.34-10.97S & 14.31 &  \nodata & \nodata &  \nodata &	 \nodata & \nodata	\\
	\hline
Orion A	&  G207.36-19.82N1 & 10.68  &      10.68  &      2.31  &  0.01  & neutral   &   \nodata	   \\   
& G207.36-19.82N2 &   10.75  &  11.06  &     0.98  &    -0.31  & blue & 10.45 \\   
& G207.36-19.82N3 &    11.09  &   11.19  &     0.80  &    -0.12  & neutral   & \nodata	 \\   
& G207.36-19.82N4 &    11.18  &   11.17  &      1.21  &  0.01  & neutral    & \nodata	  \\   
& G208.68-19.20N1 &  11.16  &      11.14  &     0.93  &    0.02  & neutral   &  \nodata	  \\   
& G208.68-19.20N2 &   11.13  &   11.15  &     0.92  &   -0.02  & neutral     & \nodata	 \\   
& G208.68-19.20N3 &   10.99  &      11.07  &      1.20  &   -0.07  & neutral    &   \nodata	\\   
& G208.68-19.20S   &   10.82  &      10.61  &      1.59  &     0.13  & neutral   &  \nodata	 \\   
& G209.55-19.68N1 &    7.03  &      7.32  &     0.45  &    -0.64  & blue     & 4.11 \\   
& G209.55-19.68N2 &    8.04  &      8.15  &     0.50  &    -022  & neutral    & \nodata  \\   
& G209.55-19.68N3 &  7.75  &      8.16  &     0.71  &    -0.58  & blue    & 23.04 \\   
& G209.55-19.68S1 &   7.42  &      7.49  &      1.08  &   -0.07  & neutral   &  \nodata	  \\   
& G209.55-19.68S2 &  7.89  &      8.03  &     0.66 &    -0.20  & blue    &  6.12\\   
&G209.77-19.40E1  &   8.18  &      8.02  &     0.50  &     0.32  & red  &  \nodata	 \\   
& G209.77-19.40E2 &   8.29  &      8.07 &     0.56  &     0.39  & red   &  \nodata	  \\   
& G209.77-19.40E3 &   8.16  &      7.94  &     0.71  &     0.32  & red   &  \nodata	  \\   
& G209.77-19.40W  &  8.36  &      8.31  &     0.33  &     0.16  & neutral     & \nodata	 \\   
& G209.77-19.61E   &    7.95  &      8.06  &     0.62  &    -0.17  & neutral    &  \nodata	 \\   
& G209.77-19.61W  &  7.20  &      7.19  &     0.59  &    0.02  & neutral     & \nodata	 \\   
& G209.79-19.80E   &  5.52  &      5.45  &     0.49  &     0.13  & neutral    &  \nodata	\\   
& G209.79-19.80W  &   5.86  &      5.89  &     0.99  &   -0.02  & neutral    &  \nodata	 \\   
& G209.94-19.52N   &   7.77  &      7.95  &     0.83  &    -0.23  & neutral   &  \nodata	  \\   
& G209.94-19.52S1  &   8.29  &      8.09 &     0.93  &     0.21  & neutral    &  \nodata	 \\   
& G209.94-19.52S2  &  7.76  &      7.70  &      1.05  &    0.06 & neutral   &  \nodata	\\   
& G210.49-19.79E1  &   8.86  &      8.82  &     0.91  &    0.04  & neutral    & \nodata	 \\   
& G210.49-19.79E2  &  9.52 &      10.22  &     0.91  &    -0.77  & blue   & 138.95\\   
& G210.49-19.79W   &  9.19  &      9.11  &     0.93 &    0.08  & neutral     & \nodata	 \\   
& G210.97-19.33N    &   3.59  &      3.91 &     0.84 &    -0.39  & blue     & 7.19\\   
& G210.97-19.33S2  &  3.25  &      3.47  &      1.68  &    -0.13 & neutral     & \nodata	 \\   
&  G211.16-19.33N1 &   3.04  &      3.27  &     0.80  &    -0.28  & blue     & 5.06\\   
&  G211.16-19.33N2 &  3.36  &      3.55 &     0.67  &    -0.28  & blue    & 2.91\\   
&  G211.16-19.33N3 &   3.36  &      3.30  &     0.59  &     0.11 & neutral   & \nodata	\\   
& G211.16-19.33N4  &  4.19  &      4.55  &     0.81  &    -0.45  & blue    & 4.33\\   
& G211.16-19.33N5  &   3.38  &      4.48 &      1.02  &     -1.09  & blue    &  17.99\\   
& G211.16-19.33S    &   4.09  &      3.31  &     0.71  &      1.10  & red     &  \nodata	\\   
& G215.44-16.38      &  11.31  &      11.38  &     0.44  &    -0.16  & neutral   &  \nodata	 \\   
& G215.87-17.62N  &  9.34 &      9.21 &     0.63  &     0.20  & neutral    &  \nodata	 \\   
& G215.87-17.62M  &  9.00  &      9.02  &     0.78  &   -0.04  & neutral  &  \nodata	  \\   
& G215.87-17.62S    &  9.61  &  \nodata &  \nodata & \nodata  &  \nodata & \nodata	   \\   
\hline
OrionB    &  G201.52-11.08 & 9.37 &    \nodata &  \nodata & \nodata  &  \nodata   &   \nodata	   \\   
&  G201.72-11.22 &  9.94  &      \nodata &  \nodata & \nodata  &  \nodata    & \nodata	 \\   
& G203.21-11.20E1  &   9.61  &      10.21  &      1.28  &    -0.47  & blue     & 38.32\\   
& G203.21-11.20E2  &    9.65  &      10.29  &      1.23  &    -0.52  & blue    &  43.44\\   
& G203.21-11.20W1  &  9.64  &      10.46  &      1.13  &    -0.73  & blue     & 58.87\\   
& G203.21-11.20W2 &   9.63  &      10.21  &     0.74  &    -0.78  & blue     &44.63 \\   
& G204.4-11.3A2E  &  1.77  &      1.56  &     0.63  &     0.33  & red      & \nodata	\\   
& G204.4-11.3A2W &   1.75  &      1.67  &     0.71  &     0.10  & neutral     & \nodata	\\   
& G205.46-14.56N2  &   9.86  &      9.84  &      1.04  &    0.02  & neutral     &  \nodata	\\   
& G205.46-14.56N3  &    9.94  &      9.90  &     0.94  &    0.04  & neutral      & \nodata	\\   
& G205.46-14.56M1   &   10.06  &      10.02  &     0.93  &    0.05  & neutral     &  \nodata	\\   
& G205.46-14.56S1  &  9.99  &      10.12  &      1.18  &    -0.11  & neutral   &  \nodata	 \\   
& G205.46-14.56S2 &   10.54  &      10.42  &     0.65  &     0.20  & neutral    &   \nodata	\\   
& G205.46-14.56S3  &  10.36  &      10.41  &     0.61  &   -0.09  & neutral    & \nodata	  \\   
& G206.12-15.76  &   8.34  &      8.51  &     0.63  &  -0.27  & blue    & 14.37\\   
& G206.21-16.17N  &  9.84  &      9.87  &     0.65  &   -0.04  & neutral   &  \nodata	  \\   
& G206.21-16.17S  & 9.55  &      9.53  &     0.33  &    0.06  & neutral     &  \nodata	\\   
& G206.69-16.60N  & 11.42  &      11.40  &     0.36  &    0.07 & neutral    &   \nodata	\\   
& G206.69-16.60S &   12.12  &      12.03  &     0.80  &     0.11  & neutral   &  \nodata	  \\   
& G206.93-16.61E1  &   10.00  &      9.83  &     0.98  &     0.17  & neutral   &  \nodata	  \\   
& G206.93-16.61E2  &   9.67  &      9.71  &     0.85  &   -0.05  & neutral     & \nodata	 \\   
& G206.93-16.61W1 &   9.02  &      9.29  &     0.96  &    -0.28  & blue    & 16.04\\   
& G206.93-16.61W3  &  9.39  &      9.64  &      1.70  &    -0.14  & neutral    & \nodata	  \\   
& G206.93-16.61W4  &   9.39  &      9.89  &      1.03  &    -0.49  & blue     & 136.04\\   
& G206.93-16.61W5  &   9.33  &      9.40  &      1.79  &   -0.04  & neutral     & \nodata	 \\   
& G206.93-16.61W6  &  9.33  &      9.39  &      2.34  &   -0.02  & neutral    & \nodata	  \\ 
\enddata
\end{deluxetable*}

\begin{deluxetable*}{lcrrrr}[b!]
\tablecaption{Blue Excess ($E$) Statistics \label{tab:11}}
\tablecolumns{6}
\tablenum{11}
\tablewidth{0pt}
\tablehead{
\colhead{Cloud \tablenotemark{}} &
\colhead{Infall tracer} &
\colhead{$N_{\rm Blue}$} &
\colhead{$N_{\rm Red}$}  &
\colhead{$N_{\rm Total}$}  &
\colhead{$E$}  \\
}
\startdata
$\lambda$ Orionis  &  \hcop & 2 & 2 & 10 &0.0  \\
OrionA & \hcop & 10 & 4 &  38 &0.13  \\
OrionB  & \hcop &7 & 2 &  25 &0.20  \\
\enddata
\end{deluxetable*}

\clearpage

\begin{longrotatetable}
\begin{deluxetable*}{lllrrrrrrrrr}
\tabletypesize{\scriptsize}
\tablecaption{Column densities of Seven Molecules and H$_{2}$ toward three molecular clouds \label{tab:12}}
\tablecolumns{12}
\tablenum{12}
\tablewidth{20pt}
\tablehead{
\colhead{Cloud \tablenotemark{}} &
\colhead{Core \tablenotemark{}} &
\colhead{Associated YSO \tablenotemark{}} &
\colhead{\nthp \tablenotemark{}} &
\colhead{\hcop \tablenotemark{}} &
\colhead{\htcop \tablenotemark{}} &
\colhead{HCN \tablenotemark{}} &
\colhead{\cth \tablenotemark{}} &
\colhead{\htco \tablenotemark{}} &
\colhead{HDCO \tablenotemark{}} &
\colhead{H$_{2}$ (30$\arcsec$)\tablenotemark{a}}  &
\colhead{H$_{2}$ (20$\arcsec$)\tablenotemark{b}}  \\
\colhead{} &\colhead{} & \colhead{} & \colhead{($10^{12}$ cm$^{-2}$)} &  \colhead{($10^{12}$ cm$^{-2}$)} & \colhead{($10^{12}$ cm$^{-2}$)} & \colhead{($10^{12}$ cm$^{-2}$)} & \colhead{($10^{14}$ cm$^{-2}$)} &  \colhead{($10^{13}$ cm$^{-2}$)} &\colhead{($10^{12}$ cm$^{-2}$)} & \colhead{($10^{22}$ cm$^{-2}$)} &\colhead{($10^{22}$ cm$^{-2}$)} 
}
\startdata
\lam	&	G190.15-13.75N	&	\nodata	&	\nodata	&	\nodata	&	\nodata	&	\nodata	&	\nodata	&	\nodata	&	\nodata	&	1.51	&	5.79	\\
	&	G190.15-13.75S	&	\nodata	&	\nodata	&	3.43	&	0.42	&	9.54	&	\nodata	&	1.53	&	0.49	&	1.71	&	4.83	\\
	&	G191.90-11.21N	&	Class 0	&	\nodata	&	7.58	&	\nodata	&	28.21	&	\nodata	&	12.49	&	\nodata	&	1.56	&	1.47	\\
	&	G191.90-11.21S	&	\nodata	&	0.70	&	3.01	&	0.69	&	78.64	&	12.40	&	3.17	&	3.11	&	0.64	&	3.13	\\
	&	G192.12-10.90N	&	\nodata	&	1.00	&	3.77	&	1.23	&	10.49	&	10.20	&	4.03	&	3.18	&	0.74	&	1.78	\\
	&	G192.12-10.90S	&	flat	&	\nodata	&	\nodata	&	\nodata	&	11.97	&	1.75	&	0.43	&	\nodata	&	0.26	&	1.22	\\
	&	G192.12-11.10	&	\nodata	&	\nodata	&	1.48	&	1.22	&	18.79	&	1.95	&	4.20	&	3.32	&	3.69	&	9.46	\\
	&	G192.32-11.88N	&	Class 0	&	1.64	&	12.57	&	2.17	&	45.33	&	8.97	&	11.38	&	2.50	&	1.69	&	4.71	\\
	&	G192.32-11.88S	&	\nodata	&	1.38	&	11.04	&	2.05	&	46.87	&	13.17	&	5.88	&	2.10	&	1.43	&	3.97	\\
	&	G196.92-10.37	&	Class 0	&	3.35	&	9.61	&	2.22	&	3.57	&	18.40	&	7.76	&	1.84	&	3.54	&	7.73	\\
	&	G198.69-09.12N1	&	\nodata	&	\nodata	&	5.37	&	1.21	&	6.81	&	2.24	&	2.25	&	1.00	&	1.25	&	3.98	\\
	&	G198.69-09.12N2	&	\nodata	&	20.22	&	5.05	&	\nodata	&	8.86	&	2.82	&	3.08	&	\nodata	&	1.08	&	4.11	\\
	&	G198.69-09.12S	&	Class I	&	2.42	&	9.44	&	1.53	&	6.57	&	7.17	&	5.79	&	1.66	&	4.03	&	10.68	\\
	&	G200.34-10.97N	&	Class 0	&	0.68	&	3.93	&	1.39	&	9.13	&	6.90	&	3.36	&	0.71	&	1.81	&	4.06	\\
	&	G200.34-10.97S	&	flat 	&	1.49	&	3.05	&	\nodata	&	13.10	&	6.63	&	1.27	&	0.67	&	1.86	&	4.20	\\
OrionA	&	G207.36-19.82N1	&	\nodata	&	1.24	&	19.87	&	\nodata	&	4.75	&	7.04	&	7.28	&	\nodata	&	4.24	&	10.30	\\
	&	G207.36-19.82N2	&	\nodata	&	1.10	&	23.84	&	2.08	&	5.01	&	5.34	&	6.99	&	1.98	&	4.69	&	8.85	\\
	&	G207.36-19.82N3	&	\nodata	&	1.02	&	10.44	&	1.49	&	5.99	&	4.79	&	6.77	&	7.26	&	3.00	&	8.19	\\
	&	G207.36-19.82N4	&	\nodata	&	1.09	&	10.84	&	2.06	&	7.47	&	6.52	&	8.31	&	0.38	&	2.29	&	6.52	\\
	&	G208.68-19.20N1	&	Class 0	&	3.15	&	165.17	&	6.15	&	16.44	&	27.30	&	28.75	&	11.56	&	28.40	&	50.17	\\
	&	G208.68-19.20N2	&	\nodata	&	7.20	&	53.55	&	7.08	&	16.74	&	25.87	&	23.22	&	9.03	&	23.57	&	51.76	\\
	&	G208.68-19.20N3	&	\nodata	&	4.81	&	86.57	&	7.83	&	16.44	&	31.09	&	28.40	&	5.56	&	23.99	&	47.32	\\
	&	G208.68-19.20S	&	Class I	&	4.55	&	64.48	&	5.08	&	16.44	&	18.23	&	26.26	&	1.32	&	12.23	&	24.60	\\
	&	G209.55-19.68N1	&	Class 0	&	1.56	&	28.82	&	1.68	&	14.29	&	7.02	&	9.16	&	\nodata	&	10.29	&	14.37	\\
	&	G209.55-19.68N2	&	\nodata	&	1.06	&	8.91	&	0.64	&	15.14	&	\nodata	&	3.56	&	\nodata	&	1.90	&	4.69	\\
	&	G209.55-19.68N3	&	Class I	&	4.08	&	9.12	&	1.50	&	20.38	&	3.91	&	6.22	&	\nodata	&	5.39	&	8.74	\\
	&	G209.55-19.68S1	&	Class I	&	2.99	&	20.36	&	1.56	&	17.31	&	2.10	&	7.27	&	\nodata	&	3.15	&	6.11	\\
	&	G209.55-19.68S2	&	\nodata	&	1.92	&	13.57	&	1.61	&	24.02	&	3.61	&	4.15	&	6.09	&	6.34	&	10.30	\\
	&	G209.77-19.40E1	&	Class I	&	1.54	&	6.65	&	0.64	&	21.95	&	11.94	&	3.41	&	\nodata	&	6.71	&	10.83	\\
	&	G209.77-19.40E2	&	\nodata	&	1.54	&	6.03	&	0.83	&	18.73	&	11.57	&	5.50	&	0.61	&	6.05	&	10.51	\\
	&	G209.77-19.40E3	&	\nodata	&	1.21	&	8.29	&	0.90	&	15.61	&	14.67	&	4.56	&	\nodata	&	4.58	&	9.09	\\
	&	G209.77-19.40W	&	\nodata	&	0.37	&	1.17	&	0.24	&	20.55	&	4.45	&	3.25	&	\nodata	&	1.75	&	3.98	\\
	&	G209.77-19.61E	&	flat	&	0.61	&	5.16	&	0.86	&	21.54	&	4.63	&	3.32	&	\nodata	&	3.27	&	6.84	\\
	&	G209.77-19.61W	&	\nodata	&	1.56	&	9.83	&	0.44	&	16.53	&	\nodata	&	3.26	&	\nodata	&	1.73	&	4.39	\\
	&	G209.79-19.80E	&	Class I	&	3.64	&	2.94	&	0.82	&	13.75	&	4.09	&	2.49	&	\nodata	&	2.25	&	4.95	\\
	&	G209.79-19.80W	&	\nodata	&	3.83	&	10.21	&	2.13	&	12.40	&	5.77	&	7.12	&	2.89	&	7.19	&	10.09	\\
	&	G209.94-19.52N	&	\nodata	&	3.67	&	6.87	&	1.61	&	25.04	&	13.43	&	8.94	&	7.24	&	4.73	&	7.28	\\
	&	G209.94-19.52S1	&	\nodata	&	2.24	&	9.33	&	1.70	&	42.64	&	10.97	&	7.24	&	10.73	&	4.14	&	7.54	\\
	&	G209.94-19.52S2	&	flat	&	1.46	&	18.15	&	1.27	&	22.69	&	13.30	&	7.49	&	\nodata	&	2.08	&	3.77	\\
	&	G210.49-19.79E1	&	flat	&	1.23	&	54.09	&	2.44	&	8.74	&	12.86	&	11.42	&	1.26	&	11.67	&	12.92	\\
	&	G210.49-19.79E2	&	Class I	&	1.44	&	50.46	&	3.19	&	9.71	&	12.35	&	17.04	&	6.52	&	23.40	&	28.41	\\
	&	G210.49-19.79W	&	Class 0	&	2.41	&	26.39	&	2.10	&	8.89	&	13.03	&	10.62	&	0.76	&	20.65	&	24.02	\\
	&	G210.97-19.33N	&	flat	&	\nodata	&	9.13	&	0.44	&	4.19	&	8.78	&	2.64	&	\nodata	&	2.61	&	7.86	\\
	&	G210.97-19.33S2	&	Class 0	&	34.28	&	24.74	&	2.18	&	3.85	&	14.12	&	8.40	&	\nodata	&	5.59	&	11.02	\\
	&	G211.16-19.33N1	&	Class I	&	0.57	&	9.58	&	1.35	&	9.04	&	8.84	&	4.20	&	\nodata	&	2.58	&	6.43	\\
	&	G211.16-19.33N2	&	Class 0	&	1.85	&	5.33	&	1.38	&	10.35	&	4.74	&	3.66	&	\nodata	&	1.98	&	5.61	\\
	&	G211.16-19.33N3	&	\nodata	&	0.92	&	11.69	&	1.33	&	64.10	&	4.77	&	4.25	&	7.07	&	1.47	&	4.51	\\
	&	G211.16-19.33N4	&	\nodata	&	3.15	&	7.22	&	1.22	&	6.07	&	2.44	&	6.48	&	\nodata	&	1.65	&	5.05	\\
	&	G211.16-19.33N5	&	\nodata	&	0.83	&	13.41	&	1.29	&	19.56	&	7.65	&	7.01	&	\nodata	&	2.11	&	6.05	\\
	&	G211.16-19.33S	&	Class I	&	\nodata	&	1.42	&	0.44	&	9.01	&	\nodata	&	4.54	&	\nodata	&	1.29	&	4.72	\\
	&	G215.44-16.38	&	\nodata	&	1.71	&	6.56	&	0.28	&	2.67	&	4.03	&	0.95	&	\nodata	&	0.69	&	2.70	\\
	&	G215.87-17.62N	&	\nodata	&	3.63	&	7.15	&	0.70	&	3.96	&	2.27	&	2.09	&	\nodata	&	3.01	&	5.98	\\
	&	G215.87-17.62M	&	\nodata	&	\nodata	&	1.30	&	0.70	&	7.22	&	0.85	&	4.21	&	\nodata	&	2.29	&	6.61	\\
	&	G215.87-17.62S	&	\nodata	&	\nodata	&	0.93	&	0.22	&	3.96	&	\nodata	&	0.65	&	\nodata	&	2.21	&	5.63	\\
OrionB	&	G201.52-11.08	&	Class 0	&	0.44	&	1.44	&	\nodata	&	8.23	&	2.15	&	0.28	&	\nodata	&	0.93	&	1.33	\\
	&	G201.72-11.22		&	\nodata	&	3.54	&	0.47	&	\nodata	&	19.73	&	3.10	&	3.34	&	0.92	&	1.15	&	2.81	\\
	&	G203.21-11.20E1	&	\nodata	&	1.00	&	0.42	&	\nodata	&	3.78	&	5.48	&	4.07	&	5.02	&	5.25	&	14.56	\\
	&	G203.21-11.20E2	&	\nodata	&	2.25	&	0.87	&	2.24	&	3.92	&	5.26	&	7.87	&	1.85	&	4.47	&	11.94	\\
	&	G203.21-11.20W1	&	\nodata	&	3.03	&	0.97	&	1.80	&	\nodata	&	6.43	&	2.02	&	5.07	&	4.41	&	13.99	\\
	&	G203.21-11.20W2	&	\nodata	&	1.04	&	0.91	&	1.52	&	\nodata	&	6.12	&	3.05	&	2.23	&	3.87	&	11.95	\\
	&	G204.4-11.3A2E	&	\nodata	&	1.15	&	5.39	&	1.19	&	37.32	&	10.45	&	9.64	&	4.16	&	7.22	&	18.17	\\
	&	G204.4-11.3A2W	&	\nodata	&	1.29	&	9.75	&	1.23	&	13.80	&	9.22	&	5.89	&	6.11	&	5.19	&	12.71	\\
	&	G205.46-14.56N2	&	Class I	&	1.15	&	46.42	&	2.68	&	6.01	&	16.45	&	18.64	&	\nodata	&	9.32	&	22.68	\\
	&	G205.46-14.56N3	&	Class I	&	1.56	&	48.24	&	2.43	&	6.24	&	13.03	&	13.97	&	2.28	&	14.16	&	30.84	\\
	&	G205.46-14.56M1	&	Class 0	&	2.39	&	45.03	&	2.42	&	6.47	&	7.89	&	10.60	&	\nodata	&	5.44	&	13.93	\\
	&	G205.46-14.56S1	&	Class 0	&	1.21	&	44.46	&	2.46	&	6.72	&	12.65	&	20.45	&	\nodata	&	9.74	&	27.21	\\
	&	G205.46-14.56S2	&	flat	&	1.54	&	39.49	&	1.65	&	7.82	&	13.08	&	12.02	&	\nodata	&	4.28	&	11.41	\\
	&	G205.46-14.56S3	&	Class I	&	0.59	&	13.79	&	1.03	&	8.26	&	8.31	&	9.59	&	3.53	&	3.68	&	10.43	\\
	&	G206.12-15.76	&	Class 0	&	0.81	&	1.85	&	0.84	&	11.21	&	4.45	&	7.29	&	1.09	&	5.93	&	14.40	\\
	&	G206.21-16.17N	&	\nodata	&	0.59	&	12.70	&	1.52	&	48.36	&	3.92	&	3.30	&	1.27	&	3.27	&	9.02	\\
	&	G206.21-16.17S	&	\nodata	&	0.92	&	9.60	&	0.34	&	46.25	&	2.40	&	1.70	&	0.96	&	2.58	&	7.60	\\
	&	G206.69-16.60N	&	\nodata	&	1.53	&	1.91	&	0.26	&	109.46	&	3.32	&	1.69	&	1.81	&	0.20	&	5.58	\\
	&	G206.69-16.60S	&	\nodata	&	2.24	&	16.85	&	1.71	&	59.12	&	4.88	&	8.08	&	4.22	&	3.25	&	7.57	\\
	&	G206.93-16.61E1	&	\nodata	&	1.69	&	108.33	&	5.24	&	39.85	&	13.04	&	12.75	&	1.27	&	10.09	&	26.10	\\
	&	G206.93-16.61E2	&	Class I	&	2.16	&	70.96	&	2.61	&	39.85	&	11.03	&	13.55	&	1.49	&	9.99	&	26.50	\\
	&	G206.93-16.61W1	&	\nodata	&	3.51	&	52.47	&	3.68	&	43.25	&	11.17	&	9.47	&	3.12	&	4.18	&	11.33	\\
	&	G206.93-16.61W3	&	Class 0	&	2.80	&	23.74	&	3.61	&	48.19	&	20.33	&	15.62	&	3.45	&	15.91	&	34.03	\\
	&	G206.93-16.61W4	&	\nodata	&	4.23	&	21.03	&	3.23	&	59.95	&	13.96	&	8.54	&	3.29	&	4.72	&	11.95	\\
	&	G206.93-16.61W5	&	\nodata	&	3.35	&	22.91	&	3.84	&	48.86	&	8.09	&	10.31	&	2.84	&	4.40	&	11.46	\\
	&	G206.93-16.61W6	&	Class 0	&	\nodata	&	22.91	&	4.08	&	39.85	&	10.14	&	15.24	&	\nodata	&	4.75	&	12.92	\\
\enddata
\tablenotetext{a}{ The 850 \micron\ dust continuum maps were smoothed to calculate the abundance ratios of \nthp, \hcop, \htcop, HCN, and \cth\ molecules.}
\tablenotetext{b}{ The 850 \micron\ dust continuum maps were smoothed to calculate the abundance ratios of \htco\ and HDCO molecules.}
\end{deluxetable*}
\end{longrotatetable}

\end{document}